%% file: c3vo_vmc.tex
\documentclass[twocolumn]{aastex631}

\usepackage{natbib}
\usepackage{amsmath, amsfonts}
\usepackage{amssymb}
\usepackage{gensymb}

\newcommand{\simgt}{\lower.5ex\hbox{$\; \buildrel > \over \sim \;$}}
\newcommand{\simlt}{\lower.5ex\hbox{$\; \buildrel < \over \sim \;$}}
\def\plotfive#1#2#3#4#5{\centering \leavevmode
    \includegraphics[angle=0,width=0.87\columnwidth]{#1} \hfil
    \includegraphics[angle=0,width=0.87\columnwidth]{#2} \hfil
    \includegraphics[angle=0,width=0.87\columnwidth]{#3} \hfil
    \includegraphics[angle=0,width=0.87\columnwidth]{#4} \hfil
    \includegraphics[angle=0,width=0.87\columnwidth]{#5} }

\def\plottwospecial#1#2{\centering \leavevmode
    \includegraphics[angle=0,width=1.03\columnwidth]{#1} \hfil
    \includegraphics[angle=0,width=0.96\columnwidth]{#2} }

\begin{document}

\title{Discovering Large-Scale Structure at $2<z<5$ in the C3VO Survey}

\author[0000-0001-7523-140X]{Denise Hung}
\affiliation{Institute for Astronomy, University of Hawai'i, 2680 Woodlawn Drive, Honolulu, HI 96822, USA}
\affiliation{Gemini Observatory, NSF NOIRLab, 670 N. A'ohoku Place, Hilo, HI, 96720, USA}

\author[0000-0002-1428-7036]{Brian C. Lemaux}
\affiliation{Gemini Observatory, NSF NOIRLab, 670 N. A'ohoku Place, Hilo, HI, 96720, USA}
\affiliation{Department of Physics \& Astronomy, University of California, Davis, One Shields Avenue, Davis, CA 95616, USA}

\author[0000-0002-9336-7551]{Olga Cucciati}
\affiliation{INAF---Osservatorio di Astrofisica e Scienza dello Spazio di Bologna, via Gobetti 93/3, I-40129 Bologna, Italy}

\author[0000-0001-6003-0541]{Ben Forrest}
\affiliation{Department of Physics \& Astronomy, University of California, Davis, One Shields Avenue, Davis, CA 95616, USA}

\author[0000-0001-7811-9042]{Ekta A. Shah}
\affiliation{Department of Physics \& Astronomy, University of California, Davis, One Shields Avenue, Davis, CA 95616, USA}

\author[0000-0001-8255-6560]{Roy R. Gal}
\affiliation{Institute for Astronomy, University of Hawai'i, 2680 Woodlawn Drive, Honolulu, HI 96822, USA}

\author[0009-0003-2158-1246]{Finn Giddings}
\affiliation{Institute for Astronomy, University of Hawai'i, 2680 Woodlawn Drive, Honolulu, HI 96822, USA}

\author[0000-0001-5796-2807]{Derek Sikorski} 
\affiliation{Institute for Astronomy, University of Hawai'i, 2680 Woodlawn Drive, Honolulu, HI 96822, USA}

\author[0000-0001-5160-6713]{Emmet Golden-Marx}
\affiliation{Department of Astronomy, Tsinghua University, Beijing 100084, People's Republic of China}
\affiliation{INAF---Osservatorio astronomico di Padova, Vicolo Osservatorio 5, 35122 Padova, Italy}

\author[0000-0003-2119-8151]{Lori M. Lubin}
\affiliation{Department of Physics \& Astronomy, University of California, Davis, One Shields Avenue, Davis, CA 95616, USA}

\author[0000-0001-6145-5090]{Nimish Hathi}
\affiliation{Space Telescope Science Institute, Baltimore, MD 21218, USA}

\author[0000-0002-2318-301X]{Giovanni Zamorani}
\affiliation{INAF---Osservatorio di Astrofisica e Scienza dello Spazio di Bologna, via Gobetti 93/3, I-40129 Bologna, Italy}

\author[0000-0001-9495-7759]{Lu Shen}
\affiliation{Department of Physics and Astronomy, Texas A\&M University, College Station, TX 77843, USA}
\affiliation{George P. and Cynthia Woods Mitchell Institute for Fundamental Physics and Astronomy, Texas A\&M University, College Station, TX 77843, USA}

\author[0000-0002-8900-0298]{Sandro Bardelli}
\affiliation{INAF---Osservatorio di Astrofisica e Scienza dello Spazio di Bologna, via Gobetti 93/3, I-40129 Bologna, Italy}

\author[0000-0001-5760-089X]{Letizia P. Cassar{\`a}}
\affiliation{INAF-IASF Milano, Via Alfonso Corti 12, 20133 Milano, Italy}

\author[0000-0002-6220-9104]{Gabriella De Lucia}
\affiliation{INAF---Astronomical Observatory of Trieste, via G.B. Tiepolo 11, I-34143 Trieste, Italy}
\affiliation{IFPU---Institute for Fundamental Physics of the Universe, via Beirut 2, 34151, Trieste, Italy}

\author[0000-0003-4744-0188]{Fabio Fontanot}
\affiliation{INAF---Astronomical Observatory of Trieste, via G.B. Tiepolo 11, I-34143 Trieste, Italy}
\affiliation{IFPU---Institute for Fundamental Physics of the Universe, via Beirut 2, 34151, Trieste, Italy}

\author[0000-0001-7455-8750]{Bianca Garilli}
\altaffiliation{Author is deceased.}
\affiliation{INAF-IASF Milano, Via Alfonso Corti 12, 20133 Milano, Italy}

\author[0000-0002-4902-0075]{Lucia Guaita}
\affiliation{Departamento de Ciencias Fisicas, Universidad Andres Bello, Fernandez Concha 700, Las Condes, Santiago, Chile}

\author[0000-0002-3301-3321]{Michaela Monika Hirschmann}
\affiliation{Institute for Physics, Laboratory for Galaxy Evolution, EPFL, Observatoire de Sauverny, Chemin Pegasi 51, 1290 Versoix, Switzerland}
\affiliation{INAF---Astronomical Observatory of Trieste, via G.B. Tiepolo 11, I-34143 Trieste, Italy}

\author[0000-0003-3004-9596]{Kyoung-Soo Lee}
\affiliation{Department of Physics and Astronomy, Purdue University, 525 Northwestern Avenue, West Lafayette, IN 47907, USA}

\author[0000-0001-7769-8660]{Andrew B. Newman}
\affiliation{Carnegie Science Observatories, 813 Santa Barbara Street, Pasadena, CA 91101, USA}

\author[0000-0002-9176-7252]{Vandana Ramakrishnan}
\affiliation{Department of Physics and Astronomy, Purdue University, 525 Northwestern Avenue, West Lafayette, IN 47907, USA}

\author[0000-0003-0898-2216]{Daniela Vergani}
\affiliation{INAF---Osservatorio di Astrofisica e Scienza dello Spazio di Bologna, via Gobetti 93/3, I-40129 Bologna, Italy}

\author[0000-0003-3864-068X]{Lizhi Xie}
\affiliation{Tianjin Normal University, Binshuixidao 393, 300387, Tianjin, People's Republic of China}

\author[0000-0002-5845-8132]{Elena Zucca}
\affiliation{INAF---Osservatorio di Astrofisica e Scienza dello Spazio di Bologna, via Gobetti 93/3, I-40129 Bologna, Italy}

\shorttitle{C3VO Large-Scale Structure}
\shortauthors{Hung et al.}

\begin{abstract}
The Charting Cluster Construction with VUDS and ORELSE (C3VO) survey is an ongoing imaging and spectroscopic campaign aiming to map out the growth of structure up to $z\sim5$ and was born from the combination of the Visible Multi-Object Spectrograph Ultra Deep Survey and the Observations of Redshift Evolution in Large-Scale Environments (ORELSE) survey. As we previously accomplished with the ORELSE survey, we apply our technique known as Voronoi tessellation Monte Carlo (VMC) mapping to search for serendipitous galaxy overdensities at $2<z<5$ in the three C3VO fields. We also apply the same technique to mock observations of simulated galaxies with properties derived from the GAlaxy Evolution and Assembly semianalytic model in order to judge the effectiveness of our search algorithm as a function of redshift, total mass, and fraction of spectroscopic redshifts. We find completeness and purity values of the order of 30-50\% for $\log (M_{z=0}/M_{\odot}) > 14$ and $2<z<4$, with a strong dependence on mass and redshift, with values as high as $\sim$80\% and $\sim$70\%, respectively, in the best-case scenario for $\log (M_{z=0}/M_{\odot}) > 14.5$. In the C3VO fields, we were able to recover many of the previously known structures in the literature as well as find hundreds of new overdensity candidates, once again demonstrating the powerful capabilities of VMC mapping when applied to wide-field optical and infrared galaxy evolution surveys at ever higher redshifts.
\end{abstract}

\section{Introduction}

Galaxy groups and clusters are among the largest gravitationally bound structures in the Universe. These large-scale structures (LSSs) at different redshifts provide a means of studying cosmology as well as the effects of environment on galaxy evolution at various stages in time. Protoclusters are the progenitors of galaxy clusters, which, at some point, will gravitationally collapse and become virialized. Clusters may be distinguished from field galaxies through a number of methods, such as the identification of a well-defined red sequence or a hot intracluster medium (ICM). Red sequence techniques to find clusters were first employed by the Red Sequence Cluster Survey \citep[RCS;][]{Gladders00, Gladders05}. The RCS made use of two filters, one in the optical and one in the near-infrared (NIR), to identify galaxy overdensities by features such as Balmer absorption or the 4000\AA\ break, which would be present in the filter wavelengths out to redshifts of $z\sim1$. With the addition of mid-infrared filters, red sequence searches can be extended up $z\sim2$ \citep[e.g.,][]{Andreon09, Muzzin09, Wilson09, Galametz12, Cooke15, Noirot16, Golden-Marx19}. 

Various cluster surveys at intermediate redshift, such as the Observations of Redshift Evolution in Large-Scale Environments \citep[ORELSE;][]{Lubin09} survey, Gemini Cluster Astrophysics Spectroscopic Survey \citep[GCLASS;][]{Muzzin12}, and the Gemini Observations of Galaxies in Rich Early ENvironments \citep[GOGREEN;][]{vanderBurg20} survey, as well as earlier cluster studies \citep[e.g.,][]{Aragon93, Kodama98, Lubin98, Stanford98}, have demonstrated how the scaling relations between color, density, morphology, and the star formation rate (SFR) change up to $z\sim1$ relative to the local Universe. Studies from these surveys have shown that quenching is likely driven by internal mass-based mechanisms as well as processes that are at play in overdense environments. There are claims that the effects of these processes are largely separable for galaxy samples at $z<1$ \citep[e.g.,][]{Peng10, Kovac14, Guglielmo15}, but some studies argue otherwise \citep[e.g.,][]{DeLucia12}. The scaling relations weaken or reverse as redshift increases, and the two modes of quenching may also no longer act fully independently at $z>1$ \citep[e.g.,][]{Balogh16, Kawinwanichakij17, Pintos-Castro19}. We therefore need to study higher-redshift protoclusters to understand how galaxies evolve into their $z\sim1$ or $z\sim0$ counterparts. 

However, even if it is possible to find such structures, it is difficult to identify which protoclusters are likely to evolve into typical lower-redshift clusters. The challenges come from both the sensitivity of cosmological simulations such as \cite{Chiang13, Chiang15}, \cite{Remus23}, and \cite{Lim24} to different stellar properties as well as the failures to reproduce the high masses and SFR observed in protocluster galaxies. It is thus imperative to ensure that overdensity characterization is as accurate as possible and the properties of their constituent galaxy populations are well constrained, which requires high-quality data and methods to obtain accurate measures of redshift, overdensity strengths, and galaxy properties.

Because protoclusters exist primarily at higher redshifts ($z\ga2$), even detecting structure with ground-based telescopes, let alone characterizing a system's mass, dynamical state, or member population, is highly challenging. In contrast, there have been many successes in finding and characterizing clusters at lower redshifts ($0<z<1.5$) from a number of different methodologies. Some notable examples among the highest-yielding surveys include the thousands or more candidate clusters found at optical and NIR wavelengths \citep[e.g.,][]{Gonzalez19, Wen22}, at X-ray wavelengths \citep[e.g.,][]{Ebeling01, Piffaretti11}, and at radio and submillimeter wavelengths by tracing the thermal Sunyaev-Zel'dovich \citep[SZ;][]{sz72} effect \citep[e.g.,][]{Planck16, Hilton21, Polletta21, Polletta22}. Lensing surveys have also seen some success, albeit finding far fewer cluster candidates relative to other methodologies \citep[e.g.,][]{Kubo09, Ford14}.

The limiting magnitude required to pick up any meaningful number of galaxies at $z>2$ is out of reach for many optical and NIR surveys (e.g., a limiting magnitude of $m_{AB} < 25$ in the $i$ band probes $\ga$0.3$L_{*}$ over $2<z<5$). This difficulty is compounded for X-ray surveys, as X-ray surface brightness falls off as $\propto(1+z)^{-4}$. Additionally, X-ray and radio/submillimeter surveys (surveys which rely on the identification of signatures from the ICM) are also less effective due to the decreased time that such processes are able to act on the member galaxies in order to build up a hot ICM. As a consequence, the ICM has only been detected in very few $z>2$ overdensities relative to the number of searches \citep[e.g.,][]{Wang16, Tozzi22, DiMascolo23}. Lensing surveys are susceptible to projection effects complicating the accuracy of mass determination. Red sequence identification is also mostly ineffective due to protocluster galaxies being predominantly blue \citep{Overzier16}. Of these shortcomings, the most feasible to overcome is in the optical and NIR, where the extreme magnitudes ($i_{AB}\la25$) needed to spectroscopically probe the member population of high-redshift protoclusters over a large area (i.e., tens of arcminutes on the sky) are within reach for the deepest ground- and space-based galaxy surveys. 

Although the number of known protoclusters at $z>2$ has increased over the past decade due to a number of systematic searches of deep observations \citep[e.g.,][]{Diener13, Chiang14, Wylezalek13, Wylezalek14, FrancknMcGaugh16, Lee16, Toshikawa16, Toshikawa18, Newman20, Lee24, Ramakrishnan24}, discovery and characterization have primarily been limited to individual spectacular structures, with some notable examples being the SSA22 protocluster at $z\sim3.1$ \citep{Chapman01}, the Spiderweb protocluster at $z\sim2$ \citep{Pentericci97, Pentericci00, Miley06, Helmut14}, the SPT2349-56 protocluster at $z\sim4.3$ \citep{Miller18, Hill22}, and the Hyperion proto-supercluster at $z\sim2.5$ (\citealt{Cucciati18}, Forrest et al. 2025, in preparation). The sensitivity offered by the James Webb Space Telescope (JWST) has also enabled protocluster searches beyond $z>5$, with protocluster candidates being found at redshifts up to $z\sim7.7$ \citep[e.g.,][]{Laporte22, Helton24, Li24}. The focus on a relatively small number of massive protoclusters, however, means that the overall data set is inhomogeneous and thus ill-suited for any broad population analyses, which limits our understanding of the environments and how they affect galaxy evolution at such high redshifts.

In this paper, we perform a search of protoclusters over 2.4 deg$^2$ in three well-studied extragalactic fields (COSMOS, ECDFS, and CFHTLS-D1). Our search is carried out using a version of Voronoi tessellation Monte Carlo \citep[VMC;][]{Lemaux17, Tomczak17, Lemaux22} mapping with limited assumptions on the underlying galaxy populations, using as seeds exquisite panchromatic imaging data as well as tens of thousands of spectroscopic redshifts from a variety of archival data sources both public and proprietary. This method has been very successful for systematically finding and characterizing hundreds of new overdensity candidates over $\sim$1.4 deg$^2$ between $0.55<z<1.37$ with the ORELSE survey \citep{Hung20}, including finding clusters of sufficient masses to offer cosmological constraints with only optical and NIR data \citep{Hung21}. Our past successes have shown that, with considerable levels of spectroscopy (see \citealt{Hung20} for a through breakdown per field), we are able to effectively search for galaxy-traced high-redshift protoclusters over a large range of masses and dynamical states.  

Although VMC mapping has also been used to find several massive higher-redshift structures in the Visible Multi-Object Spectrograph (VIMOS) Ultra Deep Survey \citep[e.g.,][]{Lemaux14b, Lemaux18, Cucciati14, Cucciati18, Shen21, Forrest23, Shah24, Staab24}, no systematic search has yet been performed within these fields. Such is the goal of this work. Armed with tens of thousands of photometric and spectroscopic redshifts, we aim to search for a large ensemble of galaxy-traced protostructures essentially free from any other bias. In this work, we further build on the methodology of \citet{Hung20} by incorporating a custom-built lightcone with a GAlaxy Evolution and Assembly \citep[GAEA;][]{DeLucia2024} semianalytics model to estimate the completeness, purity, and masses of the protostructure candidates we find.

This paper is organized as follows. The extensive imaging and spectroscopic data set used is described in Section \ref{data}. The methodology regarding how we find protocluster candidates is described in Section \ref{methodology}. We describe in Section \ref{mocks} how we use simulated data drawn from a custom-built lightcone to quantify the effectiveness of our method at finding and characterizing protocluster candidates. We go over the full catalog of candidates in Section \ref{discussion} as well as draw comparisons with the previously known structures in the literature in the fields we studied. Finally, a summary of our findings is given in Section \ref{conclusion} along with some possible future avenues on expanding upon this work. 

Unless otherwise specified, we adopt a flat cold dark matter ($\Lambda$CDM) cosmology throughout this paper, with $H_{0}$ = 70 km s$^{-1}$ Mpc$^{-1}$, $\Omega_{m}$ = 0.27, and $\Omega_{\Lambda}$ = 0.73, and reported distances are generally given in proper units (e.g., $h_{70}^{-1}$ proper kiloparsecs or megaparsecs). In some cases, where explicitly noted, distances are reported in comoving units but still follow the convention of $h_{70}^{-1}$ comoving kiloparsecs or megaparsecs.

\section{Data}\label{data}

The VIMOS Ultra Deep Survey \citep[VUDS;][]{LeFevre15, Tasca17} was a spectroscopic redshift survey of roughly 10,000 galaxies at $2<z<6$ with the goal of studying galaxy assembly. The ORELSE \citep{Lubin09} survey was a multiwavelength photometric and spectroscopic campaign designed to map out LSSs over $0.6<z<1.3$. These two surveys were combined into the Charting Cluster Construction with VUDS and ORELSE \citep[C3VO\footnote{\url{https://www.orelsesurvey.com/c3vo.html}};][]{Shen21, Lemaux22} survey, which seeks to map out the growth of structure at $0.5<z<5$. The higher-redshift end ($2 \leq z \leq 5$) is an ongoing campaign to obtain additional visible and NIR wavelength photometry and spectroscopy of three well-studied extragalactic fields: the Cosmic Evolution Survey \citep[COSMOS;][]{Scoville07} field, the Extended Chandra Deep Field South \citep[ECDFS;][]{Lehmer05}, and the first field of the Canada--France--Hawai’i Telescope Legacy Survey (CFHTLS-D1\footnote{\url{https://www.cfht.hawaii.edu/Science/CFHTLS/}}). 

In addition to the data obtained with C3VO, the data used in this work also include the extensive imaging and spectroscopic data sets of these fields from various publicly available sources. We briefly describe the basic properties of these data in this Section, but further details regarding the observations, reduction, source detection, magnitude measurements, and various validation methods can be found in \citet{Lemaux14, Lemaux14b}, the references therein, and the references cited below.

\subsection{Archival Data}
\subsubsection{Imaging Data and Photometry}
\label{sec.photdata}
The photometry in the fields at wavelengths relevant to this work generally span 10 or more bands ranging from the ultraviolet (UV) to the NIR. Many bands are extremely deep, reaching magnitudes of $m_{AB} \sim24--27$ for completeness limits of 5$\sigma$. We provide a brief summary of the archival data used in this work below, but we refer the reader to \citet{Laigle16}, \citet{Cardamone10}, and \citet{Lemaux14b} and the references therein for more details on the COSMOS, ECDFS, and CFHTLS-D1 fields, respectively.

In the COSMOS field, we utilize the ``COSMOS2015'' catalog compiled by \citet{Laigle16} as our basis catalog. The COSMOS2015 catalog consists of near-UV imaging from the GALaxy Evolution eXplorer \citep[GALEX;][]{Martin05}, ground-based $u^{\ast}BVri^{+}z^{++}$ UV/optical imaging from MegaCam on the Canada--France--Hawai'i Telescope \citep[CFHT;][]{Boulade03} and Subaru/Suprime-Cam \citep{Miyazaki02}, ground-based $YJHK$ imaging from Subaru/Hyper-Suprime-Cam  \citep{Miyazaki12}, the UltraVISTA survey \citep{McCracken12}, and CFHT/WIRCam \citep{Puget04, McCracken10}, as well as 3.6--8.0 $\mu$m imaging from Spitzer with the InfraRed Array Camera \citep[IRAC;][]{Fazio04}. The catalog contains photometric redshifts for over half a million objects over an area of 2 deg$^2$. We used v2.0 of the \citet{Capak07} catalog as an additional source for photometric redshifts that were absent in the COSMOS2015 catalog. For the objects shared between the two catalogs, no systematic offset has been observed between the derived parameters for the two different sets of photometry \citep{Lemaux22}. 

Note that while a newer version of the COSMOS photometric catalog exists, namely the COSMOS2020 catalog \citep{Weaver22}, this catalog did not exist at the inception of this study, and, thus, all mock observations and spectroscopic selection functions used to train and characterize the search algorithm in this study were based on the COSMOS2015 catalog. However, all data in this study are cut at an IRAC channel 1 (hereafter IRAC1) magnitude of IRAC1$<$24.8, a limit to which both the COSMOS2015 and COSMOS2020 catalogs are complete. While there are likely higher-order improvements that would come as a result of instead adopting the COSMOS2020 catalog as our base photometric catalog for COSMOS, we do not expect its usage to alter our findings. Moreover, given that the COSMOS2015 catalog is complete to the depth adopted for this study and the methodological approach used to generate the COSMOS2015 catalog is similar to that of the Classic/LE PHARE COSMOS2020 catalog, we chose to keep the COSMOS2015 catalog as our basis for studies in the COSMOS field. 

The photometry in the ECDFS field is sourced from optical Subaru/Suprime-Cam imaging over 18 medium bands \citep{Cardamone10}, taken as a part of the Multiwavelength Survey by Yale-Chile \citep[MUSYC;][]{Gawiser06}. The field also includes ancillary data by way of $UBVRI$ imaging from the Garching-Bonn Deep Survey \citep[GaBoDS;][]{Hildebrandt06} and $zJHK$ imaging from earlier MUSYC observations. Imaging in each of the four Spitzer bands from 3.6--8.0 $\mu$m is also included from the Spitzer IRAC/MUSYC Public Legacy in the ECDF-S \citep[SIMPLE;][]{Damen11} survey. These data overall include over 40,000 photometric redshifts over a 0.25 deg$^2$ area. We note that while this field also has observations taken with the Hubble Space Telescope (HST) for the Cosmic Assembly Near-infrared Deep Extragalactic Legacy Survey \citep[CANDELS;][]{Grogin11, Koekemoer11}, the area observed only intersects a small section of the spectroscopic coverage.

The CFHTLS-D1 field's UV/optical/NIR imaging data is described extensively in \citet{Lemaux14, Lemaux14b}. As part of CFHTLS, the data include ground-based $u^{\ast}g^{\prime}r^{\prime}i^{\prime}z^{\prime}$ imaging with CFHT/MegaCam. In NIR, $JHK_{s}$ imaging was taken with CFHT/WIRCam for the WIRCam Deep Survey \citep[WIDS;][]{Bielby12} and at 3.6 and 4.5 $\mu$m with Spitzer/IRAC for the Spitzer Extragalactic Representative Volume Survey \citep{Mauduit12}. The field includes over 150,000 photometric redshifts over a 1 deg$^2$ area.

\subsubsection{Spectroscopic Data}
\label{sec.specdata}
The spectroscopic data in our work is discussed in depth in \citet{Lemaux22}. We summarize the salient details below.

For our spectroscopic redshifts drawn from other surveys, we only adopted values that were denoted as secure by their reporting survey for our redshift range of interest ($2<z<5$). Over half (50.5\%) of these spectroscopic redshifts originated from VUDS \citep{LeFevre03}, which primarily used photometric redshifts in the catalogs described above to select targets for spectroscopic follow-up with the VIMOS on the Very Large Telescope (VLT). Details regarding the survey design, observations, data reduction, and redshift determination can be found in \citet{LeFevre15}, and a full description of the first VUDS data release can be found in \citet{Tasca17}. We consider galaxies with flags X2, X3, X4, and X9 as secure, where X may equal 0, 1, 2, or 3, according to the flagging code employed in \citet{LeFevre15}. Such flags indicate a redshift that is reliable at the 75--99.3\% level.

The next largest source (24.5\%) of our spectroscopic redshift sample came from the Bright and Deep phases of the zCOSMOS survey \citep{Lilly07, Lilly09, Diener13, Diener15}. Other spectroscopic redshifts in the COSMOS field were taken from \cite{Casey15}, \cite{Chiang15}, and \cite{Diener15} at $z\sim2.5$. The 3D-HST, VANDELS, VVDS, and ACES surveys (including \citealt{LeFevre04, LeFevre13, Vanzella09, Hathi09, Straughn09, Balestra10, Cooper12, Kurk13, Trump13, Morris15, McLure18, Pentericci18, Talia23} among others) that cover the ECDFS field make up 17.0\% of the spectroscopic redshift sample (\emph{N. Hathi 2025, private communication}). The field galaxies in the CFHTLS-D1 field are sourced from the Deep and UltraDeep phases of the VIMOS VLT Deep Survey \citep[VVDS;][]{LeFevre05, LeFevre13}, which make up 5.5\% of the spectroscopic redshift sample. Although each spectroscopic survey has its own unique flagging system, we chose galaxies with flags as close as possible to what would be considered secure in VUDS. If a galaxy appeared in multiple surveys, we adopted the redshift with the highest confidence flag, with the VUDS redshift preferred in the case of a tie. 

\subsection{C3VO Observations}\label{sec.C3VOobs}

The remainder of the redshifts were drawn from dedicated C3VO spectroscopy, which was obtained with the Multi-Object Spectrometer For Infra-Red Exploration \citep[MOSFIRE;][]{McLean12} and the DEep Imaging Multi-Object Spectrograph \citep[DEIMOS;][]{Faber03} on the Keck I/II telescopes. The C3VO survey's Keck observations were designed with the goal of a near complete magnitude-limited survey to $i^{\prime}<25.3$ of the six most significant overdensities found in VUDS, including those reported in \cite{Lemaux14, Lemaux18}, \cite{Cucciati14, Cucciati18}, \cite{Shen21}, \cite{Forrest23}, \cite{Shah24}, and \cite{Staab24}. The observations targeted star-forming galaxies of all types to a magnitude of $i_{AB} < 25.3$, magnitudes at which we could hope to recover continuum redshifts, and fainter sources ($25.3 \le i_{AB} < 26.5$) for which we would only be able to spectroscopically confirm Ly$\alpha$ emitting galaxies. Details of the C3VO observations of all structures can be found in the references given above.

For this work, we do not incorporate all C3VO observations, but rather only those from select overdensities, as our goal is primarily to assess our ability to recover structure from broad-breadth spectroscopic surveys rather than those aimed at dedicated follow-up. For consistency with some of the previous C3VO works, we include the initial DEIMOS follow-up observations of the Taralay protostructure at $z\sim4.57$ in the COSMOS field as described in  \citet{Lemaux22} and the MOSFIRE observations of PClJ0227-0421 at $z\sim3.3$ in the CFHTLS-D1 field used in \citet{Shen21}. Both of these observations add considerably to the spectroscopic sampling at the high-redshift end of their respective fields. Additionally, we incorporate spectroscopic observations of the ECDFS field described in \citet{Shah24}, which includes both C3VO observations and a newly compiled catalog of public redshifts made by one of the coauthors (N.P.H.). As can be seen in Figure 1 of \citet{Shah24}, the former comprise only a small fraction of the redshifts in the field, and are primarily targeted at the \emph{Smruti} protostructure at $z\sim3.5$. The details of these observations and the reduction of these data can be found in the corresponding reference listed above as well as \citet{Forrest23, Forrest24}.

Spectroscopic redshifts from DEIMOS data were determined with the publicly available Deep Evolutionary Extragalactic Probe 2 \citep[DEEP2;][]{Davis03, Newman13} redshift measurement program \texttt{zspec} \citep{Newman13}. Since this program was intended primarily to evaluate the spectral redshift of galaxies at $z<1.4$, empirical high-redshift galaxy templates from VUDS and VVDS were added to the base template set in the program, as were high-resolution empirical Ly$\alpha$ templates from \citet{Lemaux09}. All observed objects targeted or otherwise were visually inspected and assigned a spectroscopic redshift $z_{\rm{spec}}$ and a redshift quality code $Q$. Under the DEEP2 convention, secure stellar ($Q = -1$) and extragalactic ($Q = 3$, 4) redshifts are scientifically usable at the $\geq$95\% confidence level \citep{Newman13}.

MOSFIRE redshifts were initially determined by applying an identical methodology to the one described above for the DEIMOS spectra following a visual inspection for intermediate to high signal-to-noise ratio (S/N) emission features. A second pass redshift estimation was performed after a more comprehensive visual inspection of all two-dimensional MOSFIRE spectra, both to identify lower S/N features and detections of non-targets that serendipitously fell in the MOSFIRE slits. A one-dimensional extraction was performed at the spatial location of emission lines identified during the visual inspection process, and redshifts were determined via model fitting (see \citealt{Forrest23} for more details). Redshift flags for MOSFIRE spectra followed the convention of those employed for the DEIMOS redshift determination. 

In addition to the high-$z$ C3VO redshifts, we also incorporated Keck/DEIMOS observations in the CFHTLS-D1 field of the $z\sim1.05$ cluster XLSS005 from the ORELSE survey (see \citealt{Lemaux19} and references therein for details). This campaign yielded $\sim$750 secure spectroscopic redshifts. While these redshifts are exclusively at $z<2$, they are useful as a means of eliminating lower-redshift interlopers in our galaxy density mapping (see Section \ref{sec.method_VMC}).

Combining all data sets and imposing the NIR apparent magnitude cuts used to define both our photometric and spectroscopic sample, as discussed in Section \ref{methodology}, resulted in a total of 7451 unique galaxies with a high-quality $z_{\rm{spec}}$ in the redshift range $2<z<5$\footnote{For C3VO observations ``high-quality'' spectral redshifts refers only to flags -1, 3, \& 4. The VVDS, zCOSMOS, and our public compilation of redshifts are placed on the VUDS flagging system, where the term high-quality spectral redshifts refers to all subclasses of flags 2, 3, 4, \& 9 (see \citealt{Lemaux22} for more details on the reliability of these flags).}.

\section{Methodology}\label{methodology}

\subsection{Photometric and Spectroscopic Redshifts}

We select the spectroscopic and photometric redshifts in this work (referred to as $z_{\rm{spec}}$ and $z_{\rm{phot}}$, respectively) in an identical manner to the methods described in \citet{Lemaux22}. In summary, the spectroscopic and photometric redshifts used were cut at per-field magnitude limits: IRAC1$<$24.8 for COSMOS, IRAC1$<$24.8 for ECDFS, and IRAC1$<$23.1 or $K_s$$<$24.1 for CFHTLS-D1. We additionally required detection in two NIR bands in the CFHTLS-D1 field due to its relatively shallow IRAC imaging. Our photometric limits correspond to the $3\sigma$ limiting depth of the IRAC/WIRCam images in each of the three fields. With these cuts, the redshift sample is estimated to be 80\% complete to a mass of log$(\mathcal{M}_{*} / \mathcal{M}_{\odot}) \sim$9.2--9.5 across all three fields at $z\sim3$ \citep{Lemaux22}. For details on photometric and spectroscopic measurements, as well as details of the spectral energy distribution (SED) fitting methodologies can be found in \citep{Lemaux22} and references therein.

Applying the photometric cuts detailed above to both the spectroscopic and photometric sample, the global spectroscopic redshift fraction\footnote{The quantity here refers to the fraction of objects brighter than the photometric limit imposed in a given field in the range $2<z_{\rm{phot}}<5$ with a secure spectral redshift. In contrast, the ``spectroscopic redshift fraction,'' which is introduced later in the paper and is abbreviated as ``S$z$F,'' refers to a multiplicative factor of the fraction of galaxies with a secure spectroscopic redshifts in different photo-$z$ and magnitude bins in the COSMOS field, i.e., a scaling up or down of our spectroscopic selection function in the COSMOS field (see Section \ref{sec.lc_mocks}.)} ranges from 0.06--0.2 across the three fields, with CFHTLS-D1 having the sparsest coverage and ECDFS having the densest. For the COSMOS, ECDFS, and CFHTLS-D1 fields, the $z_{\rm{phot}}$ scatter is $\sigma_{NMAD, \Delta z / (1 + z_{\rm{spec}})} =$ 0.013, 0.017, and 0.043, with catastrophic outlier rates of 8.8\%, 12.0\%, and 14\%, and biases (i.e., ($z_{\rm{phot}}$-$z_{\rm{spec}}$)/(1+$z_{\rm{spec}}$)) of -0.0019, 0.0009, and -0.0149, respectively, as measured for galaxies in the redshift range $0\leq z_{\rm{spec}} \leq 6$ subject to the same photometric cuts as above.

\subsection{Voronoi Tessellation Monte Carlo Mapping}\label{sec.method_VMC}

\citet{Hung20} successfully used a powerful technique known as Voronoi tessellation Monte Carlo (VMC) mapping to find hundreds of candidate galaxy clusters and groups over a combined spectroscopic footprint of $\sim$1.4 deg$^2$ in the ORELSE survey at $0.55<z<1.37$. Despite the increase in data quality from the ORELSE survey in terms of overall number of spectroscopic redshifts, as well as the depth and breadth of the imaging data, we expect that the same structure search algorithm will have a more challenging time producing results that are as fruitful at higher redshift. This expectation is due partially to the lower \emph{fraction} of spectroscopy available, as the number counts of galaxies increase precipitously with deeper imaging, as well as the larger photometric redshift errors that accompany higher redshift galaxies. Additionally, the density contrast of forming structures at high redshift is less than that of massive clusters, and more resembles the density contrast of groups (see, e.g., \citealt{Forrest24}). Details on the construction of the VMC maps for both $z\sim1$ and at $2<z<5$ can be found in \citet{Tomczak17, Lemaux17, Lemaux18, Hung20, Lemaux22}, but we will summarize the process here. For this work, we adopt the particular approach of \citet{Lemaux22}.

Voronoi tessellation has broad applications that also include being capable of acting as a type of density field estimator. It defines a polygonal cell for every object in a two-dimensional plane where the area inside the cell is closer to the host object than any other object. The cell size is thus indicative of the density at a given location; the smaller the cell size, the higher the density. The survey data cover a broad range in redshift, and the data are cut into redshift slices, each 7.5 Mpc in width, or $\Delta z \sim$ 0.015--0.08 for $z\sim$2--5. We choose this approach rather than a three-dimensional tessellation primarily because peculiar velocities complicate the interpretation of line-of-sight positions. By breaking the data up into two-dimensional slices of this size, we are able to marginalize over velocity substructure in forming structures.  A tessellation is performed on each slice and the resultant density field is sampled onto a grid of 75 square kpc pixels. In order to minimize the chances of splitting individual structures across slices, we define the slices such that two sequential slices will have 90\% overlap in redshift.

In every redshift slice, there are galaxies with spectroscopic and photometric redshifts. Photometric redshifts have much larger uncertainties than spectroscopic redshifts, and we account for this through a Monte Carlo method. For each Monte Carlo realization, we sample the reconstructed probability density function (PDF) of every $z_{\rm{phot}}$ using an asymmetric Gaussian, which will shift some photometric redshifts in or out of the redshift boundaries of a particular slice and accordingly change the measured density values. Spectroscopic redshifts are also treated statistically according to their confidence flags, with reliability thresholds adopted from those reported in \cite{Lemaux22}. We then perform the Voronoi tessellation on all of the $z_{\rm{spec}}$ and $z_{\rm{phot}}$ objects that fall in a given slice for a given realization. The Gaussian sampling on all photometric redshifts and subsequent Voronoi tessellation is run a total of 100 times for each slice, and the final VMC density map for that slice is obtained by median combining the densities from all realizations. We can then obtain the local overdensity in a pixel ($i,j$) with

\begin{equation} \label{eq.overdense}
\rm{log}(1 + \delta_{\rm{gal}}) = \rm{log}(1+(\Sigma_{i,j}-\tilde{\Sigma})/\tilde{\Sigma})
\end{equation}

where $\delta_{\rm{gal}}$ is the overdensity of galaxies, $\Sigma_{i,j}$ is the given pixel's density, and $\tilde{\Sigma}$ is the median density of all pixels in the central 80\% of the slice. These local overdensities have been shown through tests to correlate well with other density metrics and trace out known structures extremely well \citep[e.g.,][]{Tomczak17, Lemaux19, Hung20}.

\subsection{Detecting Coherent Structure}\label{sec.find_struct} 

\begin{figure}
\includegraphics[width=\columnwidth]{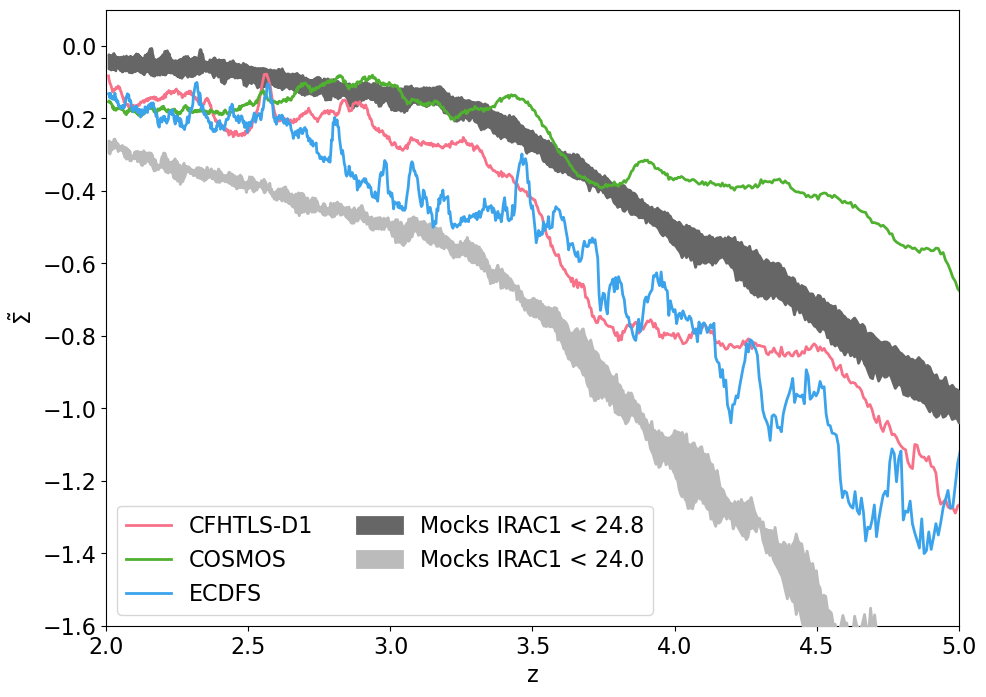}
\caption{Median galaxy (surface) density ($\tilde{\Sigma}$) as a function of redshift for the C3VO and mock fields (see Section \ref{sec.lc_mocks} for a description of the latter) as measured in the individual slices of the VMC maps. The 13 mock fields at IRAC1 $<24.8$ and the three mock fields at IRAC1 $<24.0$ are each depicted as single shaded bands showing the maximum and minimum extent of the median density between all respective mocks at each redshift. While the median galaxy density as a function of redshift in the IRAC1 $<24.8$ mocks appears to generally track that of the real data, the variation among all of the mock fields at a given magnitude cut is much smaller than the variation within each of the C3VO fields. This disparity is mainly due to the higher-order effects in the data that are not encoded in the construction of the mocks. To mitigate this difference, we do not employ a fit of the median density as a function of redshift, and, instead, take the measured median densities at each redshift as a background value (see Section \ref{sec.find_struct}).}
\label{fig:meddens}
\end{figure}

With the VMC maps in hand, in each field over the redshift range $2<z<5$, we apply our structure search algorithm. Significant detections in each slice of the VMC maps, which represent potential signal from a structure, are found using the standard photometry software package Source Extractor \citep[SExtractor;][]{Bertin96}. All slices of the VMC maps in each field are fed into an initial run of SExtractor and the outer 20--30\% of the area of the maps in each redshift slice is masked to excise areas of the map where no data exist and to mitigate edge effects. In this initial run, median density of each slice as measured in the inner 70--80\% of the maps and is used as a background level. This run is also used to set the rms fluctuations in the density field (see below), which, along with the median density, is used to set the detection threshold for structure in the maps. Unlike previous works \citep[e.g.,][]{Hung20}, we do not employ a fit of the median density as a function of redshift to determine the background. This choice was made to mitigate issues with artificial clustering in redshift space of $z_{\rm{phot}}$ that spuriously cause wide peaks or troughs in the density distribution (Fig. \ref{fig:meddens}), which lead to the suppression of of real structure and enhancement of false positives if a fit is instead employed.

\begin{figure}
\includegraphics[width=\columnwidth]{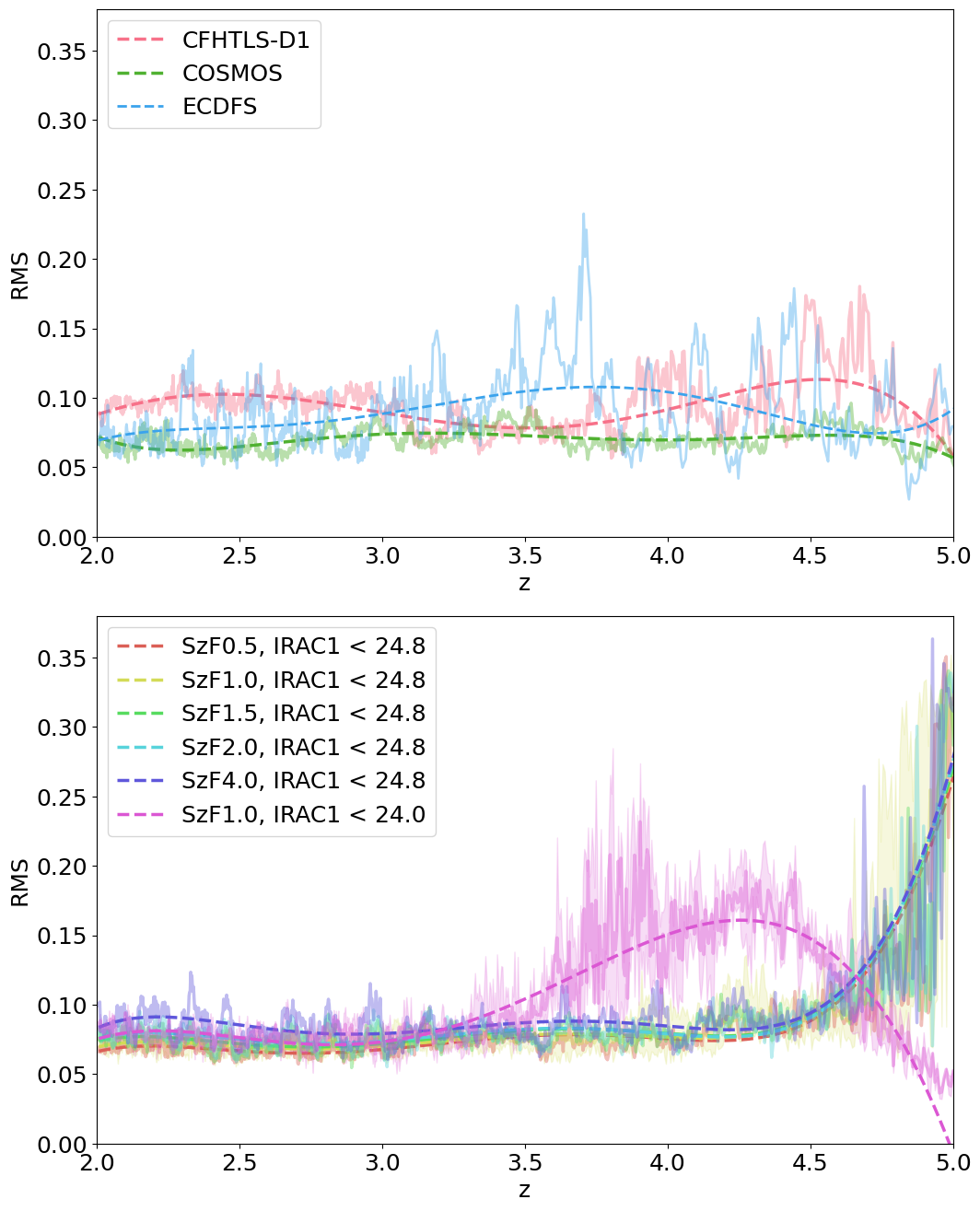}
\caption{The rms as measured by SExtractor for each of C3VO (top) and mock (bottom) fields as a function of redshift. The dashed lines indicate the outlier-clipped fifth-order polynomial fit of the rms as a function of redshift for each field. The two sets of S$z$F1.0 mocks were composed of multiple maps of different regions of the lightcone. For these mocks, the solid lines represent the median rms across all mocks as a function of redshift, with the maximum range of values is shown in the shaded regions. With a few exceptions, the rms of the mocks generally tracks that of the real data extremely well; except for the very highest redshifts, the fitted mock rms is always within 50\% of that of the real data.}
\label{fig:rms_z}
\end{figure}

The rms of the density field, which is also measured in the inner 70-80\% of the maps, varies much more rapidly and is much more sensitive to the presence of real structure (Fig. \ref{fig:rms_z}). As a result, we employ a fifth-order polynomial fit, iteratively clipped at $\pm2.5\sigma$, of the rms as a function of redshift for each field and adopt the fitted values per slice when running SExtractor on the VMC maps. In most cases, the structures only occupy a small fraction of the VMC map's area. The iterative clipping is most beneficial for the ECDFS field, where the structures can be $\sim$30--40\% of the area of the map, leading to increases in the average local density and the rms driven by astrophysical reasons. Note that the rms measured in SExtractor is different than a traditional measure of the rms of all of the pixels in a given slice. Rather, the SExtractor rms is measured on the iteratively 3$\sigma$-clipped distribution of the average density values in regions whose sizes are dictated by the BACK\_SIZE input parameter (64$\times$64 pixels in our case). Once determined for each slice, the median density and rms are then used by SExtractor, along with several other parameters, to detect the signal from potential structure in each density slice. Note that in this final run of SExtractor, we no longer apply masking, but rather identify those candidates in our final protostructure candidate catalog whose overdensity-weighted barycenters lie within 2 Mpc of the masked region in the maps, as their authenticity is likely more suspect and the strength of the detection is possibly affected by their proximity to the edges of the maps.

Four SExtractor parameters in particular are especially pertinent for this work. The DETECT\textunderscore THRESH and DETECT\textunderscore MINAREA parameters define the base thresholds for what constitutes a detection in SExtractor. DETECT\textunderscore THRESH sets how much higher the density floor must be relative to the rms noise in the background. DETECT\textunderscore MINAREA sets the minimum number of contiguous pixels. The detection parameters are best chosen to maximize completeness while also minimizing spurious detections due to noise. Tests with the ORELSE data have suggested that DETECT\textunderscore THRESH values of around 4$\sigma$ is a reasonable choice to balance maximizing the purity and completeness of detections \citep{Hung20}. While our overall spectroscopic catalog is larger with the C3VO data, the global fraction of spectroscopic redshifts is generally lower, primarily due to the extreme faintness of the galaxy population that is adopted for our density mapping (IRAC1 $<$ 24.8). In addition, the density contrast of forming clusters (i.e., protoclusters) is considerably less than the cores of established clusters by an order of magnitude or more even in the core regions (see, e.g., \citealt{Forrest24}). As such, in this work we adopt a less restrictive threshold of 3$\sigma$. Additionally, we adopt a DETECT\textunderscore MINAREA value of 100 pixels. Such an area is equivalent to 0.5625 Mpc$^2$ (corresponding to, e.g., $\sim$9 comoving Mpc$^2$ at $z=3$). This is roughly two times smaller than the sizes of lower-mass protoclusters (i.e., $M_{z=0}\sim10^{14}$ M$_{\odot}$) at these redshifts \citep[e.g.,][]{Chiang13, Lim24}, which allows us to probe down to the protogroup scale.

The DEBLEND\textunderscore NTHRESH and DEBLEND\textunderscore MINCONT parameters define the deblending parameters that can split single contiguous detections into multiple pieces. DEBLEND\textunderscore NTHRESH sets the number of deblending subthresholds that are exponentially spaced between the detection floor and the peak of the detection. DEBLEND\textunderscore MINCONT sets the minimum contrast, which is the fraction of the overdensity in a substructure relative to the overdensity in the entire structure needed to be considered a separate detection. Deblending is desirable if there are blended systems in a field that may consist of two or more substructures, but setting the deblending too high can split single structures. As we are more interested in detecting structure than in resolving the individual subcomponents of structures, we adopt a less aggressive deblending scheme that uses a DEBLEND\textunderscore NTHRESH value of 12 and DEBLEND\textunderscore MINCONT value of 0.3. Such a choice will only break up the most conspicuous of multicomponent systems, though in practice we had no cases of deblending with this choice of parameters.

If the deblending were more aggressive, we could more effectively separate the subcomponents of more complex structures, such as Hyperion, Elent\'{a}ri, Smruti, Ruchi, and PClJ0227--0421 to take a few examples from the literature. However, by effectively turning off the deblending, we can ensure greater consistency of detections between redshift slices rather than risk having subcomponents deblended in some slices but not in neighboring slices. Through this, we maximize our chances of detecting the overall structure for a given system, but as we will necessarily blur out subcomponents of complex systems, we expect a poorer performance in our detection algorithm's ability to detect all subcomponents of complex structures known from the literature (see Section \ref{litcomp}).

Once the individual detections in each redshift slice are identified with SExtractor, we link the detections across redshift slices. SExtractor reports the density-weighted barycenter (hereafter simply barycenter) in R.A. and decl. for every detection. We can check if a pair of detections across two sequential slices are part of the same chain by calculating the transverse distance between their barycenters. If they are within a linking radius, which we set to 2 Mpc (chosen to be slightly longer than the optimal length found through tests with the $z\sim1$ ORELSE data in \citealt{Hung20} to account for the greater sizes of protoclusters relative to clusters), we combine the two detections into one linked chain and obtain a new barycenter from the positions of the individual detections weighted by their isophotal flux as defined by SExtractor. It is important to note that because we applied SExtractor to density maps, the flux values are not traditional photometric values, but rather represent the level of galaxy overdensity. The linking continues until no more detections are found that can be added to the chain. 

We find all possible linked chains starting from the lowest-redshift slice. Consequently, we will have many linked chains that are either wholly subsets of larger linked chains that were started at lower redshifts or differ by a few detections in the middle of the chain. In order to avoid potentially double-counting the same protostructure candidate, we first remove any linked chain that is a complete subset of another chain. We obtain the protostructure candidates by fitting a Gaussian to the isophotal fluxes of the linked detections as a function of redshift.

In some cases, the linked chain extends over very wide redshift ranges ($\Delta z > 0.5$) and contain multiple peaks. These long linked chains contain about 10\% of all peaks we identify. We identify peaks in a linked chain by first sorting all isophotal flux values in the chain, taking note of the redshift of the highest flux point. We then move on to the next highest flux point and examine its redshift. If it is within $\Delta z < 0.02$ of the first flux point, we associate the two points together into the same peak. This $\Delta z < 0.02$ threshold was chosen as it is the expected line-of-sight extent of a protocluster plus induced motion, i.e., 10--30 comoving Mpc \citep{Chiang13, Lim24}. Otherwise, the point is categorized as a second peak. We continue this from high to low flux values, checking whether each point is within the minimum and maximum redshift range of each identified peak and associating the points accordingly, until we reach 20\% of the original highest flux point of the linked chain. We set this threshold in order to reduce complications with noise around the flux floor of the linked chain. 

\begin{figure*}
\centering
\includegraphics[width=2\columnwidth]{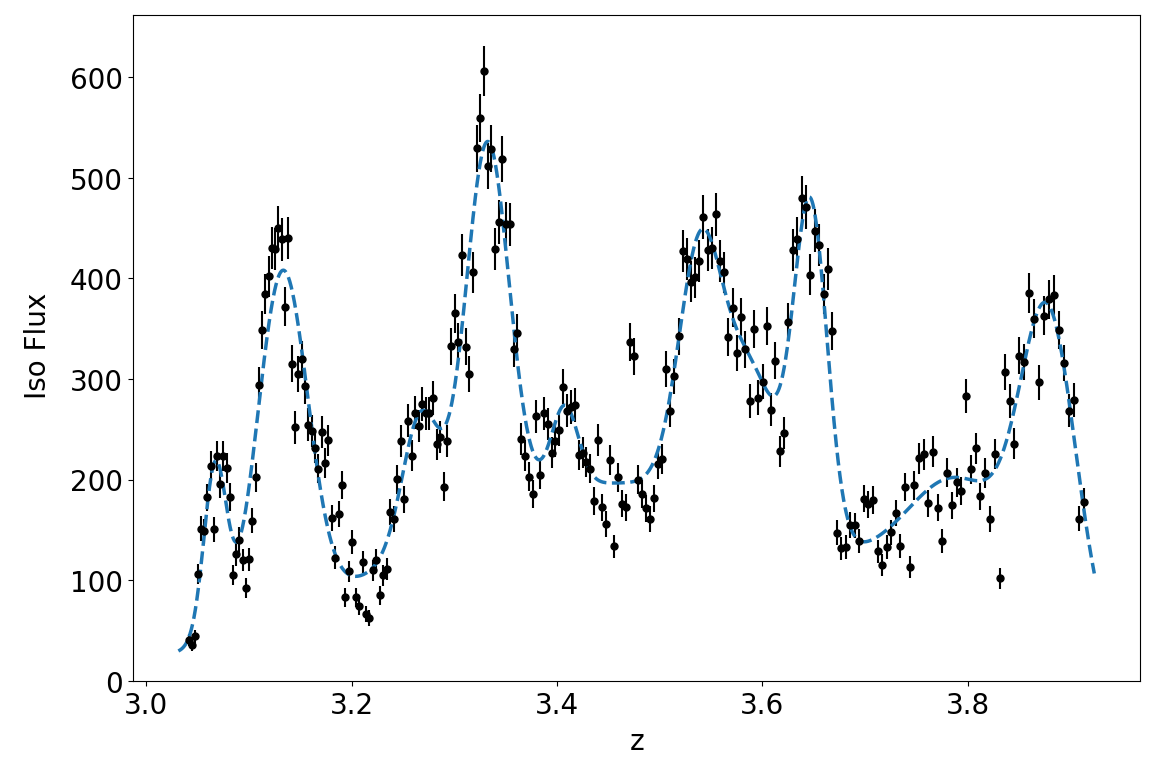}
\caption{An example of a long linked chain stretching over $\Delta z \sim 0.9$ and containing nine identified peaks. Data points represent the SExtractor detections in individual slices of our VMC overdensity maps (here we use an example from the mock 1 field, see Section \ref{sec.lc_mocks}) which are linked together using the search algorithm described in Section \ref{sec.find_struct}. The isophotal flux on the y-axis is the value directly reported by SExtractor and is a measure of density. Each peak is considered a unique protostructure candidate in our final catalog. Note how peaks close together in redshift space will contribute integrated flux to each other over the multiple Gaussians in the fit, though the fit amplitudes are relatively unaffected.}
\label{fig:long_linked_chain}
\end{figure*}

We obtain the final number of peaks by counting how many peaks have at least five associated points and perform a parametric fit characterized by an identical number of Gaussians to the linked chain (Fig. \ref{fig:long_linked_chain}). If the linked chain happens to yield no peaks with at least five points, we simply fit a single Gaussian to the entire linked chain. The initial fitting coefficients for each Gaussian are based on the parameters of the peak, where the mean and amplitude are, respectively, adopted from the redshift and flux value of the highest flux point in the peak, and initial guess for the redshift dispersion is set by the total redshift range of the points associated to the peak. If no peaks are identified due to an insufficient number of associated points, the initial fitting coefficients for the mean and amplitude of the Gaussian are set by the mean and highest flux value, respectively, of the data points in the linked chain, and we assume a redshift dispersion of 0.01. In order to avoid any likely spurious detections, we only keep the peaks where the converged Gaussian fit satisfies the following conditions: (1) an amplitude not more than 120\% of the highest flux value in the peak, (2) a mean offset of $\Delta z < 0.02$ from the median redshift of all flux points, and (3) a redshift dispersion of $\sigma_z < 0.06$.

Because of how we compile the linked chains, we have many close duplicates, which are mainly composed of the same SExtractor detections with a few different detections somewhere along the chain. These close duplicates then end up with averaged positions that are largely the same in transverse and redshift space. We thus need a method to filter out such cases to avoid possibly double counting the same structures. Once the peaks are identified, we compute the residuals with the Gaussian fit (the sum of squares of the difference between the fit and flux point divided by the total number of points within the redshift dispersion of the Gaussian fit) and remove all other peaks within $\Delta z < 0.02$ and a transverse distance of 2 Mpc from the peak with the lowest residuals. Peaks are checked starting with those with the lowest residuals and progressively moving those with higher residuals.

We generate the final list of potential overdensities, referred to hereafter as ``protostructure candidates,'' by sorting the peaks per field from high to low Gaussian amplitude and subsuming together any candidates that fall within a transverse distance of 2 Mpc and $\Delta z < 0.04$. In the case of protostructure candidates that result from two or more subsumed components, the amplitudes of the constituent protostructure candidates from the individual fits are added together in order to determine the Gaussian amplitude of each aggregate candidate, and an amplitude-weighted average of the individual positions and redshifts is additionally calculated to determine the location of the aggregate candidate. Entries in our catalog that are composed of two or more subsumed candidates make up about 10\% of our total catalog.

The choice of the various parameters described above will affect the completeness and purity of protostructure candidates we find. Less restrictive parameters will yield more protostructure candidates, but will result in a large number of spurious candidates. Ideally, we would like to find the set of values that would maximize both our completeness and purity, as was explored in depth in \citet{Hung20}. However, the parameter space is enormous and would be time-consuming to investigate completely. More care would be necessary for studies that rely on accurate structure number densities, such as cluster mass functions \citep[e.g.,][]{Hung21}. As we are only interested in a general structure search in this work, we adopt values in line with what we previously found to be optimal in \citet{Hung20} and \citet{Hung21}, finding negligible differences in the narrow parameter space we investigated in this work.

\section{Tests with Simulated Data}\label{mocks}

To test the efficacy of our search algorithm, we perform mock observations of simulated data and apply the same density mapping and structure search algorithm to the mock data for which we have absolute knowledge of the structure locations. As we describe in the following Sections, these mock observations are designed in such a way as to mimic the quality and approach of the actual observations employed in this study. In this way, in principle, subject to the limitations of the simulations, semianalytics, and our ability to impose all nuances of our own data on the mock observations, we are able to determine the ability of our data, mapping, and search algorithm to detect structures of differing properties in the early Universe. 

\subsection{Construction of the Lightcone}\label{sec.lc_const}

For our study, we use the predictions of the GAEA \citep{DeLucia2024} semianalytic model (SAM) to generate simulated galaxy catalogs. The GAEA model, using the version described in \citet{Xie2017}, was applied to the dark matter (DM) merger trees of the Millennium Simulation \citep{Springel2005}. This DM simulation is based on $N = 2160^{3}$ particles, each with a mass of $8.6 \times 10^{8}h^{-1} M_{\odot}$ and distributed within a box of size $500 h^{-1}$ comoving Mpc on a side. The simulation is run on a slightly different cosmology from the rest of this work, assuming a $\Lambda$CDM model with $\Omega_{m}=0.25$, $\Omega_{b} = 0.045$, $h = 0.73$, $\Omega_{\Lambda} = 0.75$, $n=1$ and $\sigma_{8} = 0.9$. 

Following \citet{Zoldan2017}, we generated a lightcone from the output of the GAEA SAM. This lightcone mimics a circular region of the sky with a radius of 2.3 degrees. The lightcone includes galaxies down to a stellar mass of $M_{*} \sim 10^7 M_{\odot}$, though is complete down to masses of $M_{*}\sim10^8--10^9 M_{\odot}$. The lightcone ranges from $z=0$ to $z=6$ and includes a number of physical and observable properties for model galaxies, such as their R.A., decl., redshift, and membership in DM halos and subhalos. For our study, we make use of the following quantities: R.A. and decl., observed redshift (i.e., the cosmological redshift plus the component of the peculiar velocity along the line-of-sight), and the dust-attenuated observed flux in IRAC1. We also make use of the known membership of each galaxy to its parent halo. In the Millennium DM Simulation, halos are identified by a friends-of-friends technique applied to the dark matter particles, using a linking length of 0.2 in units of the mean interparticle separation. For each halo, the virial mass is also provided as an output of the simulation.

Note that the area of the lightcone is larger than the DM simulation. As such, the lightcone is constructed of a patchwork of simulation cubes taken at single snapshots with different orientations at varying redshifts in order to populate the entire redshift range of the lightcone. Because of the finite size of the cubes, there may be issues where structure is cut off at the interface between two pasted cubes, which results in the possibility of massive systems with only a few subhalos as members. We identify and eliminate these systems from consideration in our analysis as described in Section \ref{sec.lc_subsume} and Appendix \ref{appA}.

\subsection{Selecting Lightcone Regions and Mock Observations}\label{sec.lc_mocks}

For this study, we selected five different, nonoverlapping 1$\times$1 deg$^2$ fields in the lightcone separated in R.A./decl. Figure \ref{fig:lightcone1} shows the location of each of these regions in the lightcone. A two-dimensional histogram of the $z=0$ masses and redshifts of the protostructures\footnote{Here we use the term ``protostructure'' to indicate the ensemble of halos at a given redshift that comprise a single $z=0$ structure in the simulation.}. in these regions is shown in Figure \ref{fig:massvsz2dhist_mocks}. Once these regions were selected, we then assign photometric and spectroscopic redshifts to the simulated galaxies in a way that mimicked the properties of the real data, a process we refer to as ``mocking up'' the lightcone, which we describe in the following subsection. For the construction of this catalog, the COSMOS field is chosen as the reference field, because it is the largest of the three C3VO fields and contains the majority of our data. We will discuss the mocking up of the other two C3VO fields (ECDFS and CFHTLS-D1) later in this Section.

\subsubsection{Construction of the Mock Galaxy Catalogs}
For each region selected in the lightcone, we constructed mock galaxy catalogs using the following prescription. We began with the combined photometric and spectroscopic catalog in the COSMOS field, which are described in Sections \ref{sec.photdata} and \ref{sec.specdata}. In order to impose conformity on the number counts of simulated and observed objects, the simulated and observational catalogs were cut at IRAC1$<$24.8. At fainter magnitudes, the number counts of simulated objects over the redshift range $0 < z < 6$ and $\log(\mathcal{M}_{\ast}/\mathcal{M}_{\odot})>8$ appeared to considerably outpace those in the observed data, which likely indicated that the data were not complete to magnitudes fainter than this limit. Objects fainter than IRAC1$=24.8$ were not considered further in any part of our analysis.

The observational catalog was then broken in bins of magnitude and redshift in order to (\emph{a}) assess the spectroscopic redshift fraction within that bin, and (\emph{b}) to assess the statistics of our photometric redshifts and their uncertainties as a function of these two parameters. For the former, objects were split into magnitude bins of IRAC1 $ < 22.4$, $22.4 \leq$ IRAC1$ < 23.4$, $23.4 \leq $IRAC1$ < 24.4$, and $24.4 \leq $ IRAC1$ < 24.8$. The magnitude bins were set in this way to roughly contain an equal number of objects in each bin. For each magnitude bin, we further broke down objects in redshift bins of $0 < z \leq 1$, $1 < z \leq 2$, $2 < z \leq 3$, $3 < z \leq 4$, $4 < z \leq 5$, and $z > 5$. 

In each redshift and magnitude bin we calculate the total number of galaxies with secure spectroscopic redshifts divided by the total number of objects in that bin, where the number of objects in the denominator is set by magnitude and photometric redshift cuts. The redshift bins are set to be wide enough to minimize the effects of scatter across bins due to photometric redshift errors. This quotient sets the global spectroscopic redshift fraction for that magnitude and redshift bin. The spectroscopic redshift fraction (S$z$F) for the COSMOS field across all redshift and magnitude bins is referred to hereafter as S$z$F1.0 and sets our fiducial mock spectroscopic observations for the mock catalogs. By definition, S$z$F1.0 carries with it the \emph{a posteriori} spectroscopic selection function of the COSMOS field, which is then applied to all mocks. 

The global spectroscopic redshift fraction in COSMOS per redshift and magnitude bin ranged from $\sim$0.3 for brighter objects (IRAC1$ < 22.4$) at lower redshift ($z<1$) to $\sim$0.001 for the faintest objects ($24.4 \le $IRAC1 $< 24.8$) at the highest redshifts ($z>5$). Overall, S$z$F1.0 corresponds to a global spectroscopic redshift fraction of 0.07 for a sample cut at IRAC1$<$24.8. The equivalent numbers for the ECDFS and CFHTLS-D1 fields were 0.2 and 0.06, respectively, though with the latter cut to a brightness limit of $Ks<24.1$ rather than IRAC1$<$24.8 due to the properties of the photometric data in that field (see Section \ref{sec.photdata}).

In order to approximate the properties of the photometric redshifts in the real data, objects were broken into the same redshift and magnitude bins as were used for the S$z$F calculation. For each magnitude and redshift bin, three quantities were calculated by comparing spectroscopic and photometric redshifts: the bias, the $\sigma_{\rm{NMAD}}$ (see \citealt{Lemaux22} and references therein), and the median photometric redshift uncertainty as calculated from the full PDF. This latter quantity was denoted $\sigma_{\rm{alt}}$. In bins that contained $>$10 secure spectroscopic redshifts, $\sigma_{\rm{NMAD}}$ was adopted as the photometric redshift scatter; otherwise $\sigma_{\rm{alt}}$ was adopted. 

\begin{deluxetable}{ll}
\tablecolumns{2}
\tablecaption{Corresponding spectroscopic fraction for each S$z$F}
\label{tab:szf}
\tablehead{\colhead{S$z$F}\hspace{3cm} & \colhead{Global Spectroscopic Fraction}}
\startdata
        0.5 & 0.035 \\
        1.0 & 0.070 \\
        1.5 & 0.105 \\
        2.0 & 0.140 \\
        4.0 & 0.280 \\
\enddata
\tablecomments{S$z$F1.0 is explicitly defined as equal to the global spectroscopic redshift fraction for the data in the COSMOS field that is adopted in this study. While the spectroscopic fractions listed here are for all objects brighter than IRAC1$<24.8$, the full spectroscopic selection function is applied in finer magnitude and redshift bins for each S$z$F. Refer to Section \ref{sec.lc_mocks_regions} for more details.}
\end{deluxetable}

\subsubsection{Mock Map Regions}\label{sec.lc_mocks_regions}
\begin{figure}
\centering
\includegraphics[width=\columnwidth]{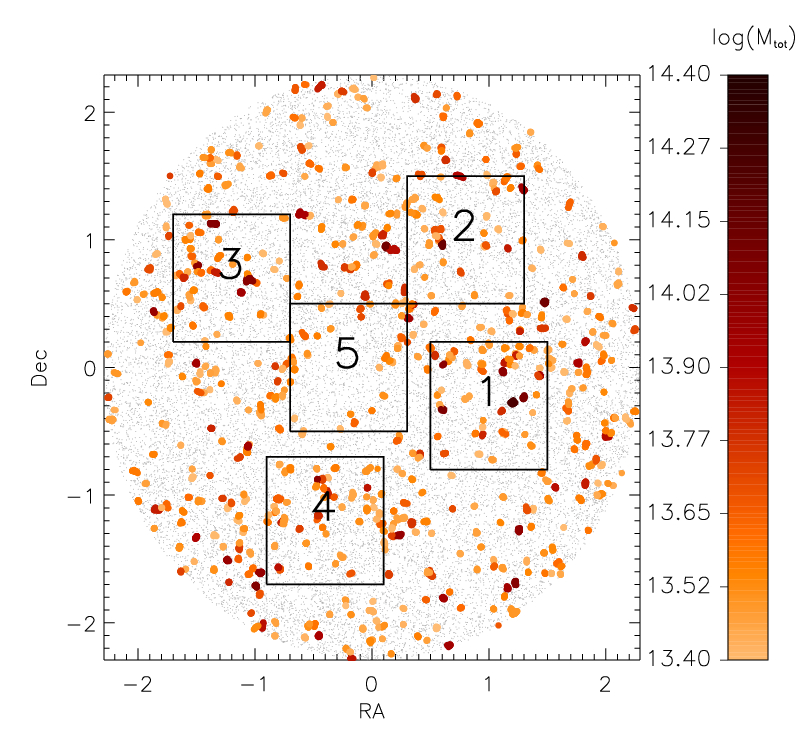}
\caption{Simulated R.A. and decl. distribution of galaxies in the lightcone used in this work. The gray dots depict a random subsample of all galaxies with $2<z<5$ and IRAC1$<$24.8. The filled circles denote galaxies with the same cut in redshift and IRAC1 as the gray dots, but belong to DM halos with total mass in the range as indicated in the color bar, in units of $\log(M/M_{\odot})$. The color bar lower limit (13.4) is chosen not to over-crowd the plot, and the upper limit is chosen to comprise the most massive DM halo at $z>2$ in this lightcone. The DM halo mass used in this figure is the one provided in the simulation. The five squares represent the five $1\times1$ deg$^2$ regions of the lightcone chosen for mock observations as described in Section \ref{sec.lc_mocks}. These regions are referred to as mocks 1-5 in the text.}
\label{fig:lightcone1}
\end{figure}

\begin{figure}
\centering
\includegraphics[width=\columnwidth]{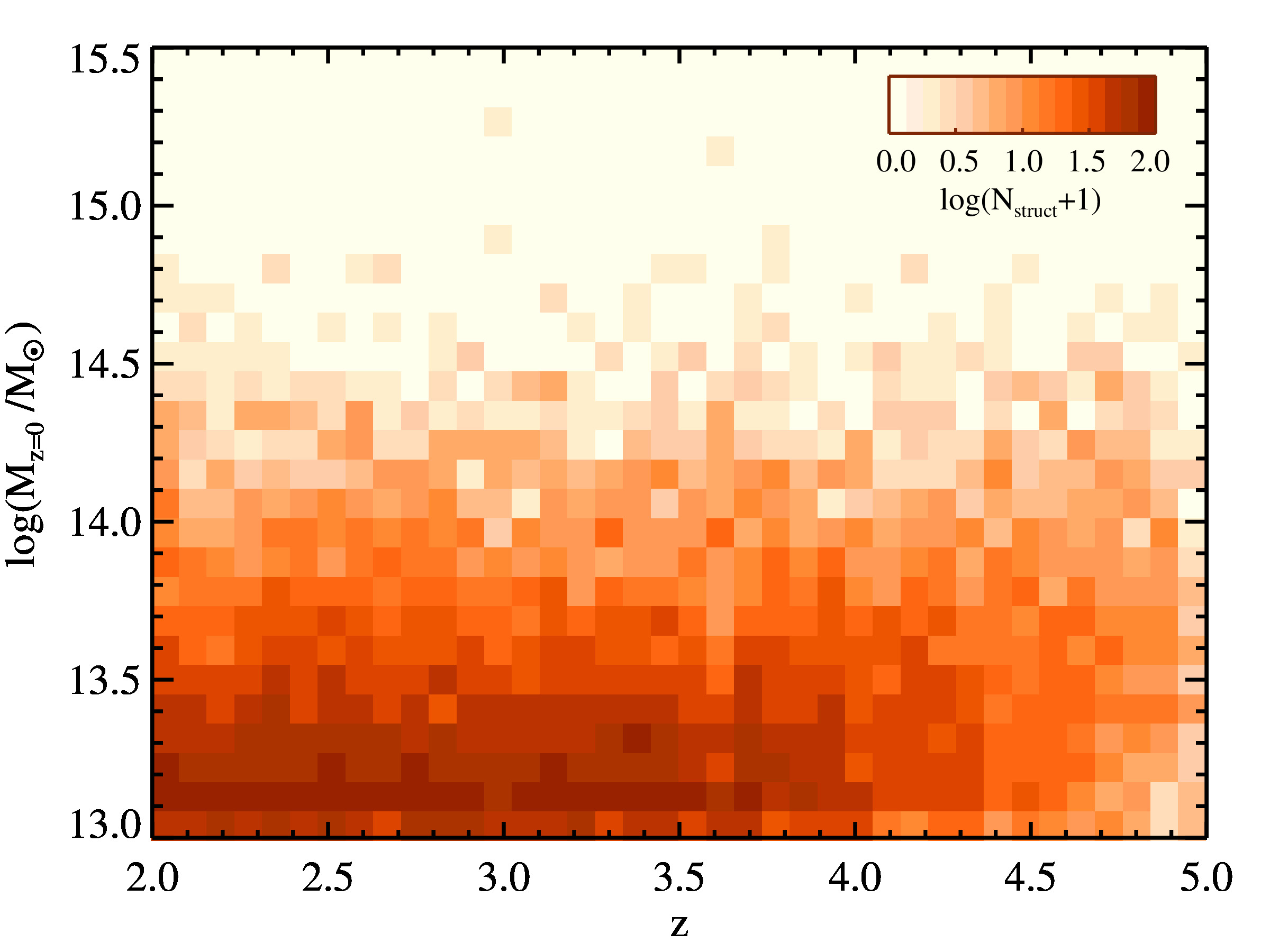}
\caption{Two-dimensional histogram of $M_{z=0}$ and redshifts for all structures that appear within the central 80\% of all five fields of the GAEA lightcone for which we generate mock catalogs. Only the parameter space shown here, i.e., structures with $M_{z=0}>10^{13}$ $M_{\odot}$ and $2<z<5$, are considered for this work. Note that the properties of the structures presented here are given prior to the subsuming process described in Section \ref{sec.lc_subsume}.}
\label{fig:massvsz2dhist_mocks}
\end{figure}

We chose five 1$\times$1 deg$^2$ regions in the lightcone to perform mock observations (Fig. \ref{fig:lightcone1}). For a mock observation of a given region of the lightcone, referred to hereafter as ``mock X,'' where X indicates the region of the lightcone as defined in Figure \ref{fig:lightcone1}, we generated a spectroscopic catalog using the S$z$F1.0 selection function as a function of magnitude and redshift as determined from the observed data. Spectroscopic redshifts were assigned randomly to galaxies in a given magnitude/redshift bin, and spectroscopic quality flags were assigned in a manner that mimicked the distribution in the real data. For each mock region other than mock 1, we generated only a single spectroscopic redshift catalog using the S$z$F1.0 selection function.

For mock 1, we computed a large number of spectroscopic catalogs to test not only our structure search algorithm as a function of S$z$F but also the effect of random sampling within a given redshift and magnitude bin on the recovery rate of structure. To this end, we generated spectroscopic catalogs using S$z$F values of 0.5, 1.0, 1.5, 2.0, and 4.0, where the fiducial spectroscopic redshift selection function from the data was multiplied by each S$z$F in order to create a new selection function (Table \ref{tab:szf}). Additionally, we created five different versions of the S$z$F1.0 catalog using five different realizations of the fiducial spectroscopic redshift selection function, with a new set of random galaxies, subject to the S$z$F1.0 selection, chosen for each realization of the S$z$F1.0 catalog in mock 1.

We also created an additional mock observation of region 1 using a brightness cut of IRAC1$<$24 using the identical S$z$F1.0 selection function as with the IRAC1$<$24.8 limit, but only applied down to the brighter magnitude cut, in order to broadly mimic the data properties of the CFHTLS-D1 field. The S$z$F0.5 mock also represents a reasonable approximation of the data quality in the CFHTLS-D1 field and is combined with the IRAC1$<$24 S$z$F1.0 mock in certain parts of the analysis in order to approximate expectations from the CFHTLS-D1 field data quality. For reference, the spectroscopic properties of the ECDFS field most closely resemble an average of that of the S$z$F 2.0 mock and S$z$F 4.0 mock. Simulated galaxies that were selected to have a spectroscopic redshift in a given realization were assigned their apparent redshift ($z_{\rm{obs}} \equiv z_{\rm{cos}} + z_{\rm{pec}}$), with $z_{\rm{cos}}$ and $z_{\rm{pec}}$ being the cosmological redshift and the peculiar redshift offset as defined by the simulations, respectively, assuming no redshift uncertainty\footnote{The spectroscopic redshift uncertainties from the data are much smaller than the size of a VMC slice and are, thus, negligible for the purposes of this exercise. The VMC approach takes into account catastrophic failures in spectroscopic redshifts as dictated by the reliability statistics for each flag, see \cite{Lemaux22}.}. 

\begin{figure*}
\plotfive{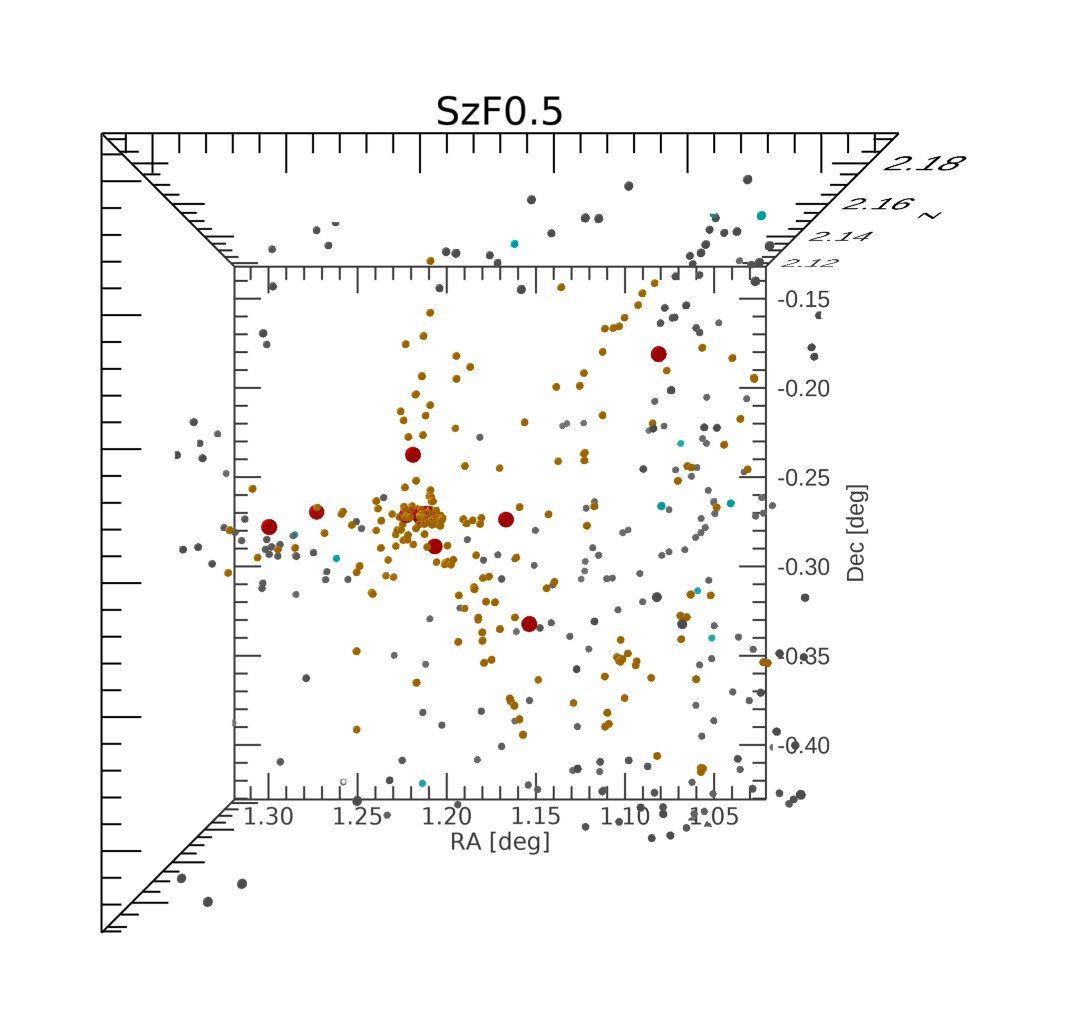}{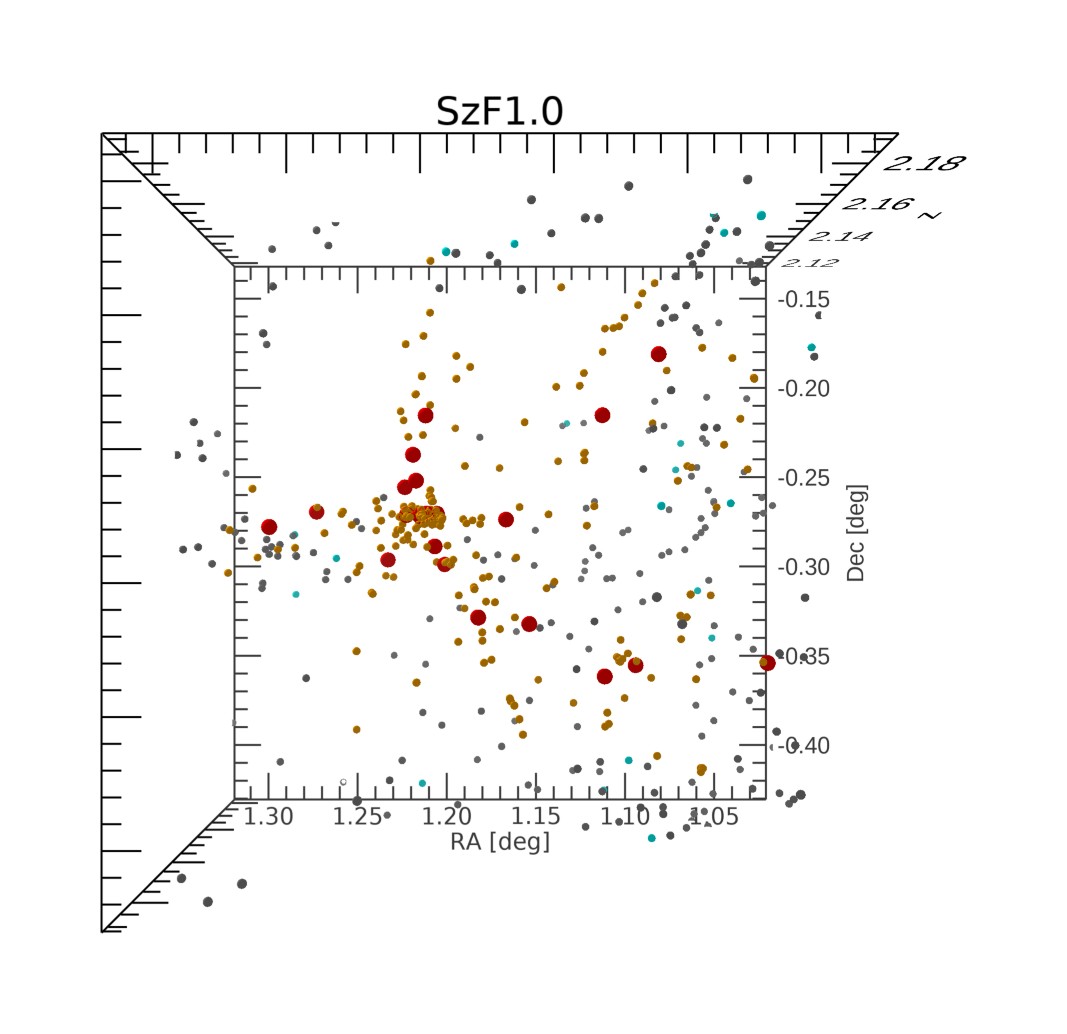}{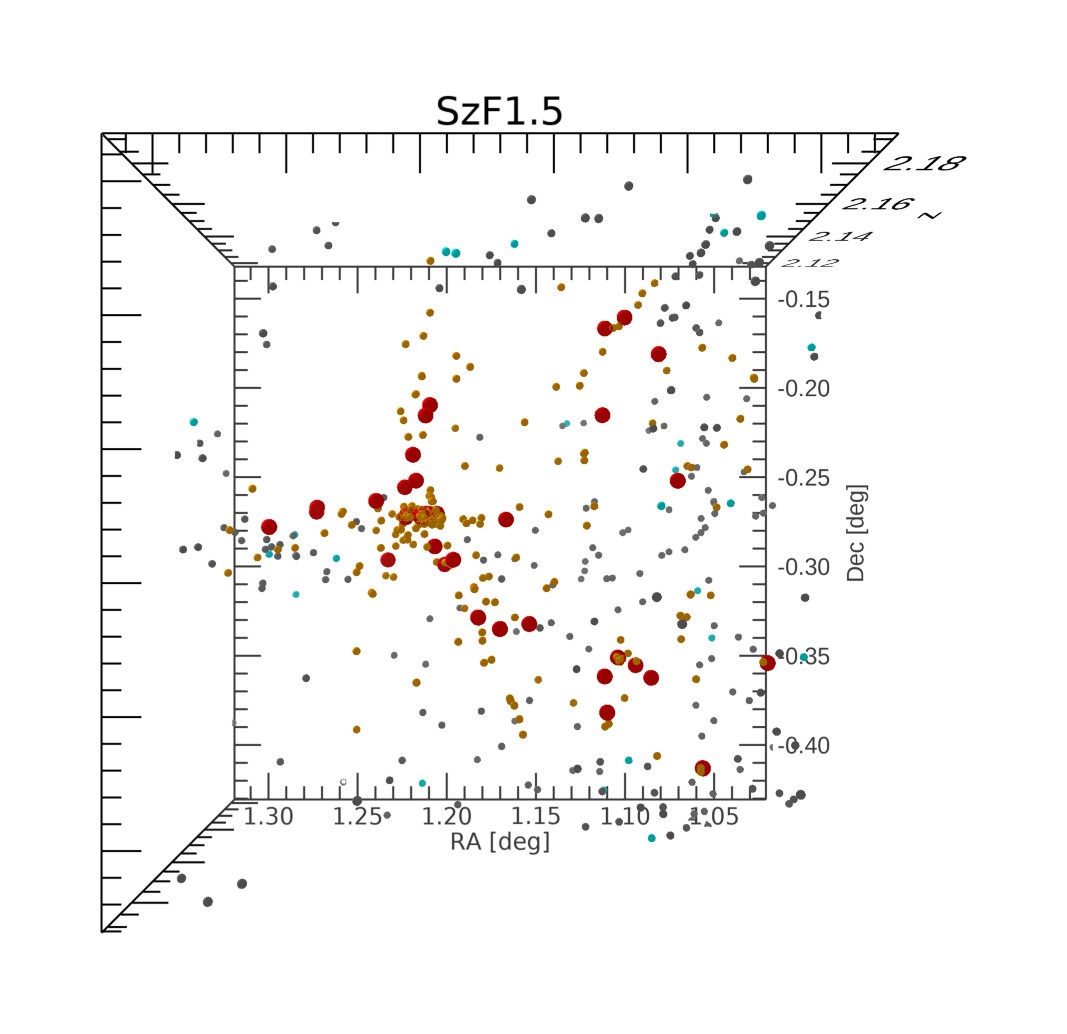}{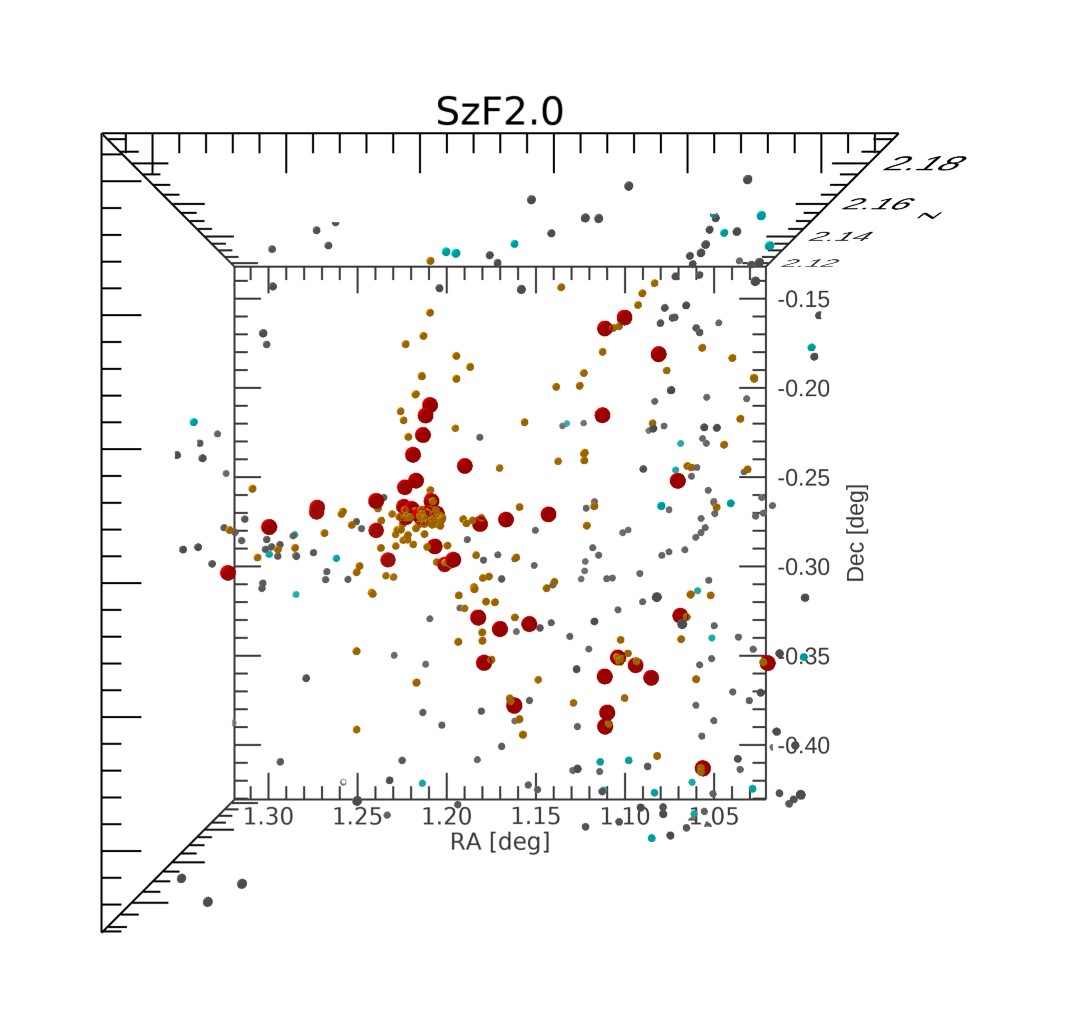}
{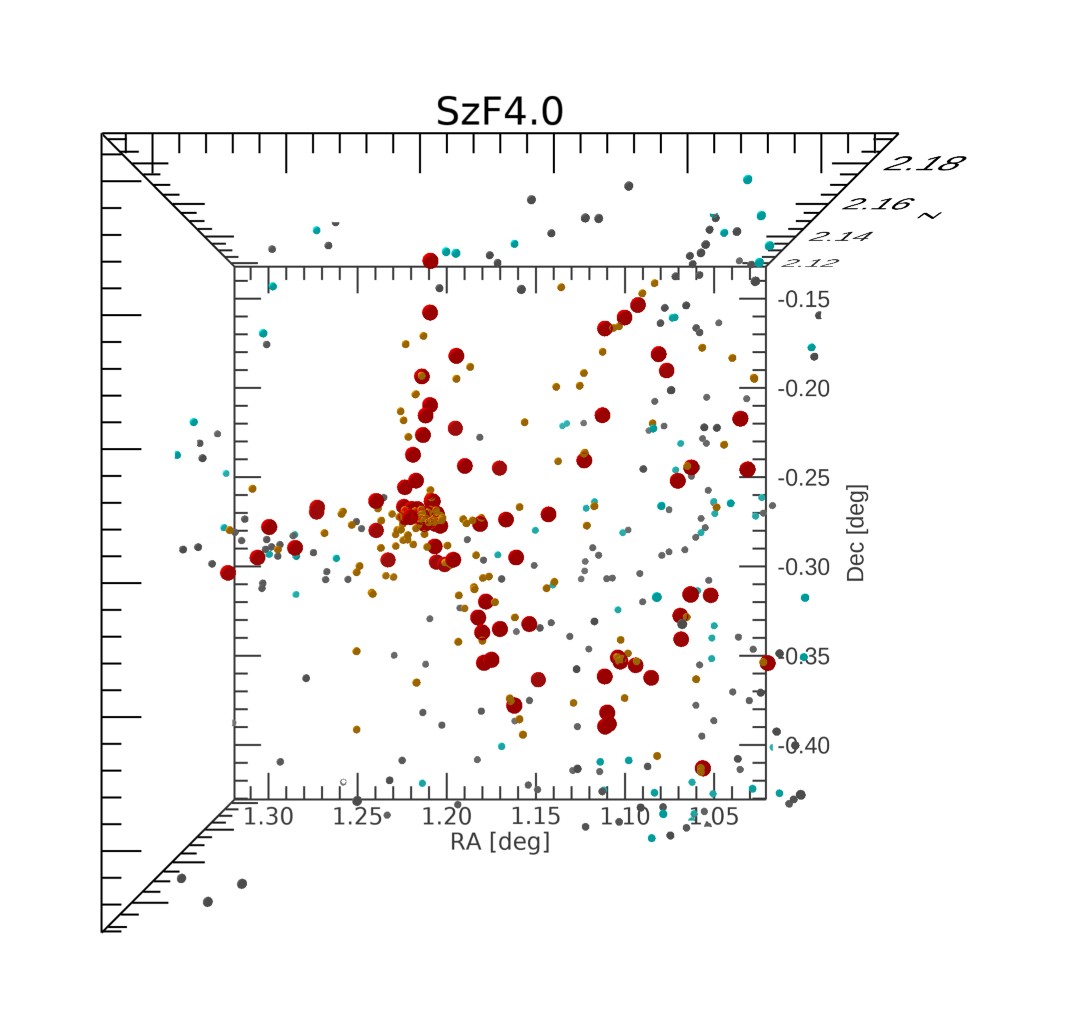}
\caption{Three-dimensional rendering of the region surrounding a simulated protocluster  at $z=2.15$ ($\log (M_{z=0}/M_{\odot}) = 14.7$) as seen in the GAEA lightcone. Each sphere represents a simulated galaxy brighter than IRAC1$<$24.8 (see Section \ref{sec.lc_mocks} for details on this cut). A total of 226 member galaxies of this protocluster brighter than IRAC1$<$24.8 are identified at this redshift, and are shown as small orange spheres. Coeval field galaxies and members of other structures are shown as small spheres of other colors. Each panel shows different mock spectroscopic observations of this region, with spectroscopic sampling increasing from the top-left panel (S$z$F0.5), in which the high-quality spectral redshift fraction is set to 0.035, to the bottom panel (S$z$F4.0, 0.28). Galaxies selected as having a high-quality spectroscopic redshift in a given mock are shown as larger red spheres.}
\label{fig:SzFlightcone}
\end{figure*}

\begin{figure*}
\plottwospecial{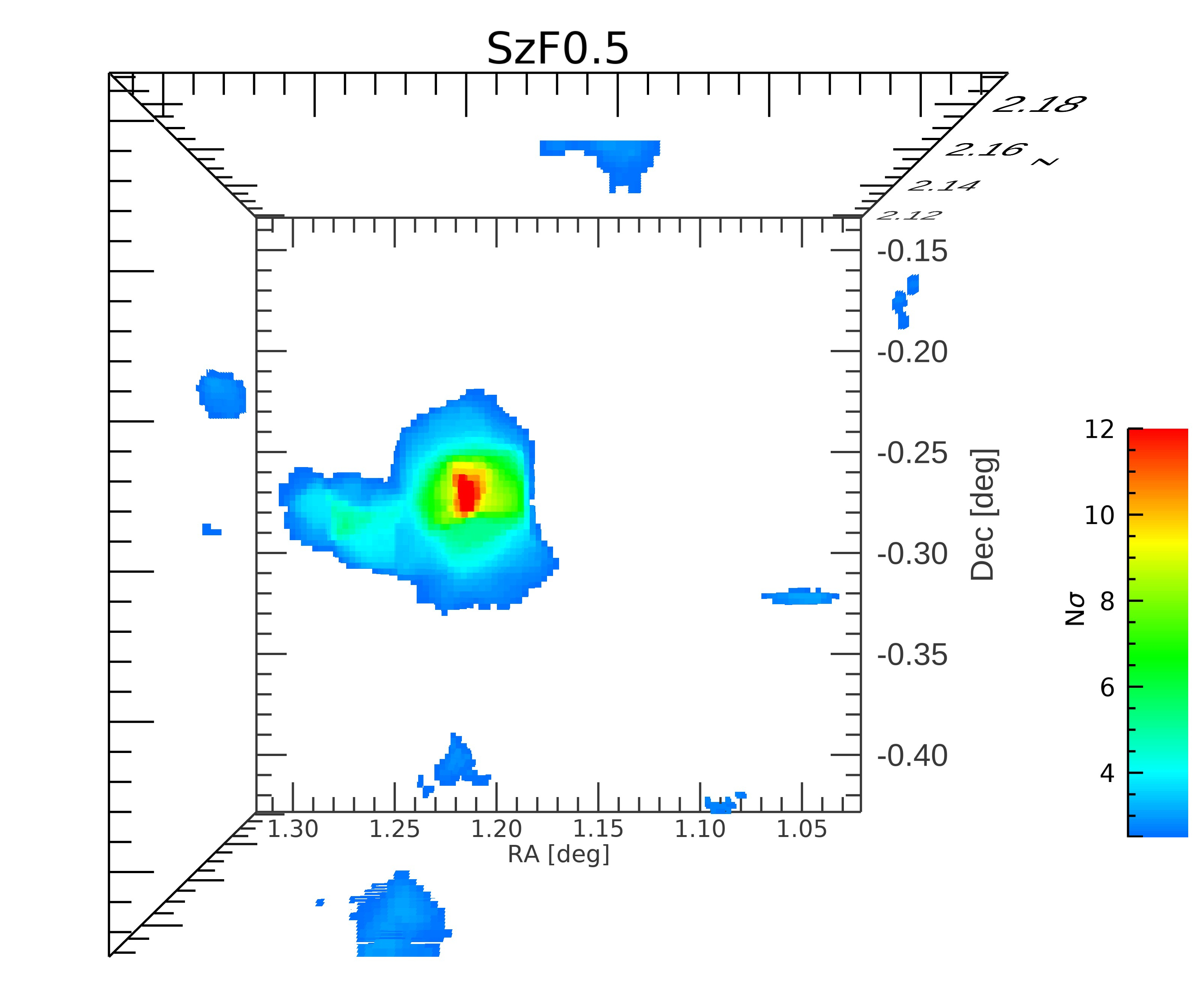}{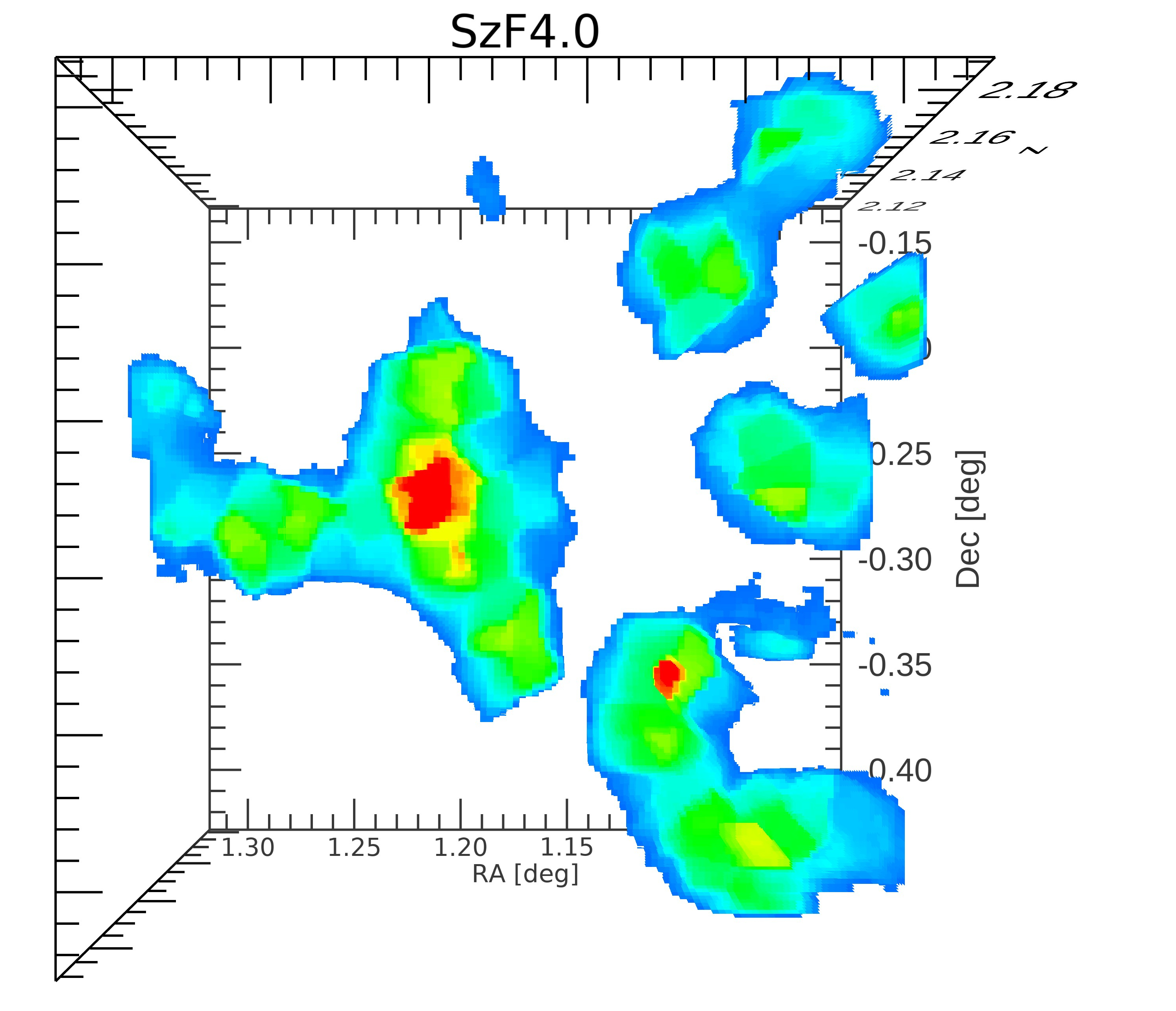}
\caption{Three-dimensional rendering of the same region of the GAEA lightcone as shown in Figure \ref{fig:SzFlightcone}, but showing our overdensity reconstruction of the region using the VMC method described in Section \ref{sec.method_VMC} for the S$z$F0.5 (left) and S$z$F4.0 (right) variants of the mock catalogs. The three-dimensional rendering is performed following the methodology of \citet{Cucciati18}. The scale bar indicates the significance above the mean density level as measured by SExtractor. While the core region of the main protocluster is detected in the S$z$F0.5 mock, a dramatic improvement in the quality of the overdensity reconstruction is seen in the S$z$F4.0 mock, including the emergence of filamentary structure and other substructure.}
\label{fig:SzFlightconeVMC}
\end{figure*}

In addition, all objects, regardless of whether they were assigned a spectroscopic redshift, were assigned a photometric redshift using the prescription: $z_{\rm{phot}} = z_{\rm{obs}} + B(1+z_{\rm{obs}}) + N\sigma_{\rm{pz}}(1+z_{\rm{obs}})$, where $B$ is the spectroscopic to photometric redshift bias in each magnitude and redshift bin, $N$ is a value sampled from a normalized Gaussian distribution for each object, and $\sigma_{\rm{pz}}$ is either set to $\sigma_{\rm{NMAD}}$ or $\sigma_{\rm{alt}}$ depending on the number of secure spectroscopic redshifts in a given bin. Effective $\pm$1$\sigma$ photometric redshift uncertainties for each simulated galaxy were determined by sampling from the PDF of the fractional photometric redshift error (i.e., $(z_{\rm{phot,1\sigma, upper}}$-$z_{\rm{phot}}$)/$(1+z_{\rm{phot}})$ and $(z_{\rm{phot}}$-$z_{\rm{phot,1\sigma, lower}}$)/$(1+z_{\rm{phot}})$) constructed from the statistics of objects in each magnitude bin in the real data. At the end of this process, each simulated galaxy had a photometric redshift with an associated $\pm$1$\sigma$ uncertainty. Additionally, those simulated galaxies selected for mock spectral observations were assigned a spectroscopic quality flag and a spectral redshift. These mock galaxy catalogs were then run through the same VMC process as was used for the real data (see Section \ref{sec.method_VMC}) to generate density reconstructions of all mock realizations of each field. Figure \ref{fig:SzFlightcone} shows an example of the region in and around a massive protocluster ($\log (M_{z=0}/M_{\odot}) = 14.7$) at $z=2.15$ located in the GAEA lightcone and the mock spectroscopic observations for five S$z$F variants. Figure \ref{fig:SzFlightconeVMC} shows the VMC reconstruction of the overdensity field in the same regions for the two extrema of the S$z$F variants, S$z$F0.5 and S$z$F4.0.

\subsection{Obtaining the Final List of Subsumed Structure Candidates}\label{sec.lc_subsume}

The lightcone is populated with structures scattered across its entire redshift extent. The spacing between structures, however, can vary, such that two structures may be so close together that we would be unable to distinguish them as separate candidates in our VMC maps. We therefore want to construct a structure catalog that encodes this confusion limit in some way so that we can better match the detections in our mock observations and have a clearer measure of the masses of the structures we detect. We did so in the following manner. Using the highest mass structure in the lightcone as a seed, we subsumed all structures that fell within $\Delta z < 0.04$ and a transverse separation of 2 Mpc of this structure. These values were chose to roughly match the confusion limit of structures present in our VMC maps. Once all nearby structures were subsumed, their masses are combined into a single protostructure with R.A./decl./$z$ coordinates dictated by a mass-weighted average of the barycenters of all constituent structures. This process continued from the highest mass structure to the lowest mass structure in the lightcone until all single or composite structures in the catalog had no companions within the above limits.

Additionally, we want to account for any artificially under-rich structures that may have a confounding effect on measurements of purity, completeness, and any mass--richness relation we attempt. As mentioned in Section \ref{sec.lc_const}, such structures likely exist in the lightcone due to structures being artificially cut off at the interface between two pasted cubes. 

We identify these structures by fitting a relation between their $z=0$ masses and numbers of members, divided into redshift bins of $\Delta z = 0.5$. Because we want to take care to not excise systems that are under-rich at our adopted magnitude limit of IRAC1$<$24.8 due to astrophysical reasons (e.g., a shifting of the luminosity function to fainter magnitudes for a given system resulting in a preponderance of member galaxies that fall below our magnitude cut), we perform this exercise with a simulated galaxy catalog cut at IRAC1$<$25.5. For each redshift bin, we fit a linear relation to the logarithmic function of both parameters, iteratively clipping out outliers that are $3\sigma$ lower in number of members than our best-fit relation. Our under-richness criterion is defined by this same $3\sigma$ threshold (see Appendix \ref{appA}). 

\subsection{Recovery of Simulated Structures}

The VMC maps generated from the mock data were then fed into SExtractor in a manner identical to that used for the VMC maps generated from the real data (see Section \ref{sec.find_struct}) with two exceptions. Because the mock observations lack the somewhat uneven sampling that is present in the real spectroscopic and photometric data, edge effects, while still present, are less of a concern. As such, we simply mask the bottom 15\% of density values in each slice of the mock VMC maps, as opposed to the outer 20--30\% area of each slice used for the VMC maps generated from the real data. This choice was made to maximize the area that we can recover a signal on. Additionally, the median density of the mocks differs somewhat from that of the real data due to higher-order differences between the data and the mock catalogs (Fig. \ref{fig:meddens}). Since the VMC maps generated from the mock data generally exhibit higher average densities than that generated from the real data, we adopt a DETECT\textunderscore THRESH value of 2.5$\sigma$ for the mock data, which roughly corresponds to 3$\sigma$ values used for the real data (see Fig. \ref{fig:meddens}).

As we did for the real data, we then generate a list of protostructure candidates from linked detections across different redshift slices in the VMC maps of the mock galaxy catalogs. These candidates then undergo an identical subsuming process to that described in Section \ref{sec.find_struct}. In order to test the efficacy of our method both in detecting subsumed structure at different masses and in recovering a signal that is commensurate in strength with the mass of each subsumed structure, we need to create a formalism to match the protostructure candidates we detect to the lightcone structures. This formalism was generated in the following manner.

\begin{figure}
\includegraphics[width=\columnwidth]{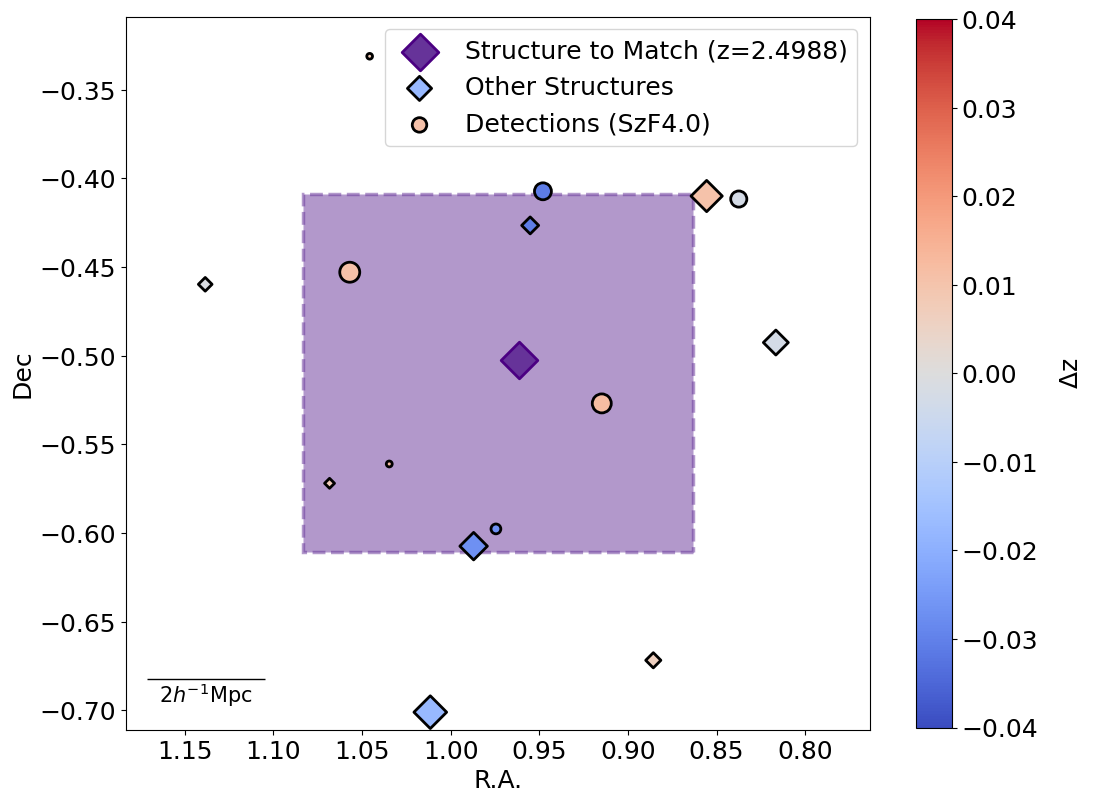}
\caption{An example of how we match the candidates we detect (denoted by the circular points) to the post-processed lightcone structures (denoted by the diamond points). The lightcone structure that we are matching to is denoted by the central violet point, and its maximum R.A./decl. extent are denoted by the shaded rectangle, showing that we have four candidate detections matched to this structure. Here, we only plot the candidates and structures that fall within the redshift extents of our matching structure. The size of each point is scaled according to the mass for structures or Gaussian amplitude for the candidates, which, respectively, range between $13.0 < \log(M_{tot}/M_{\odot}) < 14.5$ in structure mass and 33--623 in Gaussian amplitude, and the points are colored by their redshift difference with the matching structure.}
\label{fig:multidetect}
\end{figure}

We have full knowledge of the extent of the member galaxies in R.A., decl., and redshift space for each subsumed structure in the lightcone. We match a subsumed structure to a candidate (i.e., consider that simulated structure recovered) if the candidate's barycenter falls within the minimum and maximum ranges of the rectangular prism of the subsumed structure's R.A. and decl. and redshift extents. If the candidate falls within range of multiple subsumed structures, we match the candidate to the subsumed structure that has the closest barycenter in R.A. and decl. and redshift. Note that this allows for the possibility of more than one candidate within range matching to a given subsumed structure, which may be a signal from an individual substructure or simply spurious detections. Typically, however, the Gaussian amplitude of one candidate will dwarf any other matches. In the cases where we have multiple candidates matched to a given subsumed structure, the Gaussian amplitudes of each candidate are added together to eventually compare to the mass of the subsumed structure to which they are matched (Fig. \ref{fig:multidetect}).

At the completion of this process, we have a list of subsumed structures and associated Gaussian amplitudes for those structures that have a matched candidate (or multiple candidates). This catalog is used in the next Section to assess the purity and completeness of our method for structures of different masses using differing levels of spectroscopic completeness. The catalog is also used in Section \ref{massrel} to set the rough mass scaling for our candidates.

\subsection{Finding a Mass Relation}\label{sec.massrel}
\label{massrel}

The protostructure candidates in the VMC maps, generated from both the real and mock data, have various measures of galaxy overdensity. These measures result from the Gaussian fit to galaxy overdensity as a function of redshift, or, equivalently, VMC slice, for a given candidate associated with a given structure (see Section \ref{sec.find_struct}). The Gaussian fit also can give us a sense of the candidate detection strength through parameters such as the integrated area or amplitude of the fit. With the total masses provided through the lightcone data and the candidates we find in the mock VMC maps, we can attempt to use one or more of the candidate detection strength parameters as a proxy for the structure total mass. 

However, some structures in the lightcone are close enough together such that we cannot distinguish them as separate structures in the mock VMC maps, and so the candidates we detect will be some mixture of the signal from multiple structures. In such cases, the candidate detection strength will not be representative of the ``known'' structure mass, resulting in additional scatter in the relation between structure mass and its proxy. In order to mitigate the effect of these ambiguous cases, we impose an isolation criterion on the subsumed structures for the exercise described in this Section. Starting from the highest mass subsumed structure, we check for any other structure within its effective radius. The effective radius is defined as the average radius that encloses 90\% of the stellar mass of a structure and is dependent on $z=0$ mass of the subsumed structure, as well as redshift. We estimate the effective radius using the relations given in Figure 2 of \citet{Muldrew15}, where the effective radii are given as a function of redshift in three different $z=0$ mass bins of $\log(M_{200}/M_{\odot}) < 14.6$, $14.6 \leq \log(M_{200}/M_{\odot}) < 15$, and $\log(M_{200}/M_{\odot}) \geq 15$. For a given redshift, the range of effective radii can be considerably large (up to $\sim$10 Mpc in comoving space for the largest mass bin), but for simplicity, we adopt the mean value reported. Note that, for this exercise, we assume that the total $z=0$ mass of the structure as reported in the GAEA lightcone is approximately equal to the $M_{200, z=0}$ value used by \citet{Muldrew15}.

We also make use of the effective radius in the redshift dimension. However, because of redshift uncertainties in our data and the natural apparent lengthening of structure in the redshift dimension due to induced motion, we scale the redshift distance threshold up such that at zero transverse separation, $\Delta z = 0.04$ is considered equal to the effective radius, with a reduced redshift window applied with increasing transverse distance. The functional form of this ellipsoid is given by:

\begin{equation}
\delta_{xy}^2 + (\delta _{z} / \gamma)^2 < R_{\rm{eff}}^2
\end{equation}

\noindent where $\delta_{xy}$ is the transverse separation, $\delta _{z}$ is the redshift separation, $\gamma$ is the scaling factor, and $R_{\rm{eff}}$ is the effective radius. For the masses and redshifts of the structures in the lightcone, the effective radii typically fall somewhere between 2 and 6 Mpc. It is worth noting that imposing this isolation criterion is a small effect, removing only about 1\% of structures between $2 < z < 5$ in the lightcone. If we do not find any other structures within this ellipsoid with masses $>$10\% of the seed structure, then we consider that structure to be isolated. Of the subsumed structures in the lightcone, the vast majority ($\sim$99\%) are considered to be isolated by this metric.

\begin{figure*}
\includegraphics[width=2.0\columnwidth]{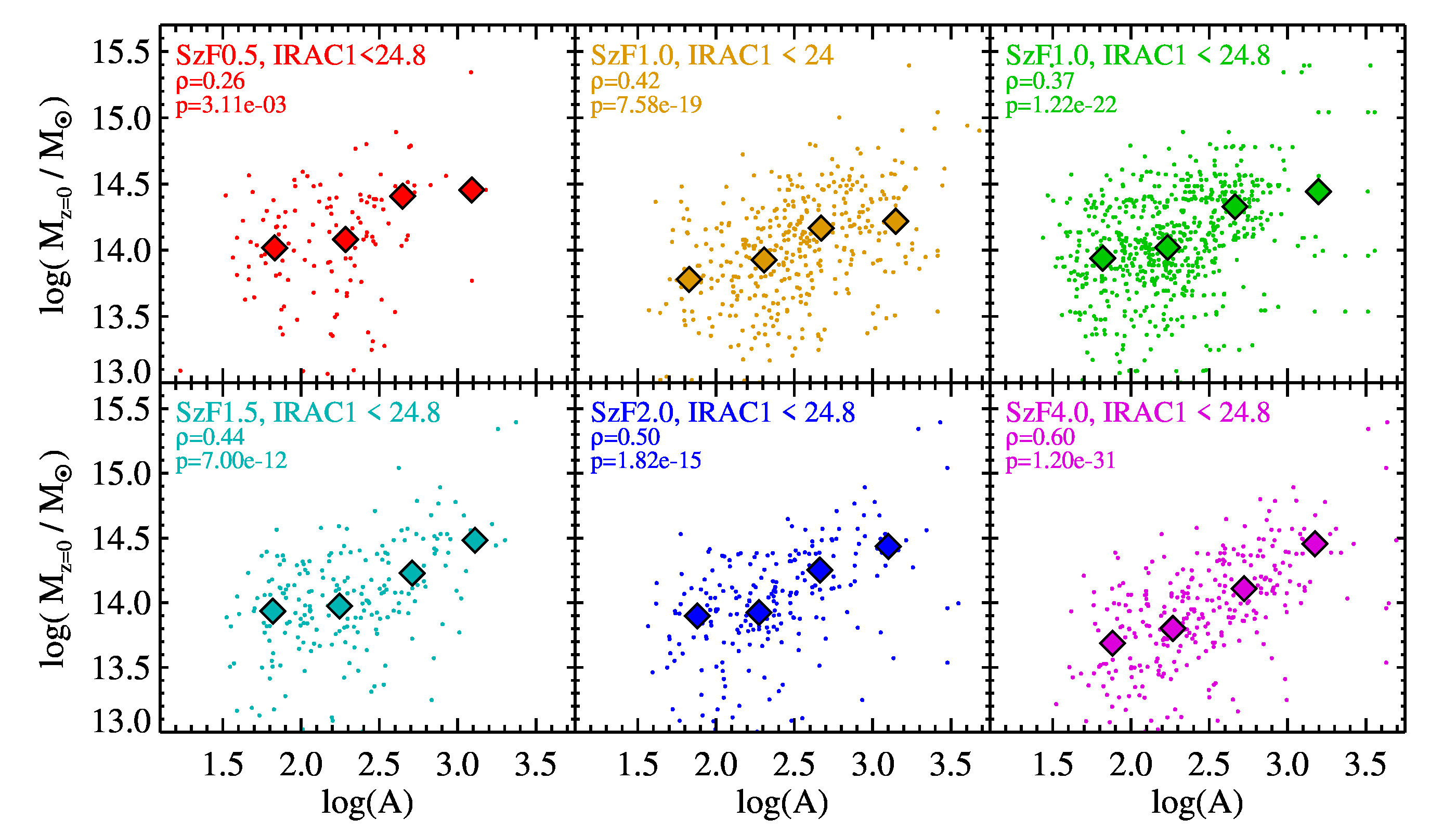}
\caption{Gaussian amplitudes ($A$) of candidates associated with simulated structure in the GAEA lightcone plotted against the $z=0$ mass of each structure. A running median is plotted as the large symbols in each panel. Only those structures in the selected mock fields (see Section \ref{sec.lc_mocks}) and only those structures with associated candidates are shown. Structures clipped from our final catalog as being under-rich due to the nature of the lightcone construction or those structures that have subsumed a clipped structure (see Section \ref{sec.lc_subsume}) are not shown. The six panels show mocks with differing levels of data quality, with the worst mock data quality appearing in the first two panels, and the best mock data quality appearing in the bottom-right panel. Note that the two S$z$F1.0 variants are mocked over several fields, rather than the single mock field used for all other variants, which results in more data points. Additionally, the S$z$F1.0, IRAC1$<$24.8 mock is also realized five times in one of the mock fields (field 1), which results in multiple instances of a candidate associated with a single structure across the different realizations. The Spearman correlation coefficient between the two parameters, $\rho$, is shown in each panel, along with the associated p-value indicating the likelihood of the null hypothesis that the two quantities are uncorrelated. While there is considerable scatter between the two quantities, the null hypothesis is rejected at the $>$3$\sigma$ level for all mocks and at the $\gg$3$\sigma$ level for all variants except S$z$F0.5, IRAC1$<$24.8.}
\label{fig:logampvsmass}
\end{figure*}
 
For each isolated subsumed structure that has a corresponding detection in the VMC maps of the mock galaxies, we compared both the integrated area of the Gaussian fit to the candidate and the amplitude of the same fit to the $z=0$ mass of that structure for each S$z$F variant in the mocks. Of the two potential proxies, we find stronger correlation between the $z=0$ mass and the Gaussian amplitude and thus adopt it as our generalized mass proxy. More specifically, we perform a Spearman correlation test on $M_{z=0}$ of isolated structure versus the Gaussian amplitudes of the associated candidates, finding correlation coefficients of $\rho_{M_{z=0}-A}=0.29, 0.47, 0.41, 0.45, 0.50, 0.60$ for the S$z$F0.5, S$z$F1.0 IRAC1$<$24, S$z$F1.0, S$z$F1.5, S$z$F2.0, and S$z$F4.0 mocks\footnote{For those S$z$F variants where multiple mocks were created, the structures and candidates across all mocks of a given variant were combined together prior to running a Spearman test. This is also the case for the fits described below.}, respectively. The Spearman test returned a rejection of the null hypothesis of no correlation at the $\gg$3$\sigma$ level for every S$z$F variant except S$z$F0.5, where the rejection of the null hypothesis was at the $\sim$3.5$\sigma$ level. In Figure \ref{fig:logampvsmass} we show the $z=0$ mass versus the Gaussian amplitude of candidates associated with isolated structures in the mocks for each of the S$z$F variants. 

\begin{figure}
\includegraphics[width=\columnwidth]{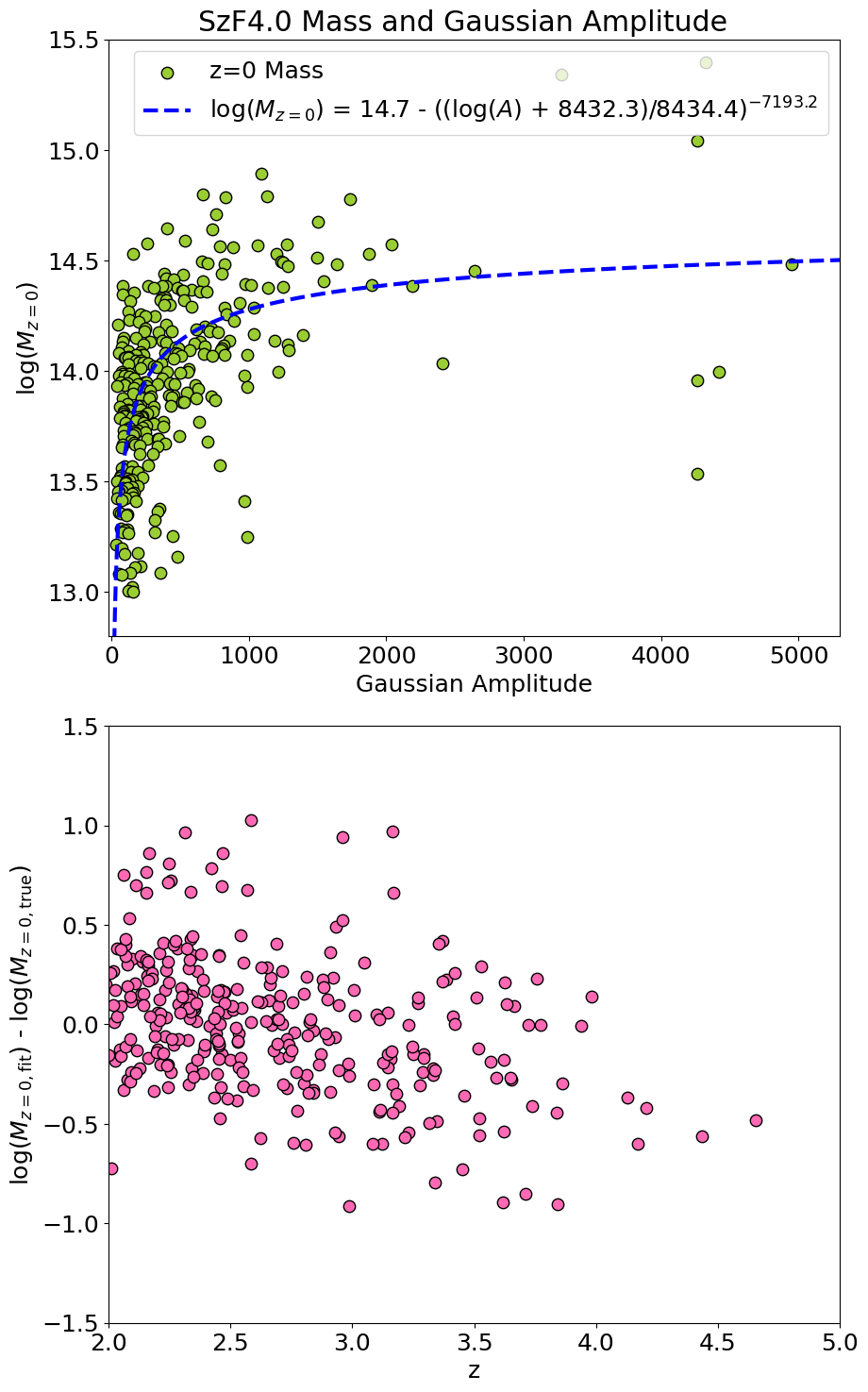}
\caption{The top panel shows the mass relation for the SzF4.0 mock in terms of Gaussian amplitude ($\rm{A}$) of the candidate and the known $z=0$ mass. The bottom panel shows a comparison of the $z=0$ and fitted masses in terms of redshift.}
\label{fig:massfit_redshift_ampl}
\end{figure}

We attempt to construct a mass relation by fitting the Gaussian amplitudes, $A$, of our candidates associated with structures in the mocks with their known $z=0$ masses in logarithmic space to the functional form $y=\alpha - ( (x-\beta)/\gamma)^{\delta}$, where $x = \log(A)$ and $y = \log(M_{z=0})$ which was previously used in \citet{Hung20}. The fit of Gaussian amplitude ($A$) versus $M_{z=0}$ is performed for each S$z$F variant, with dramatic differences in both the fit parameters and the scatter around the relation for the different variants (see Table \ref{tab:massfit}) for the resultant parameters. The large scatter in the data in all of the S$z$F variants tested here (0.3 dex) results in lower-mass structures having their masses overestimated by these relations and higher-mass structures having their masses underestimated (Fig. \ref{fig:logampvsmass}). While this issue is the least acute in the S$z$F4.0 case, the fit still precludes the possibility of returning masses in excess of $\sim$10$^{14.6}$ $M_{\odot}$, despite $\sim$10 structures in this region of the lightcone having masses that exceed that threshold (Fig. \ref{fig:massfit_redshift_ampl}). Even in the S$z$F4.0 case, the scatter is considerable, with mass estimates being no less certain than $\sim$0.3 dex.

\begin{deluxetable}{l|ccccc}
\tablecaption{Mass fit parameters}
\label{tab:massfit}
\tablehead{\colhead{Mock}\hspace{0.1cm} & \colhead{$\alpha$} & \colhead{$\beta$} & \colhead{$\gamma$} & \colhead{$\delta$} & \colhead{$\sigma_{\rm{NMAD}}$\tablenotemark{a}}}
\startdata
        S$z$F0.5 & 141.9 & -995.9 & 4655.9 & -3.2 & 0.32 \\
        S$z$F1.0\tablenotemark{b} & 14.5 & -2867.4 & 2869.1 & -2867.0 & 0.35 \\
        S$z$F1.0 & 14.3 & -10313.5 & 10314.7 & -15714.5 & 0.36 \\
        S$z$F1.5 & 235.3 & -1286.6 & 7260.7 & -3.1 & 0.34 \\
        S$z$F2.0 & 14.8 & -9761.1 & 9763.1& -6476.6 & 0.31 \\
        S$z$F4.0 & 14.8 & -4431.6 & 4433.72 & -3731.6 & 0.32 \\
\enddata
\tablenotetext{a}{Scatter given in units of $\log(M/M_{\odot})$, as defined in \citet{Lemaux22} and references therein.}
\tablenotetext{b}{Cut at IRAC1$<$24, all other mocks are cut at IRAC1$<$24.8}
\end{deluxetable}

Although we find some utility in performing this mass fit, the estimated masses resulting from the fit are of limited use as they lack both accuracy and precision, especially at the lower S$z$F values. As such, we do not report masses in our final protostructure candidate catalog but simply provide the Gaussian amplitude of the candidates in our catalog as well as in most of the Figures in this paper. The conversions in Table \ref{tab:massfit} are provided for readers wishing to assign $M_{z=0}$ values to our candidates, with the understanding that the resultant values are likely biased at some level and lacking in precision. Zoomed-in three-dimensional mapping of each structure, like those performed in \cite{Cucciati18}, \cite{Forrest23}, \cite{Shah24}, and \cite{Staab24}, would be required in order to recover more robust mass estimates.

\begin{figure}
\includegraphics[width=\columnwidth]{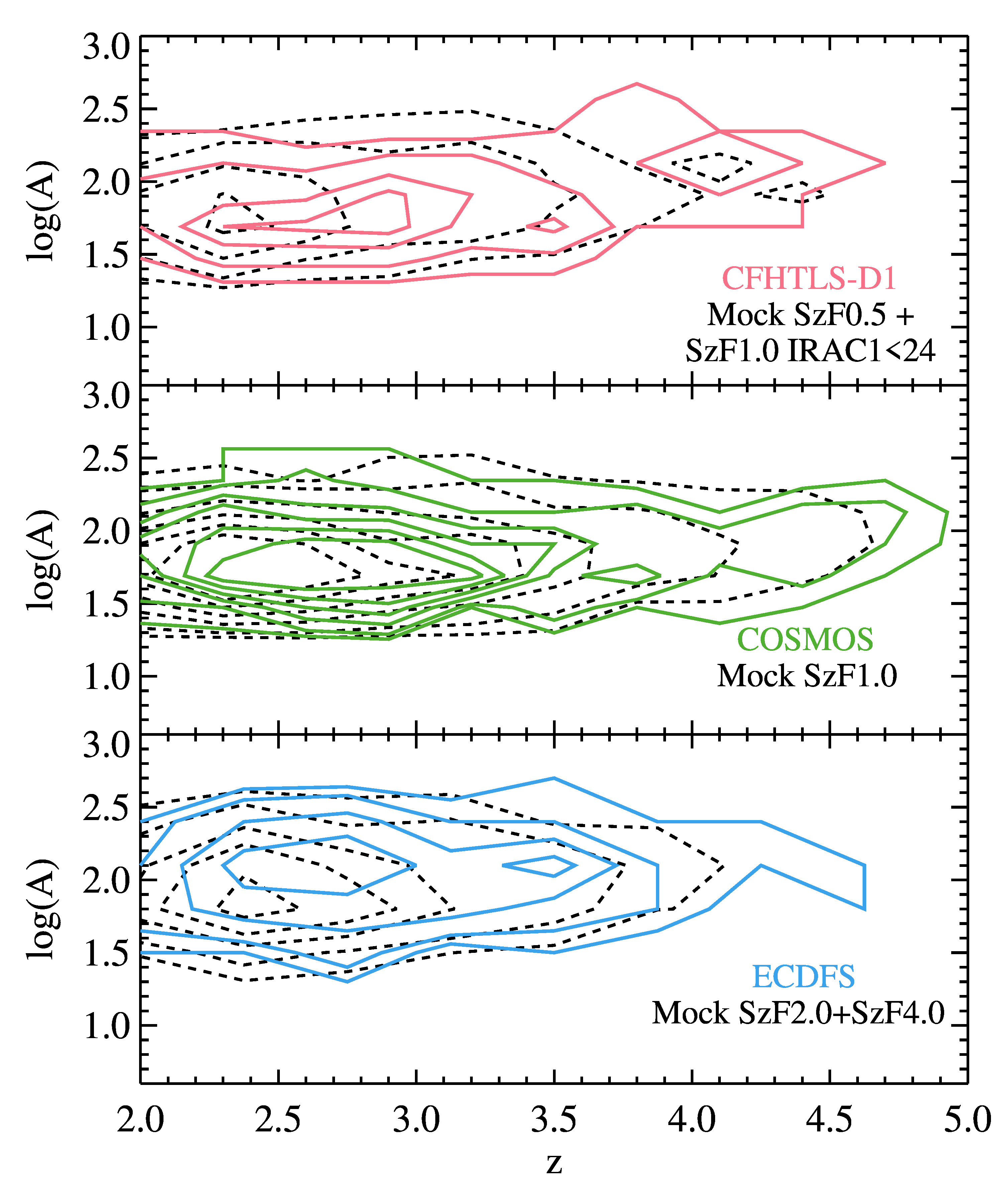}
\caption{Distribution of the Gaussian amplitudes ($A$) and redshifts for the final list of protostructure candidates in CFHTLS-D1 (top), COSMOS (middle), and ECDFS (bottom) shown as the colored solid contours. Also shown in the black dashed contours is the distribution in the mocks that most closely resemble the data properties of each field. The distribution of protostructure candidates in the real data appear nearly identical to those of the associated mocks, implying that the simulated density field tracks reality reasonably well.}
\label{fig:logampnz_mocksvsdata}
\end{figure}

Figure \ref{fig:logampnz_mocksvsdata} shows the distribution of Gaussian amplitudes and redshifts for our final set of candidates in all three C3VO fields as compared to candidates in the mocks appropriate for each field. For the CFHTLS-D1 field, we combine all combinations of the S$z$F1.0 mocks cut at IRAC1$<$24 and the S$z$F0.5 mock. For the COSMOS field, we use all combinations of the S$z$F1.0 mocks cut at IRAC1$<$24.8. And for the ECDFS field, we use a combination of the S$z$F2.0 and S$z$F4.0 candidate catalogs as the ECDFS spectroscopic redshift fraction is roughly halfway in between that of the two mocks. Overall, a high degree of concordance is seen between the distributions of the candidates in each field versus those of the appropriate mocks indicating that likely (a) the LSS in the lightcone is largely similar to the LSS in the observed fields, and (b) our mapping methodology is equally effective at detecting structure in the mocks as in the real data.

Of course, there are several degeneracies affecting this argument on various levels. Chief among them is that we cannot fully rule out the possibility that the properties of LSS in the observed Universe are considerably different than those in our simulations. Such differences could result in discrepancies in the purity and completeness between the real and mock data that then, by chance, produce similar distributions. However, given all of the tests performed on the consistency between simulated galaxies and the observed galaxy catalogs and the care taken to reproduce our spectroscopic and photometric data quality and analysis strategy in the mocks, we take this as the less likely scenario and consider the high degree of concordance between the mock candidates and the candidates in the real data as lending credence to the efficacy of our approach.  

\subsection{Completeness and Purity}

Ideally, the catalog of candidates we find with the simulated data should be able to recover as many subsumed structures as possible, corresponding to high completeness, while minimizing spurious detections, corresponding to high purity. The completeness $C$ is given by:
\begin{equation}
C = \frac{S_{\rm{rec}}}{S_{\rm{total}}}
\end{equation}

\noindent where $S_{\rm{rec}}$ and $S_{\rm{total}}$ are the number of recovered and total subsumed structures, respectively. For this calculation, we limit the search area to the central 0.91 $\times$ 0.91 deg$^2$ of each mock field to avoid complications with edge effects. In order to further reduce ambiguities, we limit our completeness calculations to only those structures that are isolated.

\begin{figure*}
\centering
\includegraphics[width=2\columnwidth]{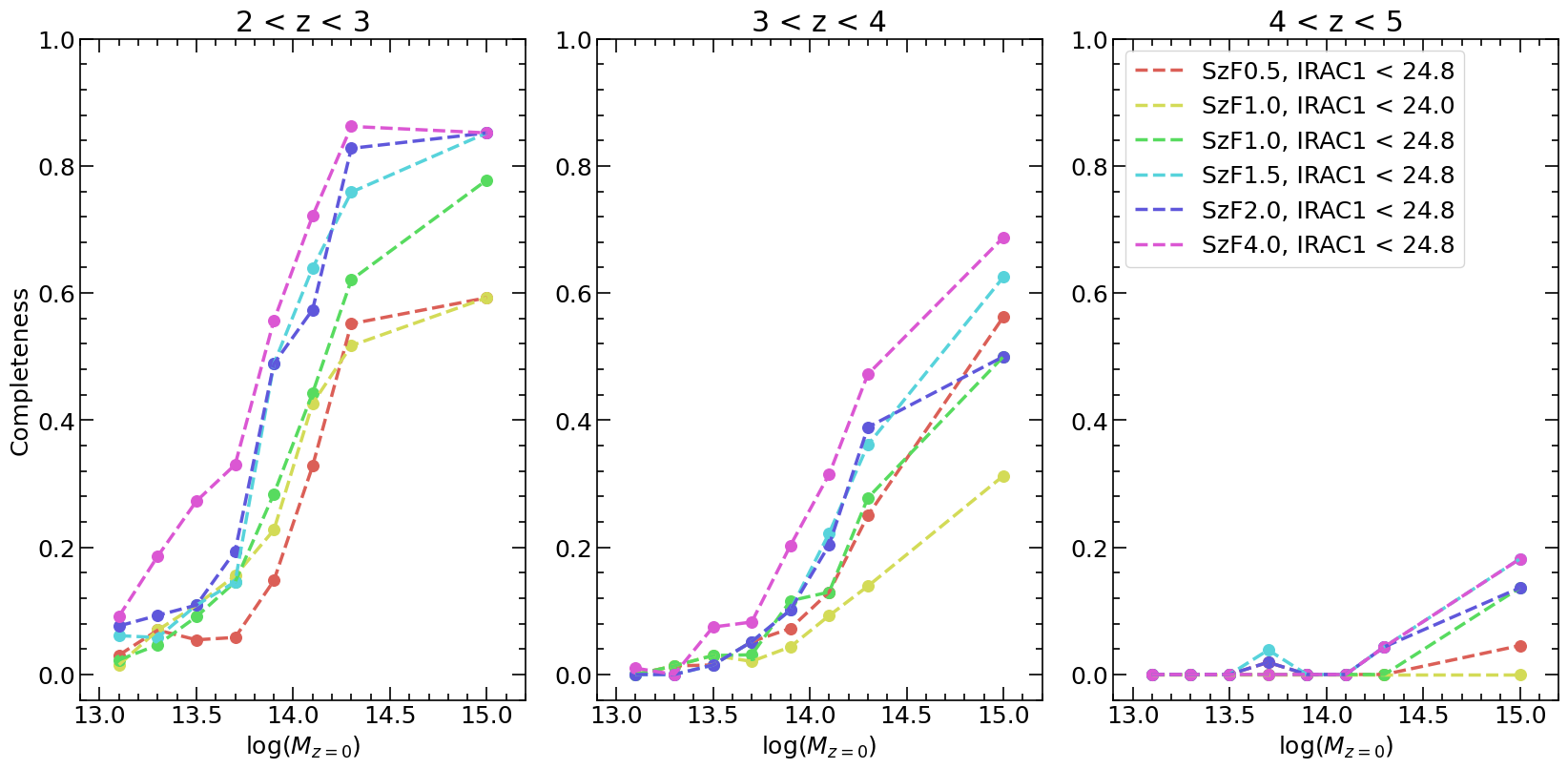}
\caption{Completeness as a function of mass, redshift, and S$z$F. The numbers for the S$z$F1.0 mocks at IRAC1 $< 24.8$ and IRAC1 $\la 24.0$, where we used multiple VMC maps, are median averages. The masses are separated into bin sizes of $\log(M_{tot}/M_{\odot}) = 0.2$ other than the highest mass bin at $\log(M_{tot}/M_{\odot}) > 14.4$, which will contain as few as 10 structures depending on the mock field. Generally, the completeness increases with mass and S$z$F and decreases with redshift, dropping to negligible numbers at $\log(M_{tot}/M_{\odot}) < 13.5$ and $z > 4$. With at least modest spectroscopy, however, we can achieve appreciable completeness at lower redshifts.} 
\label{fig:comp}
\end{figure*}

In Figure \ref{fig:comp} we show the completeness in three redshift bins, $2<z<3$, $3<z<4$, and $4<z<5$, as a function of $z=0$ mass for the different S$z$F variants. The completeness generally improves with decreasing redshift, higher mass structures, and higher levels of spectroscopy, reaching over 80\% at $\log(M_{tot}/M_{\odot}) > 14.0$ for $2<z<3$ in the S$z$F4.0 mock. These levels of completeness are comparable to those of $z\sim1$ structures in the ORELSE survey for similar levels of spectroscopic completeness using a similar methodology (see \citealt{Hung20}), which speaks to both the quality of the $z_{\rm{phot}}$ employed in this work and the efficacy of the lightcone training method.

There is a noticeable drop-off in completeness between the $2<z<3$ and $3<z<4$ redshift bins, approximately 30\% on average, though the completeness still generally remains high for the most massive structures and most heavily spectroscopically sampled mocks. A much larger drop in completeness is observed in our highest redshift bin at $4<z<5$, where we recover, at most, 20\% completeness at $\log(M_{tot}/M_{\odot}) > 14.4$ (S$z$F4.0). The completeness in the IRAC1 $\la 24.0$ mocks are roughly equivalent to the S$z$F0.5 IRAC1 $< 24.8$ mock for $2<z<3$, but the completeness numbers for the former mock are noticeably worse at $3<z<4$ and fall to zero at $4<z<5$. The latter is a natural consequence of the shallower photometric cut in that mask given how few galaxies are brighter than this apparent magnitude limit at such redshifts.

For the purity analysis, we impose no isolation criterion, as we are mainly concerned with recovering signal no matter what form it takes. The purity $P$ is given simply as

\begin{equation}
P = \frac{D_{\rm{matched}}}{D_{\rm{total}}}
\end{equation}

\noindent where $D_{\rm{matched}}$ and $D_{\rm{total}}$ are the number of matched and total candidates, respectively.

\begin{figure}
\centering
\includegraphics[width=\columnwidth]{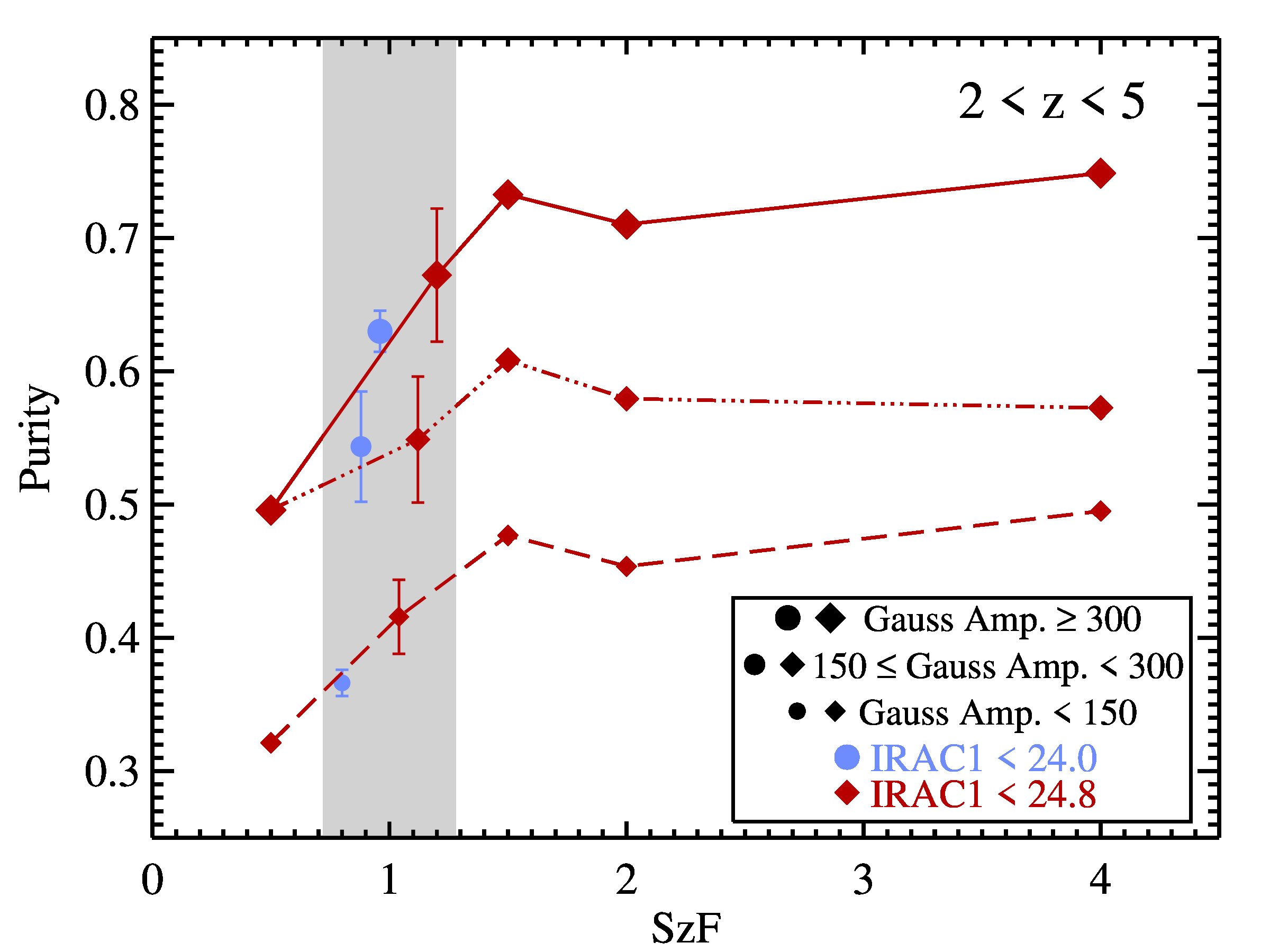}
\caption{Purity of our structure search algorithm at different levels of spectroscopic redshift completeness and with different photometric limits. The connected red diamonds indicate the purity level for mocks for which the photometric data are cut at IRAC1 $<$ 24.8, while the blue circles indicate the purity level for those mocks cut at IRAC1 $<$ 24. The size of the symbol indicates the Gaussian amplitudes considered for a given purity limit. The shaded region shows those purity levels for our nominal spectroscopic redshift fraction, S$z$F1.0, with a slight offset in the horizontal position of the points for clarity. The scatter in each S$z$F1.0 point indicates the $\sigma_{\rm{NMAD}}$ across all S$z$F1.0 mocks for a given photometric cut, and the point indicates the median purity value. For all other mocks, we simply plot the purity calculated for the single field that was mocked at that level of spectroscopic redshift completeness.}
\label{fig:pur}
\end{figure}

In Figure \ref{fig:pur} we plot the recovered purity as a function of S$z$F and as a function of Gaussian amplitude of the candidates. The purity is seen to improve precipitously with spectroscopic fraction up until it levels off at S$z$F1.5. Higher levels of spectroscopic sampling do not appear to improve the purity beyond about 80\%. However, the protostructure catalog that is used to assess the purity is cut at $log(M_{z=0}/M_{\odot})>13$, and it is likely that we are detecting lower-mass structures at higher S$z$F levels and spuriously considering them as false positives. While there does appear to be a dip in purity between S$z$F1.5 and S$z$F2.0, we note that we only analyzed multiple mocks for the S$z$F1.0 cases. The scatter seen in the purity values across the different mocks for the S$z$F1.0 case, represented by the error bars in the shaded region of Figure \ref{fig:pur}, would likely be the similar for higher S$z$F variants. Thus, it is likely that the purity numbers for S$z$F1.5, 2.0, and 4.0 are all statistically consistent. While our purity numbers noticeably improve when restricted to higher-amplitude candidates (i.e., generally higher mass structures), we impose no cut on our candidate catalog in order to maximize the number of structures we can potentially find.
 
\section{Results and Discussion}\label{discussion}

\subsection{The C3VO Protostructure Candidate Catalog}

We provide in Table \ref{tab:catalog} an abbreviated list of all 561 unique protostructure candidates in the galaxy density maps of the COSMOS, CFHTLS-D1, and ECDFS fields that fulfill all criteria outlined in Section \ref{sec.find_struct} over the redshift range $2<z<5$\footnote{Due to decreased data quality, in particular the shallower IRAC data in the field, the candidate catalog for the CFHTLS-D1 field is truncated to the redshift range $2 < z < 4.5$.}. The abbreviated 623 individual peaks and their parameters are reported in Table \ref{tab:peaks}. The full tables can be found online. Judging from our comparisons with simulations, the protostructure candidates listed likely have $z=0$ masses in excess of $>10^{13} M_{\odot}$, i.e., the progenitors of modern-day groups and clusters.

\input{catalog_short}
\input{catalog_peaks_short}

The candidates in this catalog vary tremendously in their strength, with Gaussian amplitudes spanning over 2 orders of magnitude from $\sim$20 to $>$2000, which likely indicates a similar dynamic range in protostructure masses, i.e., from $M_{z=0,\rm{tot}}\sim10^{13} M_{\odot}$ to $M_{z=0,\rm{tot}}\sim10^{15} M_{\odot}$. Of the 561 protostructure candidates, 343 are in the COSMOS field, 138 are in CFHTLS-D1, and 80 are in ECDFS. This distribution is reasonable given the size of each field and the relative data quality. In Table \ref{tab:catalog} we provide the field associated with each candidate, the barycenter in the R.A., decl., and redshift dimensions, the Gaussian amplitude of the associated candidate, the uncertainty in the Gaussian amplitude, the number of subsumed peaks and type of candidate (i.e., single versus in a multipeak chain, see Section \ref{sec.find_struct}), and a unique identifying designation. Though we urge caution, the Gaussian amplitudes can be translated into $M_{z=0,\rm{tot}}$ masses by adopting the appropriate parameters from Table \ref{tab:massfit} and the formula presented in Section \ref{sec.massrel} for each field (see Section \ref{sec.massrel} for the appropriate mocks for each field). Better methods exist for estimating the mass of candidate protostructures through dedicated zoom-in mapping (see, e.g., \citealt{Cucciati18, Lemaux18, Shen21, Forrest23, Shah24, Staab24}). However, these methods are highly parameter-sensitive and require each structure to be examined on an individual basis, with careful consideration for various factors like the spectral sampling, the tracer population, and the galaxy bias. Thus, they cannot be broadly applied to a large sample such as ours.

\subsection{Literature Comparisons}
\label{litcomp}
\begin{figure*}
\centering
\includegraphics[width=\columnwidth]{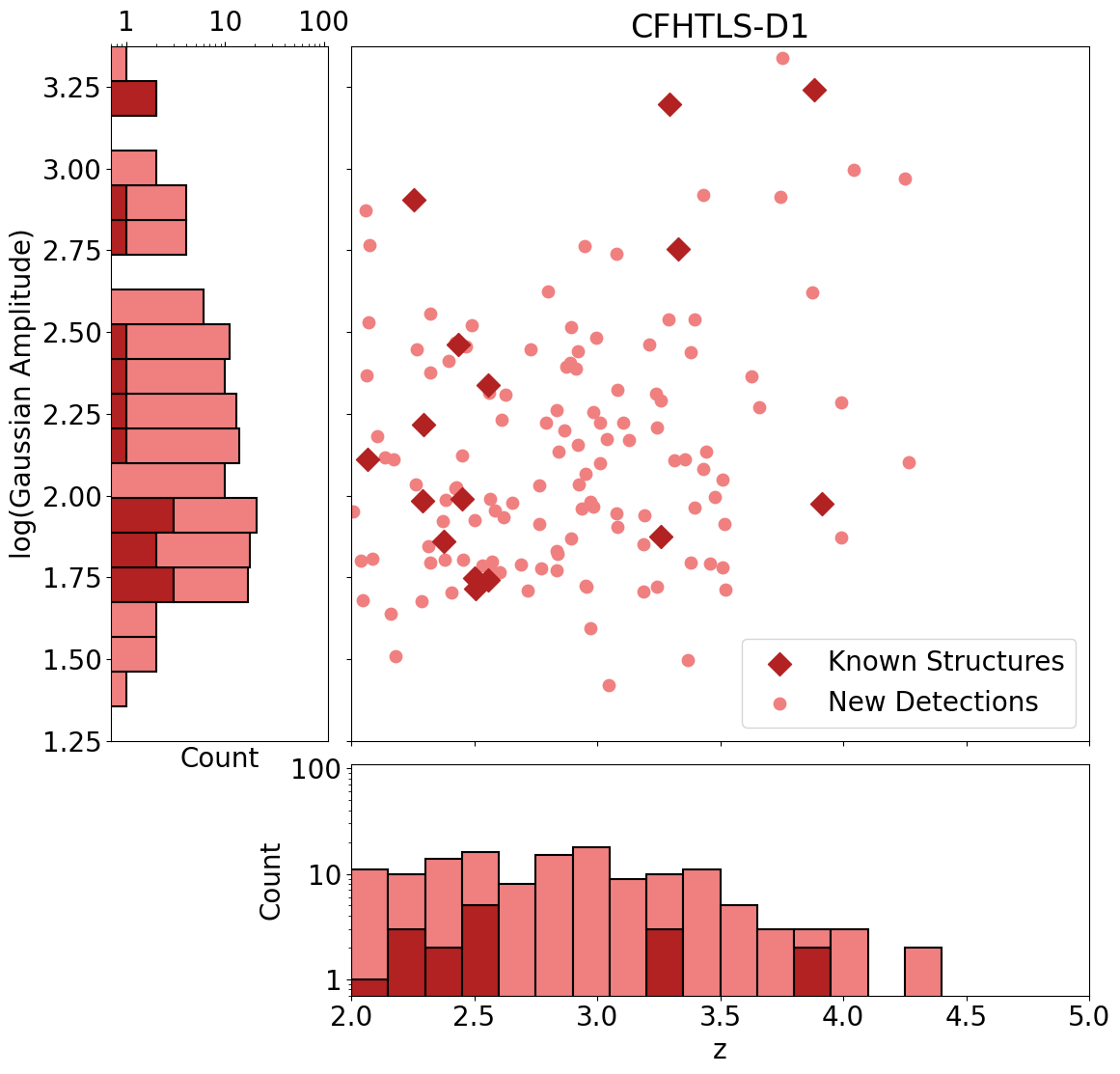}\includegraphics[width=\columnwidth]{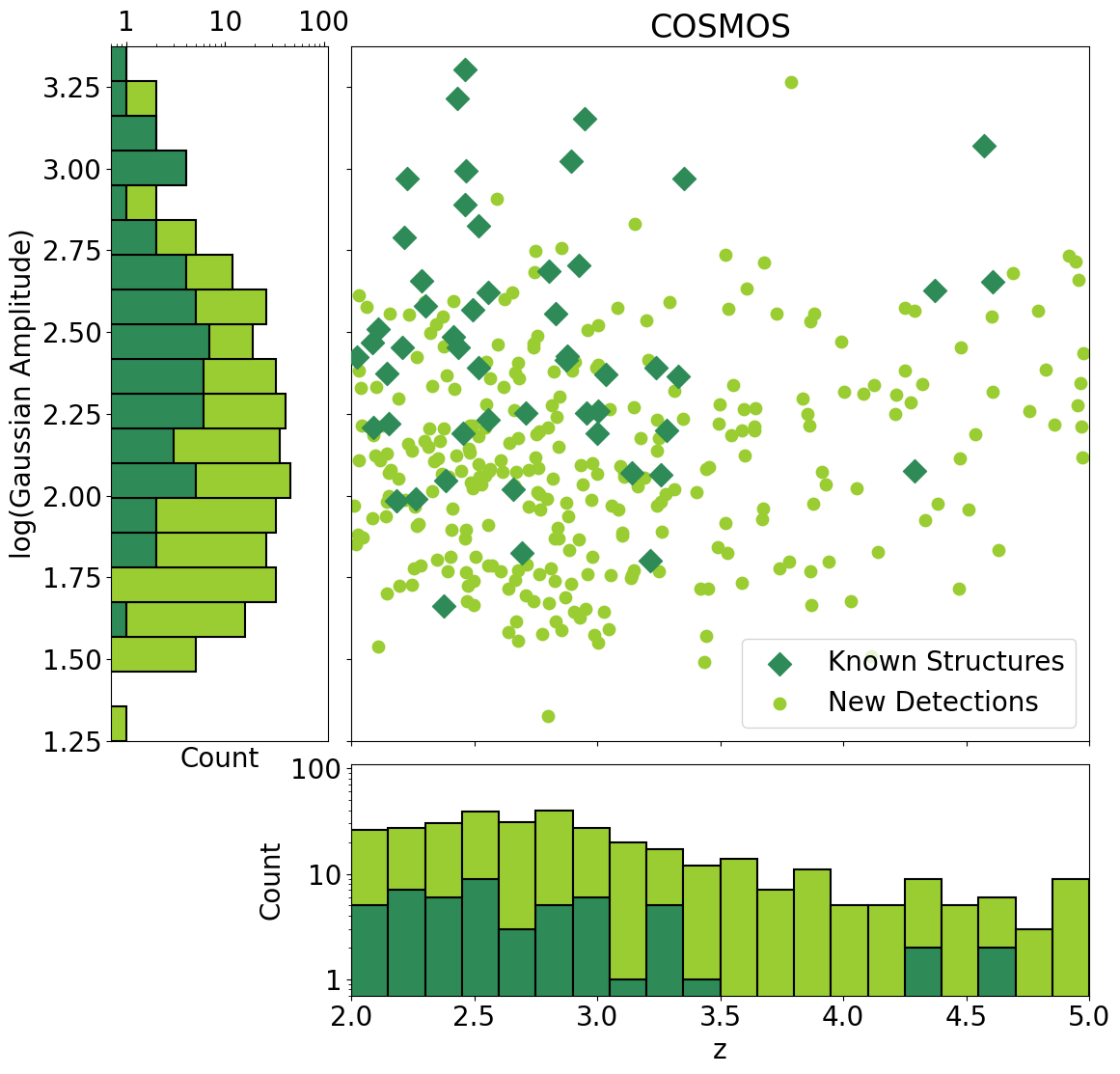}
\includegraphics[width=\columnwidth]{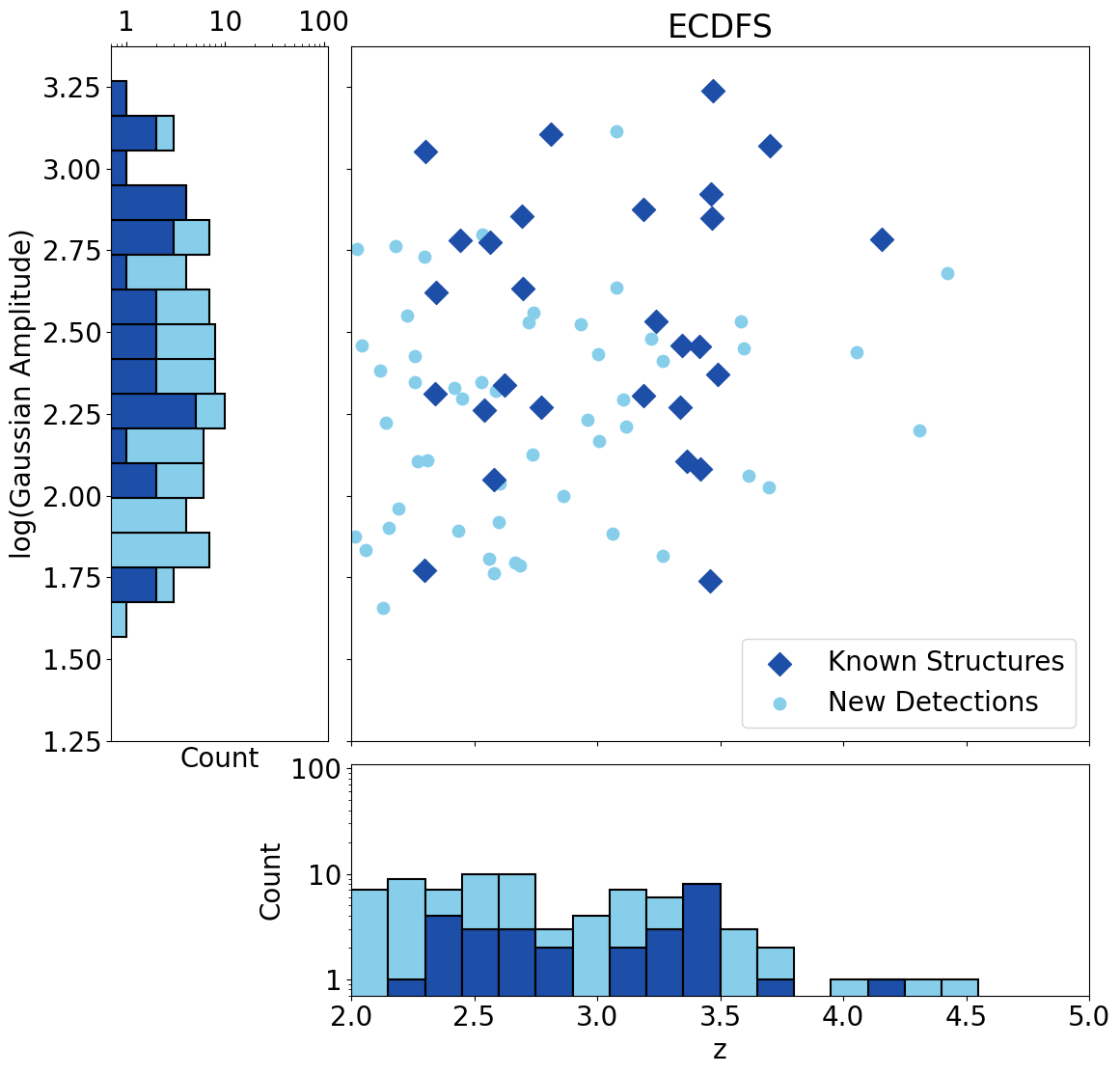}\includegraphics[width=\columnwidth]{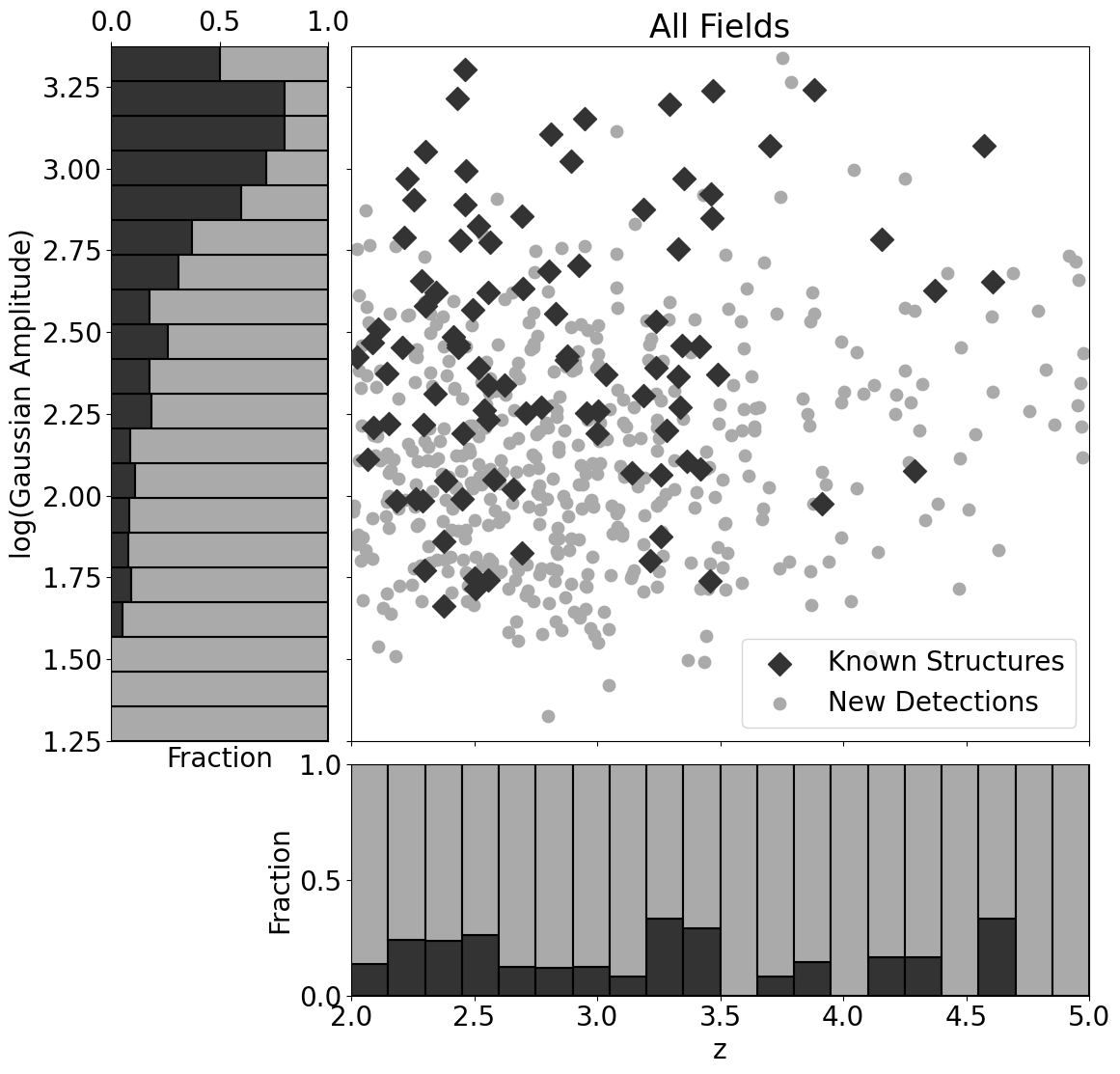}
\caption{The mean redshifts and Gaussian amplitudes of our protostructure candidates in all C3VO fields, differentiated by whether we could match them to previously known structures (denoted by the diamonds) or not (denoted by the circles). Due to the lack of data at higher redshifts, we limited our data to $2 < z < 4.5$ in the CFHTLS-D1 field. }
\label{fig:recovered_zamp}
\end{figure*}

In Tables \ref{tab:litsearchC}, \ref{tab:litsearchE}, and \ref{tab:litsearchV} we list the currently known protostructures in the three high-$z$ C3VO fields as reported in the literature. These lists contain only protostructures that have been spectroscopically confirmed to some level, either through galaxy-traced methods or through intergalactic medium (IGM) tomographic methods. In addition to the spectroscopically confirmed protostructures reported in the literature, we also include all spectroscopic overdensity candidates from the VUDS survey that were found via the spectroscopic overdensity methods described in \cite{Lemaux14b} and \cite{Lemaux18}. These candidates are also referred to as protostructures for the remainder of the paper for simplicity. Central coordinates and redshifts are taken directly from the literature for literature-reported structures and from a unit-weighted positional mean of all potential member galaxies for VUDS spectroscopic candidate overdensities. For protostructures reported in C3VO studies \citep{Cucciati14, Cucciati18, Lemaux18, Shen21, Forrest23, Shah24, Staab24}, we list the galaxy overdensity-weighted barycenter associated with each $5\sigma$ peak of a given protostructure when available. The various peaks are denoted either by the number appended to each structure name or by the proper name of the peak. The numbers of each peak match those in the referenced study. The references associated with each protostructure are listed in the Tables.

As we did for the matching process between simulated structure and candidates in the mock observations of the lightcone, we search for candidates for each protostructure listed in Tables \ref{tab:litsearchC}, \ref{tab:litsearchE}, and \ref{tab:litsearchV} within a line-of-sight distance of $\Delta z\le0.04$ and a transverse radius of d$_{\rm{trans}}\le2$ h$_{70}^{-1}$ Mpc. Since many of these protostructures are observed over much larger transverse scales, and since some protostructures (e.g., those from \citealt{Newman22}) have fairly large positional uncertainties, we additionally search on scales of d$_{\rm{trans}}\le5$ h$_{70}^{-1}$ Mpc if no associated candidates are found within d$_{\rm{trans}}\le2.5$ h$_{70}^{-1}$ Mpc (though still employing a line-of-sight criterion of $\Delta z\le0.04$). If multiple candidates are found within the search volume, the candidate with the largest associated mass is reported. We note that, beyond initial testing of the overdensity search algorithm using the VUDS candidate overdensities and a few of the published C3VO protostructures, the search algorithm was not optimized to find overdensities in the literature. Further, with the exception of a few pointed C3VO observations in the ECDFS field, in which we included DEIMOS/MOSFIRE observations targeting the protostructure ``Smruti'' at $z\sim3.5$, and in the COSMOS field, in which we included two DEIMOS masks targeting the protostructure ``Taralay'' at $z\sim4.57$, we do not include any spectroscopic observations specifically targeting any of the protostructures included in this comparison in the construction of our maps. The inclusion of such observations would only serve to increase our recovery fraction. 

In total, 97 protostructures are listed in Tables \ref{tab:litsearchC}, \ref{tab:litsearchE}, and \ref{tab:litsearchV}. Of these, 74\% (72/97) have a match to a candidate in our catalog within $\Delta z\le0.04$ and d$_{\rm{trans}}\le2.5$ h$_{70}^{-1}$ Mpc and 91\% (86/97) have a match within $\Delta z\le0.04$ and d$_{\rm{trans}}\le5$ h$_{70}^{-1}$ Mpc. Of the 14 protostructures that have no associated candidate(s) within d$_{\rm{trans}}\le2.5$ h$_{70}^{-1}$ Mpc but do have associated candidates within the larger radial search of d$_{\rm{trans}}\le5$ h$_{70}^{-1}$ Mpc, a large fraction ($\sim$40\%) are of peaks in known multicomponent systems (e.g., Hyperion, Elent\'{a}ri, PClJ0227-0421) where blending of the signal from the various peaks is likely to be an issue for our search algorithm. These blending issues are also likely to be at least partially responsible for the nondetections associated with four of the peaks of the known C3VO protostructures. We emphasize here that the search algorithm developed in this paper is optimized to \emph{detect} overdensities at high redshift, not to parse out their morphology or distribution of substructure at a detailed level. Once structure is detected, methods such as those used in \cite{Cucciati18}, \cite{Shen21}, \cite{Forrest23}, \cite{Shah24}, and \cite{Staab24} can be used to further zoom in on a given protostructure and map its density distribution on a finer scale. Of the nine protostructures that have no associated candidate(s), four (LATIS-D2-1, QPC.28, PG.4.53, and Tosh20.2) lie at the boundaries of our maps in the transverse or line-of-sight dimensions. 

As a reminder, most of the redshifts obtained from Keck/DEIMOS and Keck/MOSFIRE as part of the dedicated follow-up of protostructures in the three C3VO fields were not used in the creation of our maps. The exceptions are those redshifts described in \cite{Shen21}, \cite{Lemaux22}, and \cite{Shah24}, which comprise only $\sim$25\% of the full C3VO-Keck spectroscopic database and are almost exclusively limited to three protostructures (Smruti in ECDFS, Taralay in COSMOS, and PClJ027-0421 in CFHTLS-D1). Additionally, no redshifts obtained as part of the dedicated follow-up of any of the literature protostructures discussed in this Section were used in the making of our density maps. 

Overall, our mapping and search algorithms show an impressive recovery of known protostructures. We note that the protostructures listed only constitute roughly $\sim$20\% of the candidates presented in our catalog. Figure \ref{fig:recovered_zamp} shows a comparison between the redshifts and Gaussian amplitudes for previously known protostructures versus those that are newly discovered in this work. As expected, the previously known structures are primarily concentrated among the candidates with the largest Gaussian amplitudes (which are presumably the most massive candidates in these fields), as can be seen in the approximately exponentially increasing fraction of known structures with detection strength across all fields. We are able to find lower strength candidates in all three fields up to about $z\sim3.5$, with the redshift boundary in CFHTLS-D1 appearing especially sharp. Due to our lack of completeness at higher redshifts, we are unable to conclusively comment on the behavior seen in these plots for candidates at $z\ga4$.

\subsection{Comparison to the LATIS Survey}

Of note is the efficacy of the search algorithm in recovering signal for massive protostructures in the Ly$\alpha$ Tomography IMACS Survey \citep[LATIS;][]{Newman20, Newman22} via IGM tomography. This survey covers three fields, two of which overlap with C3VO: the CFHTLS-D1 field and the COSMOS field (i.e., CFHTLS-D2). The eight protostructures reported in \cite{Newman22}, of which we compare to seven in our study\footnote{The eighth protostructure is in the CFHTLS-D4 field---a field that is not covered by C3VO.}, are expected from simulations to be very massive ($\mathcal{M}_{tot}\sim10^{14}-10^{15.5}$) but are generally not well traced by rest-frame UV-selected galaxies. Despite an apparent paucity of such galaxies in many of these protostructures, six out of the seven strongest absorbing systems in LATIS as reported in \cite{Newman22} are recovered in our search. The only undetected protostructure (LATIS-D2-1) is near the edge of our VMC map and is likely prevented from being detected due to its proximity to the region where we consider our density reconstruction to be reliable. However, despite having successfully detected protostructures associated with most of the strongest LATIS absorption systems, the strength of the C3VO galaxy-traced candidates is moderate in general, with a median Gaussian amplitude of $\tilde{\rm{A}}$$\sim$200 for the detected LATIS structures versus a median Gaussian amplitude of $\tilde{\rm{A}}$$\sim$420 for all recovered structures taken from the literature. 

The results presented in \cite{Newman22} were based on a partial spectroscopic data set with only a portion of the planned survey depth, breadth, and sampling rate. The survey has since been completed, and IGM maps have been recomputed in all three fields covered by the survey (A. Newman 2025, private communication). For brevity, we refer to the second-generation maps, those computed with the full survey data, as ``LATIS2'' and the first-generation maps that formed the basis of \cite{Newman22} as ``LATIS1''. The strongest large-scale IGM detections in the LATIS2 maps are reported in Newman et al. (2025, in preparation). This sample includes the 19 strongest detections in COSMOS and the eight strongest detections in the CFHTLS-D1 field. Of these detections, we find counterparts for 17 of the 19 ($\sim$89\%) in the COSMOS field and four of the eight (50\%) in the CFHTLS-D1 field within $R_{\rm{proj}}<5$ Mpc and $\Delta z<0.05$\footnote{Eleven of the 20 counterparts are found within $R_{\rm{proj}}<2.5$ Mpc and $\Delta z<0.05$.}. Only one of the detections in the LATIS2 maps, LATIS2-D2-01 (formerly LATIS1-D2-1), lies at the edge of our maps and is likely undetectable. All other detections are well contained within the areal and redshift coverage of our maps. Interestingly, we do not detect any associated structure in our maps for either of the two strongest detections in the LATIS2 map in the CFHTLS-D1 field. This absence, and the generally decreased efficacy of our method at detecting LATIS structures in the CFHTLS-D1 field, is likely due to the decreased quality of data in that field in terms of the spectroscopic sampling, the photometric redshift uncertainties, and the shallower NIR data. The average Gaussian amplitude for LATIS2 counterparts detected in our maps is $\tilde{\rm{A}}$$\sim$210, slightly higher than that of the LATIS1 counterparts ($\tilde{\rm{A}}$$\sim$200), but only roughly half the strength of that of our typical literature counterpart ($\tilde{\rm{A}}$$\sim$420).

Since our density mapping is based on a combined spectroscopic and photometric sample that is limited in the rest-frame optical, making it effectively a stellar-mass-limited sample, the decrease in strength of LATIS detections may indicate that such structures are not only deficient in rest-frame UV-selected galaxies but rather in galaxies in general. However, the relative contribution of the spectroscopic sample in our mapping decreases with increasingly red colors, where redder star-forming galaxies and quiescent galaxies primarily enter the mapping through their photometric redshifts, which complicates interpretation. Ultimately, a full census in the rest-frame optical is necessary to make any definitive conclusions about the overall paucity of galaxies in LATIS structures.

\begin{figure}
\centering
\includegraphics[width=\columnwidth]{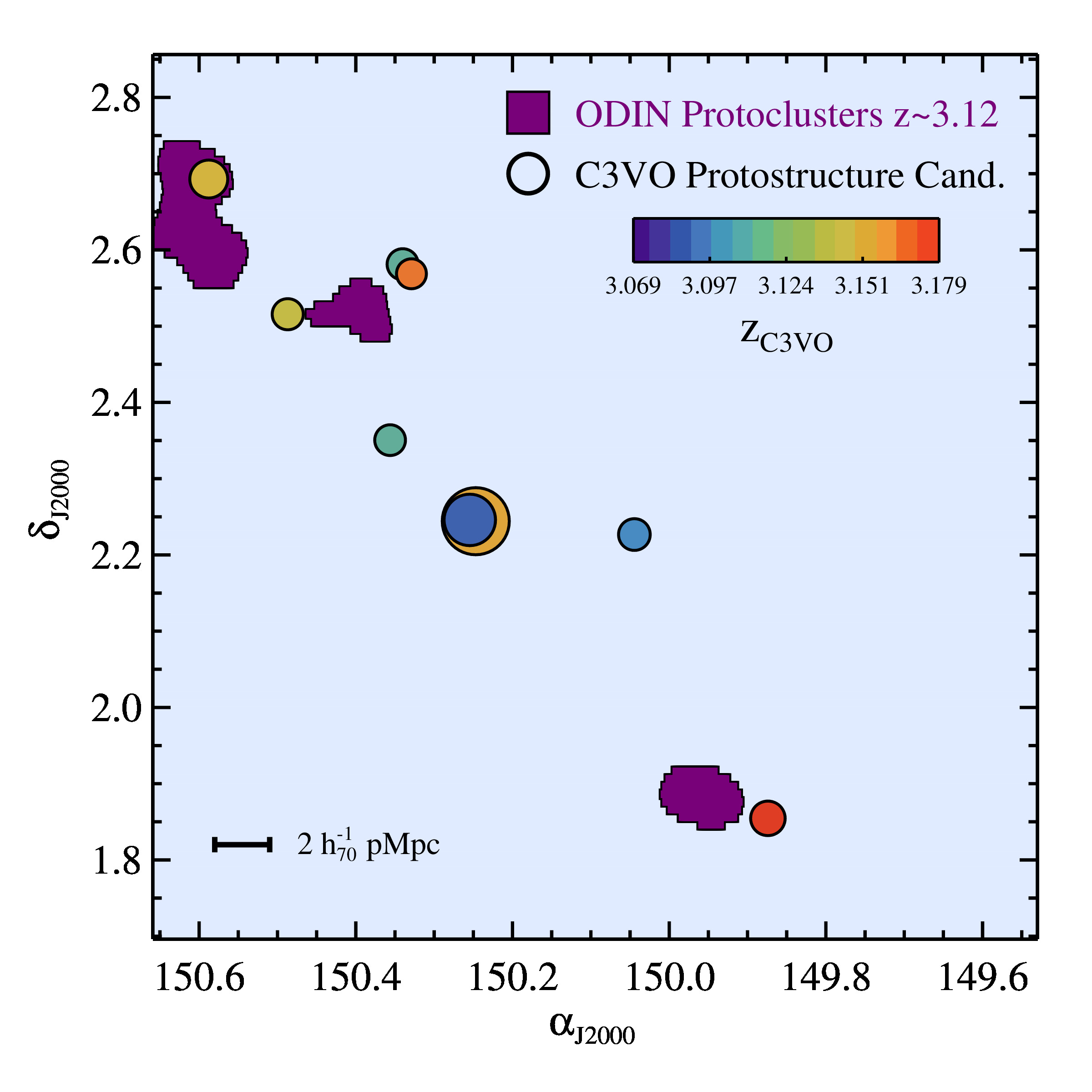}
\caption{Comparison between protoclusters detected as overdensities of LAEs at $z\sim3.1$ by the ODIN survey (dark purple regions), taken from \cite{Ramakrishnan23}, and C3VO protostructure candidates (colored circles) taken from this work. Candidates from C3VO are limited to the redshift range $3.068 \leq z \leq 3.180$, which matches the redshift selection of ODIN at these redshifts, and only those candidates with Gaussian amplitudes $\ge$100 are shown. Circles are logarithmically scaled by Gaussian amplitude of the C3VO detection, and the color bar indicates the redshift of C3VO protostructure candidates. All three ODIN protoclusters appear fairly well traced by C3VO protostructure candidates, though the barycenters of the two sets of structures appear offset by 2--2.5 Mpc in general. Approximately 50\% of the C3VO protostructure candidates in this volume, including the two strongest detections, do not have counterparts in the ODIN maps. This may imply large-scale suppression of Ly$\alpha$ emission in some high-redshift protostructures, though some of these structures have systemic redshifts outside the bounds of the ODIN redshift sensitivity function.}
\label{fig:ODIN-C3VO}
\end{figure}

\subsection{Comparison to the ODIN Survey}

The One-hundred-deg$^{2}$ DECam Imaging in Narrowbands \citep[ODIN;][]{Lee24} survey is designed to target Lyman$\alpha$-emitting galaxies (LAEs) in three redshift windows centered at $z\sim$2.4, 3.1, and 4.5. Observations are taken using custom-built narrowband filters installed in the Dark Energy Camera, which is mounted on the Cerro Tololo Inter-American Observatory Blanco 4 m telescope. This wide-field survey has overlap with several legacy extragalactic fields, including all three high-redshift fields targeted by C3VO. 

A comparison between C3VO overdensities at $z\sim2.45$ in the COSMOS field, using maps generated in a similar manner to what was used here, and ODIN precursor NB422 narrowband observations taken with the One-Degree Imager mounted on the Wisconsin--Indiana--Yale--NOIRLab (WIYN) Observatory targeting LAEs at the same redshift that was already presented in \cite{Huang22}. These observations broadly mimic those of ODIN, though they are slightly shallower, target only the lowest redshift that ODIN targets ($z\sim2.45$), and are taken in a filter with an FWHM roughly twice as wide as that of the ODIN observations ($\sim$75 \AA\, versus 170\AA). In this Section, we concentrate on a comparison between protostructure candidate detections in C3VO and those detected in ODIN in the COSMOS field at $z\sim3.1$. Specifically, we focus on maps generated by \cite{Ramakrishnan23} of the extended COSMOS field from early release data. These maps cover nearly an order-of-magnitude larger area than the $\sim$1.3 deg$^2$ C3VO maps in the COSMOS field. A comparison of overdensities between the two surveys at $z\sim2.45$ and $z\sim4.55$ in the COSMOS field, as well as the comparisons in other fields will be the subject of future work.

As discussed extensively in \cite{Huang22}, LAEs appear to trace overdensities in a complicated manner. While certain structures appear to be well traced by LAEs, with some structures exhibiting higher levels of LAEs overdensity relative to other galaxy types, other structures appear to be largely devoid of LAEs \citep[e.g.,][]{Apostolovski24}. In the Hyperion proto-supercluster \citep{Cucciati18}, \cite{Huang22} found that LAEs generally trace the overall galaxy density distribution well, with several of the most prominent matter peaks in Hyperion exhibiting strong LAE overdensities, and a significant correlation was observed between the density distribution of LAEs and spectroscopically confirmed member galaxies. However, at least one of the peaks (peak 5; \citealt{Wang16}) was not well traced by LAEs, perhaps due, at least in part, to the advanced evolutionary stage of that peak. Similar variation between the density of LAEs and the overall galaxy population has also been observed in several case studies of overdensities at high redshift \citep[e.g.,][]{Shi19a, Shi19b}/ Additionally, it potentially originates from variations in the density of large-scale, cool H I gas reservoirs, which have the effect of diffusing or, in the presence of dust, destroying Ly$\alpha$ photons, causing an overall suppression of LAEs. Indeed, \cite{Huang22} also found, by cross-correlating IGM tomographic maps from the COSMOS Lyman-Alpha Mapping And Mapping Observations (CLAMATO, \citealt{Lee16, Lee18}) and LAEs, that while LAEs appear to trace structure moderately rich in large-scale, cool H I gas well, LAEs tend to avoid regions with the highest H I column densities (see also \citealt{Momose21}). A multitude of other considerations could also modulate the overdensity of LAEs within LSSs, including the prevalence of outflows due to potential increased AGN activity \citep[e.g.,][]{Shah24b} or star formation activity \citep[e.g.,][]{Lemaux22, Staab24}, changes in the age and/or dust content of constituent member galaxies, and small number statistics due in part to LAEs being fairly rare populations at these redshifts \citep[e.g.,][]{Cassata15}. 

In Figure \ref{fig:ODIN-C3VO} we show a comparison between ODIN-detected protoclusters and C3VO protostructure candidates in the region of COSMOS where the ODIN maps and our VMC maps overlap. We limit the C3VO protostructure candidates to those with Gaussian amplitudes $\ge$100 and a redshift range $3.068 \leq z \leq 3.180$. The amplitude cut was made in order to include only those C3VO candidates with a $\ga$50\% chance of being a genuine overdensity. The redshift range was chosen to order to roughly match the redshift selection function of ODIN N501 narrowband observations ($3.095 \leq z \leq 3.154$) with an additional ($\pm\Delta z=0.025$) buffer in order to account for uncertainties associated with C3VO systemic redshifts and the wings of the NB501 throughput curve. In total, three ODIN protoclusters lie within the overlap volume, all three of which have at least one C3VO protostructure candidate whose barycenter falls within a projected distance of 2.5 Mpc of the ODIN protocluster barycenters. Both ODIN protoclusters in the northeast of the overlap volume appear to be potentially associated with multiple C3VO protostructure candidates at a variety of different redshifts. The ODIN protocluster at [R.A., decl.] = [150.54, 2.52] also appears to be associated with MAGAZ3NEJ1001, which is an overdensity that, perhaps not surprisingly given its presence in the ODIN map, appears to consist of primarily bluer star-forming galaxies, demographics that hold even for its most massive members \citep{McConachie24}. All C3VO protostructure candidates that are within a proximity of 2.5 Mpc of ODIN protoclusters have moderate Gaussian amplitudes in the range of $A\sim100--200$.

In contrast, the inverse comparison reveals that the two most strongly detected, and presumably most massive, protostructure candidates in the C3VO map in this volume ($A\sim375--675$), which are located at [R.A., decl.] = [150.25, 2.25], do not have counterpart overdensities in the ODIN maps. Additionally, several protostructure candidates with Gaussian amplitudes that are similar to those associated with the ODIN protoclusters do not have associated detections in the ODIN maps, with $\sim50$\% of C3VO candidates lying farther than 2.5 Mpc from an ODIN protocluster. However, three out of the four of the C3VO candidates that are not associated with an ODIN protocluster lie either outside or at the edge of the redshift selection window for ODIN LAEs. The one C3VO candidate with a systemic redshift that lies well within the ODIN redshift selection window (located at [R.A., decl.] = 150.36, 2.35) has a barycenter that lies within 5 Mpc of an ODIN-selected protocluster in projection. Furthermore, the two strongest C3VO detections in this volume were detected as potential overdensities in \cite{Ramakrishnan23}, but did not exhibit a strong enough overdensity over a sufficiently large area to pass their fiducial cuts for classification as a genuine protocluster. As both of these C3VO candidates lie at the extrema of the ODIN redshift selection window, it is likely that the ODIN observations are only able to recover a small fraction of member galaxies in these systems. 

Generally, there does appear to be some level of variation in the observed overdensity of typical star-forming galaxies, as probed by C3VO, and LAEs, as probed by ODIN, with $\sim1.5--2.5$ Mpc projected spatial offsets seen between the barycenter of C3VO candidates and those of associated ODIN protoclusters as well as the additional C3VO structure without an obvious ODIN counterpart. While some of this variation is likely attributable to differences in the redshift selection function between the two surveys as well as various systematics, it is possible that some of this variation is due to large-scale processes that modulate the emission and/or escape of Ly$\alpha$ photons among member galaxies in high-redshift protoclusters. A systematic comparison between C3VO protostructure candidates and ODIN LAE-selected overdensities in all three high-redshift C3VO fields at all three redshift ranges for which ODIN is sensitive is necessary in order to better understand processes serving to suppress Ly$\alpha$ emission in some protoclusters and enhance it in others. 

\startlongtable
\begin{deluxetable*}{ccccccccccc}
\tablecolumns{11}
\tablecaption{Comparison of the C3VO Candidate Protocluster Catalog to the Literature in the COSMOS Field}
\label{tab:litsearchC}
\tablehead{
\cutinhead{\textbf{COSMOS}}
\colhead{Structure} & \colhead{R.A.$_{\rm{lit}}$} & \colhead{Decl.$_{\rm{lit}}$} & \colhead{$z_{\rm{lit}}$} & \colhead{R.A.$_{\rm{C3VO}}$} & \colhead{Decl.$_{\rm{C3VO}}$} & \colhead{$z_{\rm{C3VO}}$} & \colhead{Amp.\tablenotemark{a}} & \colhead{d$_{\rm{trans}}$} & \colhead{ID$_{\rm{C3VO}}$$^{b}$} & \colhead{ref.} \\
\colhead{} & \colhead{[$^{\circ}$]} & \colhead{[$^{\circ}$]} & \colhead{} & \colhead{[$^{\circ}$]} & \colhead{[$^{\circ}$]} & \colhead{} & \colhead{} & \colhead{[h$_{70}^{-1}$ Mpc]} & \colhead{} & \colhead{}
}
\startdata
C.PS4  &   150.4500  &  2.3710  &    2.035  & 150.4395 & 2.3726 & 2.024 & 265 & 0.33 & C\textunderscore SP38 & 1,2 \\
COSTCO2  &   149.8710  &  2.2290  &   2.047  & 149.9323 & 2.1311 & 2.089 & 162 & 3.57$^{\ast}$ & C\textunderscore SP93 & 3 \\
ZFIRE  &   150.0940  &  2.2510  &   2.095  & 150.0459 & 2.2415 & 2.112 & 322 & 1.51$^{+}$ & C\textunderscore SP28 & 4 \\
G237  &   150.5070  &  2.3120  &  2.160  & 150.4889 & 2.3617 & 2.146 & 237 & 1.62 & C\textunderscore SP54 & 5,6 \\
COSTCO3  &   150.1290   &  2.2750  &  2.160  & 150.0957 & 2.2449 & 2.208 & 284 & 1.38 & C\textunderscore SP33 & 3 \\
C.PS6  &   150.1900  &  2.1330  &  2.174  & 150.1786 & 2.1887 & 2.153 & 166 & 1.74 & C\textunderscore SP87 & 1,2 \\
C.PS8     &  149.8640    &   2.0900   &  2.216  & 149.8603 & 2.0950 & 2.215 & 615 & 0.19 & C\textunderscore SP7 & 1,2 \\
BD2.2  &   150.1970  &  2.0030   &  2.232  &  150.1848 & 2.0176 & 2.226 & 929 & 0.58$^{+}$ & C\textunderscore SP6 & 7 \\
C.PS9  &   149.9200  &  2.2210  &  2.277  &  149.8851 & 2.1652 & 2.284 & 455 & 2.00 & C\textunderscore SP14 & 1,2,8 \\
COSTCO5  &   149.9380   &  2.0910  &  2.283  & 150.0112 & 2.1181 & 2.262 & 98 & 2.38 & C\textunderscore SP175 & 3 \\
COSTCO1  &   150.1100  &  2.1610  &  2.298  & 150.1021 & 2.2218 & 2.301 & 380 & 1.86 & C\textunderscore SP34 & 3 \\
COSTCO4  &   149.7060  &  2.0240  &   2.391  & 149.7715 & 2.1389 & 2.416 & 305 & 3.99$^{\ast, +}$ & C\textunderscore SP29 & 3 \\
Theia  &   150.0937  &  2.4049  &  2.468  & 150.0951 & 2.4086 & 2.468 & 988 & 0.12$^{+}$ & C\textunderscore SP5 & 9, 10, 11 \\
Eos  &   149.9765  &  2.1124  &  2.426  & 150.0125 & 2.1124 & 2.432 & 1640 & 2.31 & C\textunderscore SP2 & 9 \\
Helios  &   149.9996  &  2.2537  &  2.444  & 150.0125 & 2.1124 & 2.432 & 1640 & 2.25 & C\textunderscore SP2 & 8, 9, 12, 13 \\
Selene  &   150.2556  &  2.3423  &  2.469  & 150.1992 & 2.3781 & 2.462 & 2013 & 2.00$^{+}$ & C\textunderscore SP1 & 9, 11 \\
Hyp5  &   150.2293  &  2.3381  &  2.507  & 150.1992 & 2.3781 & 2.462 & 2013 & 1.50$^{+}$ & C\textunderscore SP1 & 9, 14 \\
Hyp6  &   150.3316  &  2.2427  &  2.492  & 150.3403 & 2.2236 & 2.496 & 370 & 0.63$^{+}$ & C\textunderscore SP22 & 9 \\
Hyp7  &   149.9581  &  2.2187  &  2.423  & 150.0125 & 2.1124 & 2.432 & 1640 & 2.01 & C\textunderscore SP2 & 9 \\
LATIS1-D2-0  &  150.1030  &  2.1750  &  2.562  &  150.1327 & 2.3380 & 2.558 &  417 & 4.93$^{\ast, +}$ & C\textunderscore SP18 & 15 \\
LATIS1-D2-1$^{\dagger}$  &  149.5320  &  1.9880  &  2.460 & \nodata & \nodata & \nodata & \nodata & \nodata & \nodata & 9 \\
LATIS1-D2-2  &  149.6910  &  2.1750  &  2.679 & 149.6475 & 2.1254 & 2.711 & 179 & 1.94 & C\textunderscore SP137 & 15 \\
LATIS1-D2-3  &  150.3470  &  2.2740  &  2.457 & 150.3403 & 2.2236 & 2.496 & 370 & 1.53$^{+}$ & C\textunderscore SP22 & 15 \\
LATIS1-D2-4  &  150.0330  &  2.1750  &  2.685 & 150.0188 & 2.2508 & 2.658 & 105 & 2.27$^{+}$ & C\textunderscore SP165 & 15 \\
QPC.2.8$^{\dagger}$  &    150.0436   &  1.6799  &  2.766  & \nodata & \nodata & \nodata & \nodata & \nodata & \nodata & 16 \\
C.PS14  &    150.1900  &  2.2900  &  2.816 & 150.1916 & 2.2928 & 2.805 & 487 & 0.09 & C\textunderscore SP12 & 1,2 \\
C.PS15  &    149.9480  &  2.3070  &  2.816 & 149.9523 & 2.3208 & 2.830 & 361 & 0.42 & C\textunderscore SP23 & 1,2 \\
PClJ1000+0200  &  150.0910  &  1.9920  &  2.911  & 150.0675 & 1.9907 & 2.895 & 1056 & 0.68$^{+}$ & C\textunderscore SP4 & 17 \\
C.PS17  &    149.9110  &  2.2680  &  2.921 & 149.9077 & 2.1100 & 2.924 & 507 & 4.56$^{\ast}$ & C\textunderscore SP11 & 1,2 \\
C.PS18  &    150.2500  &  2.2790  &  2.953 & 150.2746 & 2.2719 & 2.951 & 1418 & 0.74 & C\textunderscore MP2$^{\ddagger}$ & 1,2 \\
C.PS19  &  149.7000  &  1.9790  &  3.002 & 149.9725 & 2.3217 & 3.004 & 181 & 0.59 & C\textunderscore SP80 & 1,2 \\
C.PS20  &  149.9980  &  2.3000  &  3.035  & 149.9725 & 2.3217 & 3.035 & 235 & 0.95 & C\textunderscore SP55 & 1,2 \\
MAGAZ3NEJ1001  &  150.4417  &  2.5192  &  3.123  & 150.4869 & 2.5157 & 3.141 & 118 & 1.28 & C\textunderscore MP52$^{\ddagger}$ & 18 \\
C.PS21  &   149.6900  &  1.8090  &  3.232  & 149.6874 & 1.8094 & 3.232 & 63 & 0.08 & C\textunderscore SP223 & 1,2 \\
C.PS22  &    149.8810  &  1.8500  &  3.232  & 149.8937 & 1.8543 & 3.238 & 246 & 0.38 & C\textunderscore SP47 & 1,2 \\
Elent\'{a}ri1  &  149.8060  &  2.4510  &  3.366  & 149.9001 & 2.3145 & 3.353 & 931 & 4.58$^{\ast}$ & C\textunderscore MP4$^{\ddagger}$ & 19, 20 \\
Elent\'{a}ri2  &  149.8774  &  2.2850  &  3.341  & 149.9001 & 2.3145 & 3.353 & 931 & 1.03 & C\textunderscore MP4$^{\ddagger}$ & 19 \\
Elent\'{a}ri3  &  149.9369  &  2.2749  &  3.269  & 149.8863 & 2.3472 & 3.282 & 159 & 2.46$^{+}$ & C\textunderscore MP46$^{\ddagger}$ & 19 \\
Elent\'{a}ri4  &  150.4021  &  2.3356  &  3.255  & 150.4004 & 2.3382 & 3.258 & 116 & 0.09 & C\textunderscore SP148 & 19 \\
Elent\'{a}ri5  &  150.2153  &  2.4941  &  3.343  & 150.2141 & 2.4612 & 3.329 & 231 & 0.91 & C\textunderscore SP57 & 19 \\
Elent\'{a}ri6  &  149.8287  &  2.4306  &  3.315  & 149.9001 & 2.3145 & 3.353 & 931 & 3.78$^{\ast,+}$ & C\textunderscore MP4$^{\ddagger}$ & 19 \\
Elent\'{a}ri7  &  150.1099  &  2.5634  &  3.354  & 150.2141 & 2.4612 & 3.329 & 231 & 4.03$^{\ast}$ & C\textunderscore SP57 & 19, 20 \\
Elent\'{a}ri8  &  149.8869  &  2.3242  &  3.248  & 149.8863 & 2.3472 & 3.282 & 159 & 0.64 & C\textunderscore MP46$^{\ddagger}$ & 19 \\
Elent\'{a}ri9  &  149.7315  &  2.5829  &  3.408  & \nodata & \nodata & \nodata & \nodata & \nodata & \nodata & 19 \\
Elent\'{a}ri10  &  150.4236  &  2.5071  &  3.409  & \nodata & \nodata & \nodata & \nodata & \nodata & \nodata & 19 \\
C.PS25     &  150.1000  &  1.9770  &  3.803 & \nodata & \nodata & \nodata & \nodata & \nodata & \nodata & 1,2 \\
TI.S1     &  150.3518 & 2.4781  & 4.327 & 150.2001 & 2.3864 & 4.290 & 119 & 4.44$^{\ast}$ & C\textunderscore SP140 & 21 \\
TI.S2     &  150.1951 & 2.3069 & 4.372 & 150.1956 & 2.3048 & 4.372 & 423 & 0.05 & C\textunderscore SP16 & 14 \\
PG.4.53$^{\dagger}$  &   150.6025   &  2.4041  &  4.531  & \nodata & \nodata & \nodata & \nodata & \nodata & \nodata & 22 \\
Taralay1  &  150.3522   &  2.3540   &  4.567  & 150.3551 & 2.3534 & 4.573 & 1179 & 0.07 & C\textunderscore SP3 & 2,21 \\
Taralay2  &  150.1771   &  2.3013   &  4.592  & 150.1372 & 2.3240 & 4.609 & 450 & 1.12 & C\textunderscore SP15 & 2,21 \\
\enddata
\tablenotetext{a}{Amplitude of the Gaussian fit to the linked candidate.}
\tablenotetext{b}{ID from Table \ref{tab:catalog}. Note that some matched candidates may be near the boundary of our maps, see corresponding entry in Table \ref{tab:catalog}.}
\tablenotetext{\ast}{No candidate within $R_{\rm{proj}}<2.5$ Mpc, but one or more candidates within $R_{\rm{proj}}<5$ Mpc.}
\tablenotetext{+}{Multiple candidates within the matching volume, the properties of the strongest candidate are reported. The ID$_{\rm{C3VO}}$ of the strongest candidate is listed first.}
\tablenotetext{\dagger}{Near the border of the VMC map, unlikely to be detected due to border masking.}
\tablenotetext{\ddagger}{Single peak of a multiply peaked chain, estimated amplitude and mass less certain.}
\tablerefs{[1] \cite{Lemaux14b}; [2] \cite{Lemaux18}; [3] \cite{Ata22}; [4] \cite{Yuan14}; [5]  \cite{Koyama21}; [6] \cite{Polletta21}; [7] \cite{Darvish20}; [8] \cite{FrancknMcGaugh16}; [9] \cite{Cucciati18}; [10] \cite{Casey15}; [11] \cite{Diener13}; [12] \cite{Lee16}; [13] \cite{Chiang15}; [14] \cite{Wang16}; [ 15] \cite{Newman22}; [16] \cite{Ito23}; [17] \cite{Cucciati14}; [18] \cite{McConachie24}; [19] \cite{Forrest23}; [20] \cite{McConachie22}; [21] \cite{Staab24}; [22] \cite{Kakimoto23}.}
\end{deluxetable*}

\begin{deluxetable*}{ccccccccccc}
\tablecolumns{11}
\tablecaption{Comparison of the C3VO Candidate Protocluster Catalog to the Literature in the ECDFS Field}
\label{tab:litsearchE}
\tablehead{
\cutinhead{\textbf{ECDFS}}
\colhead{Structure} & \colhead{R.A.$_{\rm{lit}}$} & \colhead{Decl.$_{\rm{lit}}$} & \colhead{$z_{\rm{lit}}$} & \colhead{R.A.$_{\rm{C3VO}}$} & \colhead{Decl.$_{\rm{C3VO}}$} & \colhead{$z_{\rm{C3VO}}$} & \colhead{Amp.$^{a}$} & \colhead{d$_{\rm{trans}}$} & \colhead{ID$_{\rm{C3VO}}$$^{b}$} & \colhead{ref.} \\
\colhead{} & \colhead{[$^{\circ}$]} & \colhead{[$^{\circ}$]} & \colhead{} & \colhead{[$^{\circ}$]} & \colhead{[$^{\circ}$]} & \colhead{} & \colhead{} & \colhead{[h$_{70}^{-1}$ Mpc]} & \colhead{} & \colhead{} 
}
\startdata
VANDELS1  &  53.0629  &   -27.7237  &  2.290  & 53.0742 & -27.7048 & 2.304 & 1131 & 0.65 & E\textunderscore SP5 & 1 \\
VANDELS2  &  53.1412  &   -27.6871   &  2.300 & 53.0742 & -27.7048 & 2.304 & 1131 & 1.88 & E\textunderscore SP5 & 1 \\
VANDELS3  &  53.1421  &   -27.8221  &  2.340  & 53.0742 & -27.7048 & 2.304 & 1131 & 4.00$^{+}$ & E\textunderscore SP5 & 1 \\
E.PS1  &   53.0925  &   -27.7158  &  2.451 & 53.0570 & -27.6994 & 2.442 & 606 & 1.07 & E\textunderscore SP20 & 2,3 \\
E.PS2  &   53.1348  &   -27.9125  &  2.559  & 53.0828 & 27.9125 & 2.569 & 490 & 1.56$^{+}$ & E\textunderscore SP13 & 2,3 \\
E.PS3  &   53.1504  &   -27.8098  &  2.622 & 53.0631 & -27.7850 & 2.613 & 265 & 2.40$^{+}$ & E\textunderscore SP39 & 2,3 \\
Drishti1  &  53.0731  &   -27.9323  &  2.674  & 53.0690 & -27.8883 & 2.692 & 715 & 1.30 & E\textunderscore SP8 & 4 \\
Drishti2  &  53.1876  &   -27.7943  &  2.694  & 53.2178 & -27.7836 & 2.699 & 431 & 0.85 & E\textunderscore SP18 & 4 \\
Drishti3  &  53.1133   &  -27.8984   &  2.697 & 53.0690 & -27.8883 & 2.692 & 715 & 1.19 & E\textunderscore SP8 & 4 \\
Surabhi  &   52.9988   &  -27.8063  &  2.795  & 52.9321 & -27.8681 & 2.774 & 186 & 2.49 & E\textunderscore SP45 & 4 \\
VANDELS6  &  53.2037   &  -27.7746  &  2.800  & 53.0948 & -27.8190 & 2.813 & 1279 & 3.09$^{\ast}$ & E\textunderscore SP3 & 1 \\
VANDELS7  &  53.1412   &  -27.8612  &  3.170  & 53.1955 & -27.7507 & 3.189 & 750 & 3.39$^{\ast,+}$ & E\textunderscore SP7 & 1 \\
VANDELS8  &  53.1346   &  -27.6954  &  3.230  & 53.1955 & -27.7507 & 3.189 & 750 & 2.16 & E\textunderscore SP7 & 1 \\
VANDELS9  &  53.1229   &  -27.7404  &  3.290  & \nodata & \nodata & \nodata & \nodata & \nodata & \nodata & 1 \\
Shrawan1  &  53.2727  &   -27.7936  &  3.343  & 53.2721 & -27.7902 & 3.344 & 288 & 0.09 & E\textunderscore SP29 & 4 \\
Shrawan2  &  53.0714  &   -27.9353  &  3.355  & 53.0346 & -27.9438 & 3.363 & 128 & 0.93 & E\textunderscore SP58 & 4 \\
Shrawan3  &  53.1552  &   -27.8959  &  3.242  & 53.1510 & -27.8919 & 3.237 & 342 & 0.15 & E\textunderscore SP24 & 4 \\
Shrawan4  &  53.2022   &  -27.9406  &  3.335   & 53.1984 & -27.9349 & 3.337 & 186 & 0.18 & E\textunderscore SP46 & 4 \\
Smruti1  &   53.0076  &   -27.7463  &  3.410  &  53.0193 & -27.7403 & 3.417 & 286 & 0.33 & E\textunderscore SP30 & 4,5,6,7 \\
Smruti2  &   53.0042  &   -27.7411  &  3.479 & 53.0159 & -27.7498 & 3.490 & 234 & 0.37 & E\textunderscore SP36 & 4,5,6,7 \\
Smruti3  &   53.0613  &   -27.8723  &  3.471  & 53.8723 & -27.8795 & 3.469 & 1731 & 1.96$^{+}$ & E\textunderscore SP1 & 4,5,6,7 \\
Smruti4  &   53.2290  &   -27.8828  &  3.462  & 53.8723 & -27.8795 & 3.469 & 1731 & 2.11$^{+}$ & E\textunderscore SP1 & 4,5,6,7 \\
Smruti5  &   53.0412  &   -27.7804  &  3.530  & 53.0159 & -27.7498 & 3.490 & 234 & 1.03 & E\textunderscore SP36 & 4,5,6,7 \\
Smruti6   &  53.1586  &  -27.6964  &  3.418  & 53.1521 & -27.7078 & 3.418 & 121 & 0.25$^{+}$ & E\textunderscore SP60 & 4,5,6,7 \\
Sparsh  &   53.0579  &   -27.8670  &  3.696  & 53.0478 & -27.8730 & 3.701 & 1173 & 0.29 & E\textunderscore SP4 & 4,8 \\
Ruchi1  &   53.2124  &   -27.8306  &  4.150  & 53.2185 & -27.8273 & 4.155 & 610 & 0.16 & E\textunderscore SP11 & 4 \\
Ruchi2  &   53.1659  &   -27.6199  &  4.109 & \nodata & \nodata & \nodata & \nodata & \nodata & \nodata & 4 \\
\enddata
\tablenotetext{a}{Amplitude of the Gaussian fit to the linked candidate}
\tablenotetext{b}{ID from Table \ref{tab:catalog}. Note that some matched candidates may be near the boundary of our maps; see corresponding entry in Table \ref{tab:catalog}.}
\tablenotetext{\ast}{No candidate within $R_{\rm{proj}}<2.5$ Mpc, but one or more candidates within $R_{\rm{proj}}<5$ Mpc.}
\tablenotetext{+}{Multiple candidates within the matching volume; the strongest candidate is reported.}
\tablerefs{[1] \cite{Guaita20}; [2] \cite{Lemaux14b}; [3] \cite{Lemaux18}; [4] \cite{Shah24}; [5] \cite{Forrest17}; [6] \cite{Ginolfi17}; [7] \cite{Zhou20}; [8] \cite{KangnIm09}.}
\end{deluxetable*}

\begin{deluxetable*}{ccccccccccc}  
\tablecolumns{11}
\tablecaption{Comparison of the C3VO Candidate Protocluster Catalog to the Literature in the CFHTLS-D1 Field}
\label{tab:litsearchV}
\tablehead{
\cutinhead{\textbf{CFHTLS-D1}}
\colhead{Structure} & \colhead{R.A.$_{\rm{lit}}$} & \colhead{Decl.$_{\rm{lit}}$} & \colhead{$z_{\rm{lit}}$} & \colhead{R.A.$_{\rm{C3VO}}$} & \colhead{Decl.$_{\rm{C3VO}}$} & \colhead{$z_{\rm{C3VO}}$} & \colhead{Amp.$^{a}$} & \colhead{d$_{\rm{trans}}$} & \colhead{ID$_{\rm{C3VO}}$$^{b}$} & \colhead{ref.}  \\
\colhead{} & \colhead{[$^{\circ}$]} & \colhead{[$^{\circ}$]} & \colhead{} & \colhead{[$^{\circ}$]} & \colhead{[$^{\circ}$]} & \colhead{} & \colhead{}  & \colhead{[h$_{70}^{-1}$ Mpc]} & \colhead{} & \colhead{}
}
\startdata
V.PS1  &   36.5850  &   -4.5200  &  2.074  & 36.5831 & -4.4711 & 2.068 & 129 & 1.51 & V\textunderscore SP41 & 1,2 \\
V.PS2  &   36.7850  &   -4.2000  &  2.230  & 36.8062 & -4.1717 & 2.257 & 802 & 1.08 & V\textunderscore MP7$^{\ddagger}$ & 1,2 \\
V.PS3  &   36.7900  &   -4.3750  &  2.302  & 36.7875 & -4.3534 & 2.295 & 165 & 0.66 & V\textunderscore SP29 & 1,2 \\
V.PS4  &   36.7350  &   -4.3330  &  2.307  & 36.7875 & -4.3534 & 2.295 & 165 & 1.71$^{+}$ & V\textunderscore SP29 & 1,2 \\
V.PS5a  &   36.6270  &   -4.3450  &  2.394  & 36.6335 & -4.3443 & 2.433 & 291 & 0.20$^{+}$ & V\textunderscore SP12 & 1,2 \\
V.PS5b  &   36.6270  &       -4.3450  &  2.448  & 36.6335 & -4.3443 & 2.433 & 291 & 0.20 & V\textunderscore SP12 & 1,2 \\
LATIS1-D1-1  &  36.2200  &   -4.3530  &  2.455  & 36.1043 & -4.4570 & 2.452 & 98 & 4.67$^{\ast}$ & V\textunderscore SP56 & 3 \\
V.PS6  &    36.6340  &   -4.2040  &  2.527  & 36.6890 & -4.1833 & 2.558 & 55 & 1.75 & V\textunderscore SP97 & 1,2 \\
LATIS1-D1-0  &  36.1550  &   -4.5340  &  2.568  & 36.1354 & -4.4592 & 2.556 & 219 & 2.30 & V\textunderscore SP25 & 3 \\
Tosh16.1  &  36.1338  &   -4.3170  &  3.131  & \nodata & \nodata & \nodata & \nodata & \nodata & \nodata & 4 \\
PClJ0227-0421.1  &  36.6375  &   -4.2381  &  3.269  & 36.6969 & -4.3563 & 3.292 & 1574 & 3.69$^{\ast}$ & V\textunderscore SP2 & 1,5 \\
PClJ0227-0421.2  &  36.7438  &   -4.2949  &  3.325  & 36.6969 & -4.3563 & 3.292 & 1574 & 2.14$^{+}$ & V\textunderscore SP2 & 1,5 \\
PClJ0227-0421.3  &  36.7775  &   -4.3457  &  3.299  & 36.6969 & -4.3563 & 3.292 & 1574 & 2.25$^{+}$ & V\textunderscore SP2 & 1,5 \\
PClJ0227-0421.4  &  36.6467  &   -4.3703  &  3.300  & 36.6969 & -4.3563 & 3.292 & 1574 & 1.44 & V\textunderscore SP2 & 1,5 \\
PClJ0227-0421.5  &  36.5817  &   -4.3877  &  3.298  & 36.6969 & -4.3563 & 3.292 & 1574 & 3.31$^{\ast,+}$ & V\textunderscore SP2 & 1,5 \\
PClJ0227-0421.6  &  36.6417  &   -4.4472  &  3.302  & 36.6434 & -4.4638 & 3.258 & 75 & 0.47 & V\textunderscore SP76 & 1,5 \\
Tosh20.1  &  36.4454  &     -4.8340  &  3.675  & \nodata & \nodata & \nodata & \nodata & \nodata & \nodata & 6 \\
V.PS9    &  36.4655  &    -4.3750  &  3.886  & 36.4027 & -4.2737 & 3.884 & 1739 & 3.12$^{\ast}$ & V\textunderscore MP2$^{\ddagger}$ & 1,2 \\
Tosh20.2  &  36.1960  &   -4.9120  &  4.898  & \nodata & \nodata & \nodata & \nodata & \nodata & \nodata & 6 \\
\enddata
\tablenotetext{a}{Amplitude of the Gaussian fit to the linked candidate}
\tablenotetext{b}{ID from Table \ref{tab:catalog}. Note that some matched candidates may be near the boundary of our maps, see corresponding entry in Table \ref{tab:catalog}.}
\tablenotetext{\ast}{No candidate within $R_{\rm{proj}}<2.5$ Mpc, but one or more candidates within $R_{\rm{proj}}<5$ Mpc.} 
\tablenotetext{+}{Multiple candidates within the matching volume; the strongest candidate is reported.}
\tablenotetext{\dagger}{Near the border of the VMC map, unlikely to be detected due to border masking.}
\tablenotetext{\ddagger}{Single peak of a multiply peaked chain, estimated mass less certain.}
\tablerefs{[1] \cite{Lemaux14b}; [2] \cite{Lemaux18}; [3] \cite{Newman22}; [4] \cite{Toshikawa16}; [5] \cite{Shen21}; [6] \cite{Toshikawa20}.}
\end{deluxetable*}

\section{Conclusions}\label{conclusion}

By applying our search algorithm to our plethora of photometry and spectroscopy in the C3VO fields, we were able to find detections of 561 protostructure candidates, many of which have not previously been reported in the literature. The main takeaways from our findings are as follows:

\begin{itemize}
\item We have successfully applied the VMC mapping technique that was previously used in \cite{Hung20} at $z\sim1$ to populations that were not only at much higher redshifts, but with a data set that had, by virtue of the increased faintness of the structure member population, both a lower spectroscopic fraction and photometric redshifts with lower levels of precision and accuracy than those data sets used at $z\sim1$, further demonstrating the powerful capabilities of VMC mapping.
\item Over 400 potentially new protostructures were found in the COSMOS, ECDFS, and CFHTLS-D1 fields, all of which have some spectroscopically confirmed members.
\item With the aid of mock observations of simulated structures, we have found completeness and purity as high as $\sim$80\% and $\sim$70\%, respectively, in the best-case scenario. These numbers correspond to $2<z<3$ and structures with $\log(M_{tot}/M_{\odot}) > 14.4$ for completeness, and a spectral redshift fraction of $\sim10$\% for purity (i.e., S$z$F1.5 or better). Typical completeness and purity are around 20--40\% for $\log(M_{tot}/M_{\odot}) > 14.0$, though the precise value sensitively depends on the mass and redshift, with a precipitous falloff to $z>4$. Purity is typically found to be $\sim$40--60\%. 
\item More than 90\% of the $\sim$100 protoclusters previously reported in the literature (which were originally selected using a variety of different methods) in these three fields were recovered with our method. Of the few protoclusters that were not detected using our methodology, the vast majority were either situated near the edge of our maps or in multicomponent systems where our deblending was not aggressive enough to discriminate different subcomponents. 
\item We also compared our protostructure candidate catalog to protoclusters selected in the LATIS IGM tomographic survey and the ODIN narrowband LAE survey. We found a large degree of concordance between C3VO candidates and protoclusters selected in LATIS and ODIN, though some interesting differences were found. 
\item The R.A./decl./$z$ coordinates of all protostructure candidates are made publicly available as an electronic table on CDS. The Gaussian amplitude ($A$) of each protostructure candidate, which represents a rough mass proxy, as well as its associated error, is also provided.
\item Fits are provided between our mass proxy ($A$) and the $z=0$ mass of simulated structures for differing levels of spectroscopic completeness. These fits may be applied to the Gaussian amplitude of our candidate protostructure detections to derive crude masses for the observed candidates, though with considerable scatter and some bias. We caution that masses of our candidates can be estimated more accurately and precisely, but it must be on a case-by-case basis considering individual parameters such as the spectral sampling, tracer population, and galaxy bias for the tracers, as has been done with other C3VO studies \citep[e.g.,][]{Cucciati18, Shen21, Forrest23}.
\end{itemize}

With our results, we have once again demonstrated the efficacy of the VMC mapping method and its utility when combined with any large and complete spectroscopic and photometric data set. While our work was on a fairly small region of the sky, the methodology is adaptable to any spectroscopic and photometric data set of sufficient size. This is especially important to consider in the context of active and upcoming large-scale surveys, such as those conducted by Euclid\footnote{\url{https://www.euclid-ec.org/}}, the Vera C. Rubin Observatory\footnote{\url{https://rubinobservatory.org/}}, Subaru's Prime Focus Spectrograph\footnote{\url{https://pfs.ipmu.jp}}, the Nancy Grace Roman Space Telescope\footnote{\url{https://roman.gsfc.nasa.gov/}}, and the Maunakea Spectroscopic Explorer\footnote{\url{https://mse.cfht.hawaii.edu/}}, which are able to reach deeper magnitudes and offer larger data sets at even higher redshifts than ever before. Of these, Euclid is the only telescope currently in operation, and the Euclid Wide Survey in particular has the potential to find tens of thousands of protoclusters with masses $\geq 10^{14} M_{\odot}$ over $1.5 < z < 4$ in its 14,500 deg$^2$ area of coverage \citep{euclid}. Methods based largely on the VMC technique are already proving to be effective at uncovering a large number of protocluster candidates in EUCLID mock data and will be used for searches in real data (Ramos-Chernenko 2025, in preparation). Thanks to the deluge of data on the way that is focused on the redshift range that marks the transition between clusters and protoclusters, the next decade promises to be a boon for protocluster science. 

\section*{Acknowledgments}

{\footnotesize
We thank the anonymous referee for their valuable comments in improving this work. Some of the material presented in this paper is supported by the National Science Foundation under grant Nos. 1411943 and 1908422. This work was partially supported by NASA’s Astrophysics Data Analysis Program under grant No. 80NSSC21K0986. This work was additionally supported by the France-Berkeley Fund, a joint venture between UC Berkeley, UC Davis, and le Centre National de la Recherche Scientifique de France promoting lasting institutional and intellectual cooperation between France and the United States. This study is based, in part, on data products from observations made with ESO Telescopes at the La Silla Paranal Observatory under ESO program ID 179.A-2005 and on data products produced by TERAPIX and the Cambridge Astronomy Survey Unit on behalf of the UltraVISTA consortium. This study is based, in part, on data collected at the Subaru Telescope and obtained from the SMOKA, which is operated by the Astronomy Data Center, National Astronomical Observatory of Japan. This work is based, in part, on observations made with the Spitzer Space Telescope, which is operated by the Jet Propulsion Laboratory, California Institute of Technology under a contract with NASA. UKIRT is supported by NASA and operated under an agreement among the University of Hawaii, the University of Arizona, and Lockheed Martin Advanced Technology Center; operations are enabled through the cooperation of the East Asian Observatory. When the data reported here were acquired, UKIRT was operated by the Joint Astronomy Centre on behalf of the Science and Technology Facilities Council of the U.K. This study is also based, in part, on observations obtained with WIRCam, a joint project of CFHT, Taiwan, Korea, Canada, France, and the Canada--France--Hawaii Telescope which is operated by the National Research Council (NRC) of Canada, the Institut National des Sciences de l'Univers of the Centre National de la Recherche Scientifique of France, and the University of Hawai'i. This paper was supported by the international Gemini Observatory, a program of NSF NOIRLab, which is managed by the Association of Universities for Research in Astronomy (AURA) under a cooperative agreement with the U.S. National Science Foundation, on behalf of the Gemini partnership of Argentina, Brazil, Canada, Chile, the Republic of Korea, and the United States of America. Some portion of the spectrographic data presented herein was based on observations obtained with the European Southern Observatory Very Large Telescope, Paranal, Chile, under Large Programs 070.A-9007 and 177.A-0837. The remainder of the spectrographic data presented herein were obtained at the W.M. Keck Observatory, which is operated as a scientific partnership among the California Institute of Technology, the University of California, and the National Aeronautics and Space Administration. The Observatory was made possible by the generous financial support of the W.M. Keck Foundation. The authors wish to recognize and acknowledge the very significant cultural role and reverence that the summit of Maunakea has always had within the Native Hawaiian community. We are most fortunate to have the opportunity to conduct observations from this mountain.}

\facility{Blanco, CFHT, GALEX, ESO:VISTA, Keck:I, Keck:II, Magellan:Baade, Max Planck:2.2m, Spitzer, Subaru, VLT:Melipal}

\bibliographystyle{aasjournal}
\bibliography{C3VO_vmc} 

\appendix
\section{Boundary Effects in the Lightcone}\label{appA}
Because of the finite size of the cubes that are used to construct the lightcones (see Section \ref{sec.lc_const} for details), there exist boundary effects at the interface between cubes that have the potential to cause under-richness in structures whose centers lie near the edge of the cubes. In addition, the lightcone is constructed in such a way that it subtends a circular region of radius of $\sim$2.3 deg, which causes additional boundary effects at the border of the lightcone along the transverse dimensions. In order to determine which structures are affected by these boundary effects, we utilize a version of the lightcone that probes 2$\times$ deeper than the version of lightcone that is used for the vast majority of our mocks (i.e., IRAC1$<$25.5 versus IRAC1$<$24.8) in order to search for simulated protostructures that are significantly under-rich in member galaxies relative to the population average. A deeper IRAC1 cut is used for this exercise in order to differentiate protostructures that are under-rich due to border effects from those that appear under-rich in the IRAC1$<$24.8 due to a shift in the luminosity function of member galaxies to fainter luminosities for astrophysical reasons.

As discussed in Section \ref{sec.lc_subsume}, we break the sample of simulated protostructures into redshift bins of $\Delta z=0.5$ over the redshift range $2<z<5$ and perform a (log) linear fit of $M_{z=0,\rm{tot}}$ versus $N_{\rm{mem}}$ with iterative $3\sigma$ clipping. All protostructures that are offset from the fit relation by $\ge$3$\sigma$ on the low end of this mass--richness relation are removed from our sample and are not considered in any of the metrics presented in this paper. More specifically, we do not use any subsumed structures that contain a clipped protostructure in the mass fit. Formally, we should also remove these structures from consideration in the estimate of the completeness and the purity calculation as well, but the inclusion of these structures have a negligible effect on the overall completeness numbers and would introduce a great deal of complexity into the purity calculation for likely the same result. 

\begin{figure*}
\centering
\includegraphics[width=0.5\columnwidth]{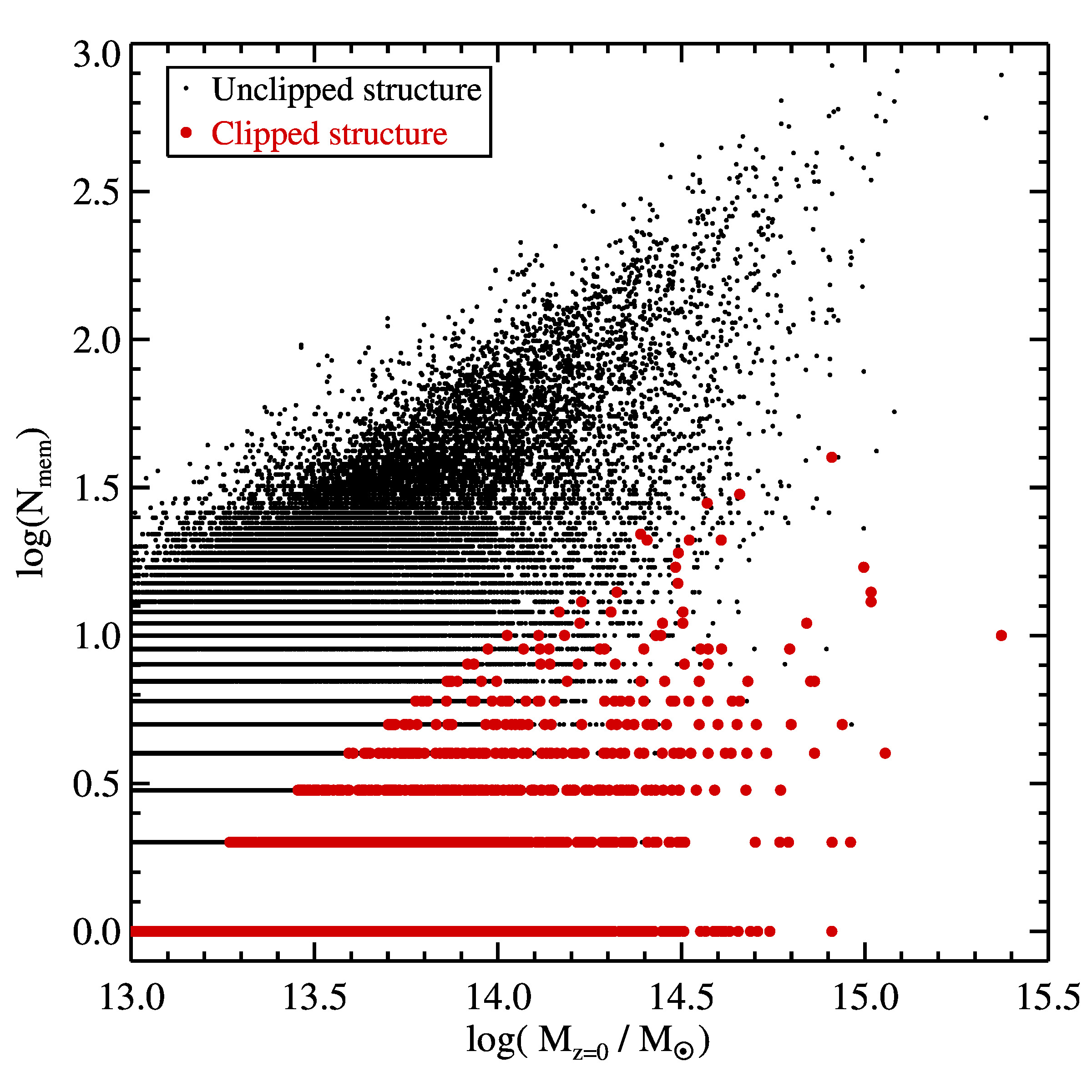}\includegraphics[width=0.5\columnwidth]{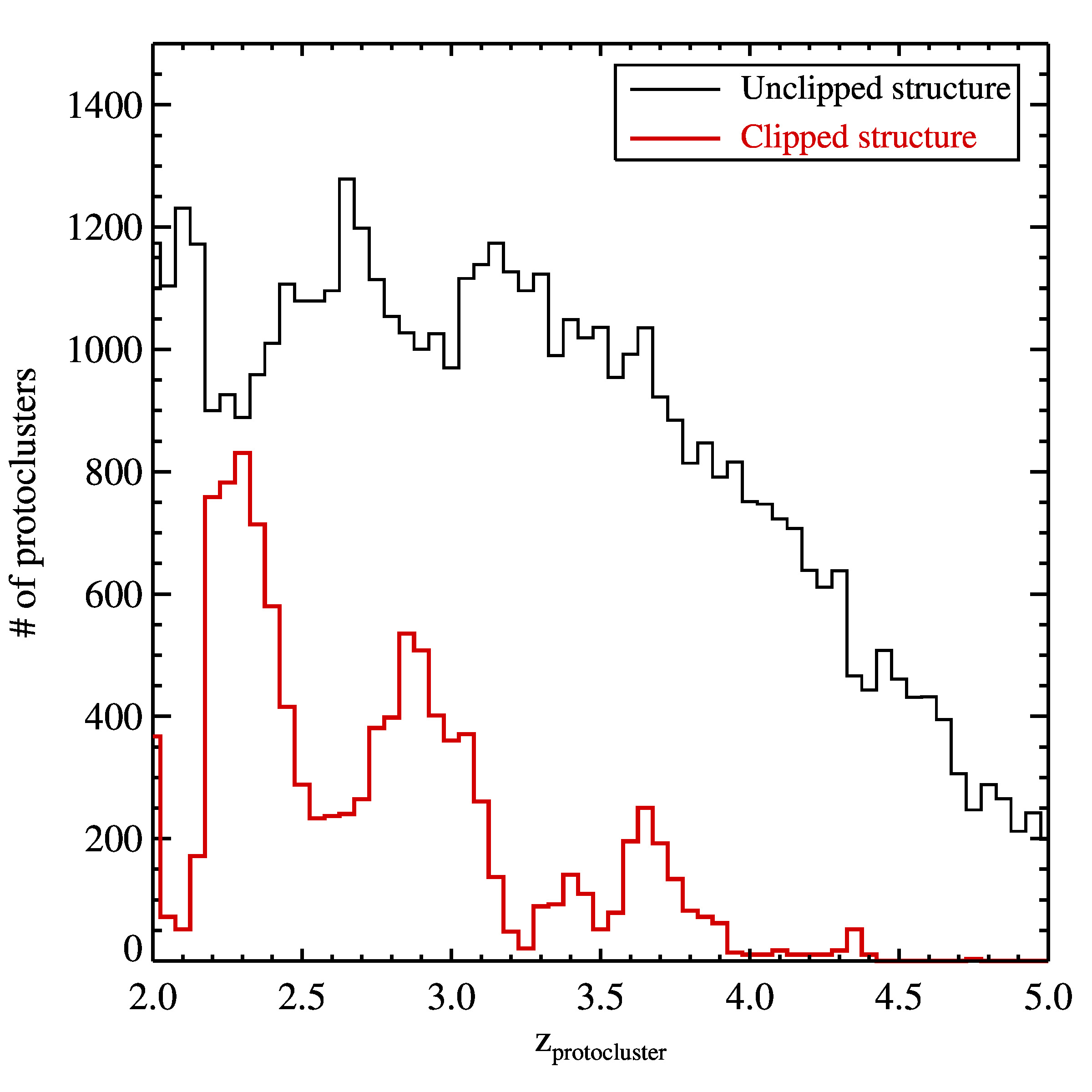}
\includegraphics[width=0.6\columnwidth]{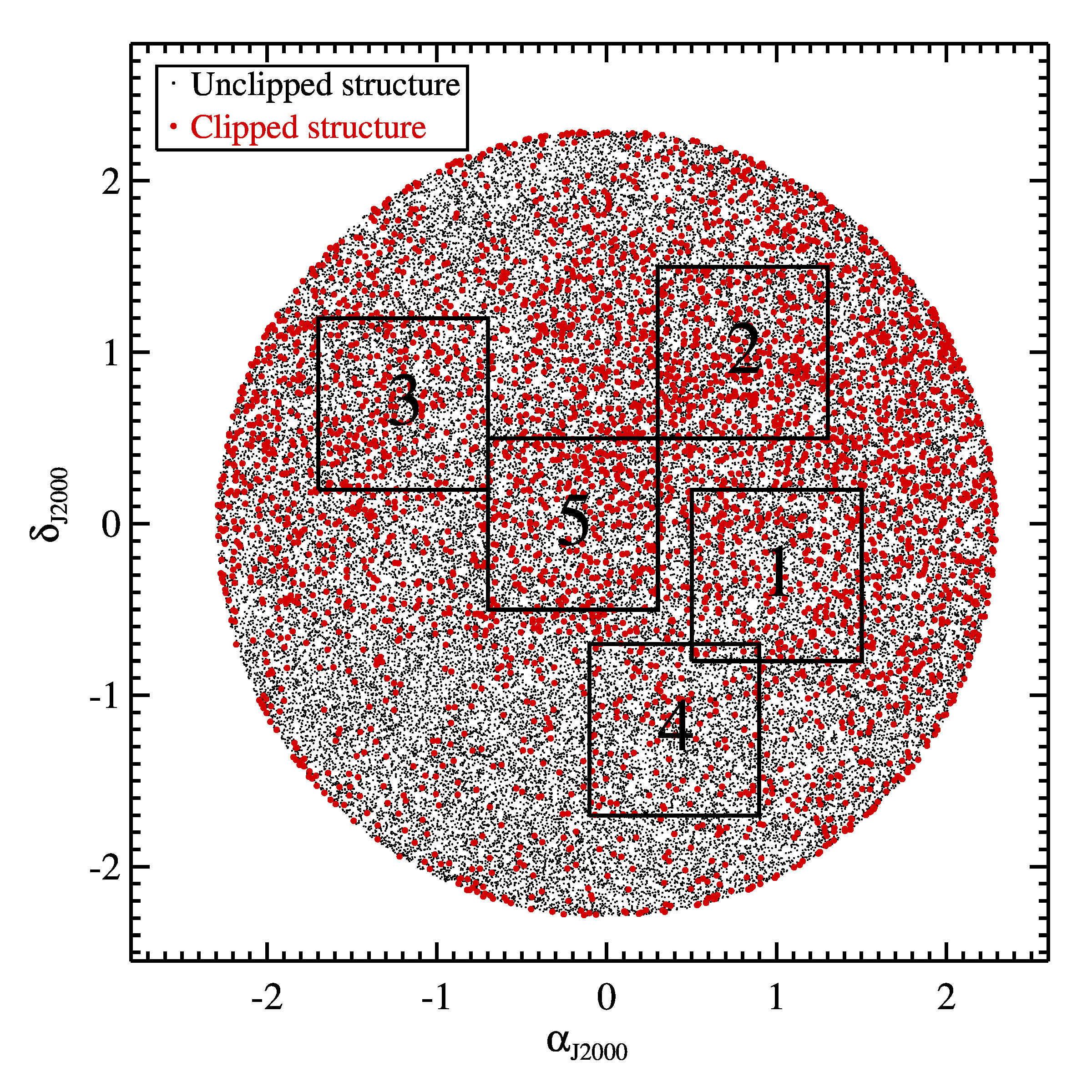}
\caption{\emph{Top Left:} $M_{z=0}$--richness relation for all structures in the GAEA lightcone. Member galaxies are assigned following the process described in Section \ref{sec.lc_const} and are limited to galaxies brighter than IRAC1$<$25.5. Protostructures identified as being artificially under-rich due to boundary effects in the lightcone, represented as red filled circles, are clipped following the method described in Section \ref{sec.lc_subsume} and constitute 0.6\% of the total structures in the lightcone. \emph{Top right:} Redshift histogram of all structures in the GAEA lightcone (black line) and an arbitrarily scaled histogram of all structures identified as being artificially under-rich (red line). Obvious clustering in redshift space is observed in the distribution of the clipped protostructures, which is a consequence of the stitching process used to create the lightcone. \emph{Bottom:} ``Sky'' plot of all protostructures in the GAEA lightcone (small black points) and all clipped protostructures (filled red circles). The regions corresponding to the five mock fields used in this work are indicated. Many of the clipped structures appear at the edge of the lightcone, however, a non-negligible fraction ($\sim$25\% of all clipped protostructures) appear in the mock regions.}
\label{fig:clipped_struct}
\end{figure*}

In total, the clipped protostructures comprise only $\sim$0.6\% of the total protostructure catalog. Note that this process will also clip protostructures that are deficient in galaxies brighter than IRAC1$<$25.5, which will potentially induce some bias in our results, as, e.g., our completeness numbers may be overestimated for massive protostructures that are observed to be under-rich in galaxies \citep[e.g.,][]{Newman22}. However, such protostructures appear, at least in our lightcones, to be rare (see also the simulation results presented in \citealt{Newman22}). Figure \ref{fig:clipped_struct} shows the $M_{z=0,\rm{tot}}$--richness relation for all simulated protostructures as well as those that are identified as artificially under-rich. Also plotted are the R.A., decl., and redshift histogram of all simulated protostructures along with those that are identified as artificially under-rich.

\end{document}

%% file: catalog_short.tex
\begin{deluxetable*}{ccccccc}
\tablecolumns{7}
\tablecaption{C3VO Candidate Protocluster Catalog}
\label{tab:catalog}
\tablehead{
\colhead{Field} & \colhead{Candidate ID\tablenotemark{a}} & \colhead{$N$ Peaks\tablenotemark{b}} & \colhead{R.A.} & \colhead{Decl.} & \colhead{$z$} & \colhead{Amp.\tablenotemark{c}} \\
\colhead{} & \colhead{} & \colhead{} & \colhead{[$^{\circ}$]} & \colhead{[$^{\circ}$]} & \colhead{} & \colhead{}
}
\startdata
COSMOS & C\textunderscore SP1 & 1 & 150.1993 & 2.3781 & 2.462 & 2013 $\pm$ 32 \\ 
COSMOS & C\textunderscore SP2 & 1 & 150.0125 & 2.1800 & 2.432 & 1640 $\pm$ 18 \\ 
COSMOS & C\textunderscore SP3 & 1 & 150.3551 & 2.3534 & 4.573 & 1179 $\pm$ 14 \\ 
COSMOS & C\textunderscore MP5$\ast$ & 1 & 150.3451 & 2.6811 & 2.750 & 559 $\pm$ 47 \\ 
\cutinhead{...}
CFHTLS-D1 & V\textunderscore MP1 & 1 & 36.3824 & -4.2238 & 3.754 & 2183 $\pm$ 427 \\ 
CFHTLS-D1 & V\textunderscore MP2 & 1 & 36.4027 & -4.2737 & 3.884 & 1739 $\pm$ 84 \\ 
CFHTLS-D1 & V\textunderscore SP2 & 1 & 36.6969 & -4.3563 & 3.292 & 1574 $\pm$ 27 \\ 
\cutinhead{...}
ECDFS & E\textunderscore SP1 & 1 & 53.1419 & -27.8795 & 3.469 & 1731 $\pm$ 58 \\ 
ECDFS & E\textunderscore SP2 & 1 & 53.1926 & -27.9161 & 3.080 & 1304 $\pm$ 18 \\ 
ECDFS & E\textunderscore SP3 & 1 & 53.0948 & -27.8190 & 2.813 & 1279 $\pm$ 17 \\ 
\enddata
\tablenotetext{a}{``MP'' and ``SP'' refer to Gaussian fits originating from multiple and single peak chains, respectively. The Gaussian amplitudes are less reliable in the case of multiple peak chains due to the blended flux values between neighboring peaks (see Fig. \ref{fig:long_linked_chain}).}
\tablenotetext{b}{Number of peaks subsumed together within a transverse radius of 2 Mpc and $\Delta z < 0.04$.}
\tablenotetext{c}{Amplitude of the Gaussian fit to the linked candidate. If the candidate consists of multiple peaks, the reported amplitude is the summed total of the amplitudes of the constituent peaks.}
\tablecomments{Candidates within 2 Mpc of the boundary of our search region are denoted by the $\ast$ symbol.}
\end{deluxetable*}

%% file: catalog_peaks_short.tex
\begin{deluxetable*}{cccccccc}
\tablecolumns{8}
\tablecaption{C3VO Overdensity Peaks Catalog}
\label{tab:peaks}
\tablehead{
\colhead{Field} & \colhead{Peak ID\tablenotemark{a}} & \colhead{Candidate ID\tablenotemark{b}} & \colhead{R.A.} & \colhead{Decl.} & \colhead{$z$} & \colhead{$\sigma_z$\tablenotemark{c}} & \colhead{Amp.\tablenotemark{d}}\\
\colhead{} & \colhead{} & \colhead{} & \colhead{[$^{\circ}$]} & \colhead{[$^{\circ}$]} & \colhead{} & \colhead{} & \colhead{}
}
\startdata 
COSMOS & $6$ & C\textunderscore SP182 & 150.3024 & 2.1370 & 2.012 $\pm$ 0.000 & 0.003 $\pm$ 0.000 & 93 $\pm$ 7 \\ 
COSMOS & $7$ & C\textunderscore SP62 & 149.9838 & 2.0783 & 2.012 $\pm$ 0.000 & 0.005 $\pm$ 0.000 & 192 $\pm$ 7 \\ 
COSMOS & $0$ & C\textunderscore SP213 & 149.9300 & 1.9182 & 2.018 $\pm$ 0.001 & 0.010 $\pm$ 0.002 & 71 $\pm$ 4 \\ 
COSMOS & $1$$\ast$ & C\textunderscore SP38 & 150.4395 & 2.3726 & 2.024 $\pm$ 0.000 & 0.010 $\pm$ 0.000 & 265 $\pm$ 6 \\ 
\cutinhead{...}
CFHTLS-D1 & $0$ & V\textunderscore SP67$\ast$ & 36.1684 & -4.8729 & 2.008 $\pm$ 0.002 & 0.004 $\pm$ 0.001 & 90 $\pm$ 9 \\ 
CFHTLS-D1 & $75$ & V\textunderscore SP84$\ast$ & 36.1129 & -4.4638 & 2.038 $\pm$ 0.001 & 0.007 $\pm$ 0.001 & 63 $\pm$ 4 \\ 
CFHTLS-D1 & $125$ & V\textunderscore SP106 & 36.6670 & -4.2870 & 2.046 $\pm$ 0.001 & 0.007 $\pm$ 0.002 & 48 $\pm$ 4 \\ 
\cutinhead{...}
ECDFS & $0$ & E\textunderscore SP21 & 53.1293 & -27.8211 & 2.005 $\pm$ 0.010 & 0.010 $\pm$ 0.008 & 168 $\pm$ 47 \\ 
ECDFS & $5$$\ast$ & E\textunderscore SP71 & 52.9692 & -27.8606 & 2.014 $\pm$ 0.011 & 0.014 $\pm$ 0.013 & 75 $\pm$ 10 \\ 
ECDFS & $22$$\ast$ & E\textunderscore SP48 & 53.0669 & -27.7158 & 2.025 $\pm$ 0.009 & 0.011 $\pm$ 0.010 & 110 $\pm$ 15 \\ 
\enddata
\tablenotetext{a}{Peaks from multiple peak chains are denoted by a superscript, which are numbered by the relative detection strength of the peak in the chain ordered from high to low.}
\tablenotetext{b}{Name of the protostructure candidate that the peak was assigned. The naming conventions of candidates follow those of Table \ref{tab:catalog}.}
\tablenotetext{c}{Redshift dispersion of the Gaussian fit to the linked candidate. Uncertainties are the formal random error derived from the covariance matrix of the fit. Value is less certain in cases where peak is embedded in a multiple peak chain.}
\tablenotetext{c}{Amplitude of the Gaussian fit to the linked candidate. Uncertainties are the formal random error derived from the covariance matrix of the fit. Value is less certain in cases where peak is embedded in a multiple peak chain.}
\tablecomments{Peaks and candidates within 2 Mpc of the boundary of our search region are denoted by the $\ast$ symbol.}
\end{deluxetable*}